\documentclass[11pt]{article}

\usepackage[margin=1in]{geometry}
\usepackage{amsmath,amssymb,amsfonts,bm,mathtools}
\usepackage{natbib}
\usepackage{booktabs}
\usepackage{enumitem}
\usepackage{hyperref}
\usepackage{xcolor}
\usepackage{graphicx}
\usepackage{subcaption}
\usepackage{algorithm}
\usepackage{algpseudocode}
\usepackage{authblk}

\hypersetup{
  colorlinks=true,
  linkcolor=blue,
  citecolor=blue,
  urlcolor=blue
}

\newcommand{\thetaVec}{\bm\theta}
\newcommand{\yVec}{\bm y}

\newcommand{\obs}{\mathrm{obs}}
\newcommand{\Normal}{\mathcal N}
\newcommand{\E}{\mathbb E}
\newcommand{\Var}{\mathbb V\mathrm{ar}}

\newcommand{\R}{\mathbb R}
\newcommand{\given}{\mid}
\newcommand{\eps}{\varepsilon}

\newcommand{\wt}{\widetilde}
\newcommand{\yobs}{\yVec_{\obs}}

\newcommand{\phiVec}{\boldsymbol{\phi}}

\title{Amortized ratio-estimation importance sampling and localized simulation-based calibration for intractable likelihoods}
\author{Umberto Picchini}
\affil{Department of Mathematical Sciences, Chalmers University of Technology and University of Gothenburg, SE-412 96 Gothenburg, Sweden}
\date{}

\begin{document}
\maketitle

\begin{abstract}
We consider simulation-based Bayesian inference (SBI) for the parameters of models with intractable likelihoods but tractable forward simulation. Building on Gaussian mixtures-of-experts surrogates, as a first contribution we develop a sequential procedure to estimate the posterior, incorporating an importance sampling step that introduces a locally amortized likelihood-to-evidence ratio estimator. The local amortization allows the ratio estimator to be evaluated at new
parameter--data pairs from a local (data-dependent) design, without re-estimation, which is
particularly convenient for our second contribution.
The second contribution is localized simulation-based calibration (SBC). Localized SBC provides a graphical diagnostic of calibration over neighborhoods that are deliberately broader than the available posterior, at low additional computational cost. Rather than repeatedly running the full inference procedure for each simulated pseudo-observation, localized SBC fits the local surrogate and ratio estimator only once for each local design, and reuses them across all  pseudo-observations. This approach is particularly appealing for SBI applications.

We evaluate both contributions on three case studies, including a real-data epidemiological application.
\end{abstract}

\section{Introduction}

We consider a simulation-based framework for parameter inference when the likelihood function is intractable, but simulation from the assumed data-generating model (DGM) is possible.
Let \(\thetaVec\in\Theta\subseteq\R^{d_\theta}\) denote model parameters, \(\yVec\in\mathcal Y\subseteq\R^{d_y}\) the output generated by the DGM simulator, and \(\yobs\) the observed data.  We assume that sampling
$
  \yVec\sim p(\yVec\given\thetaVec)
$
is possible, where $p(\yVec\given\thetaVec)$ denotes the likelihood function, while pointwise evaluation of \(p(\yVec\given\thetaVec)\) is unavailable or impractical.  The inferential target is the posterior
\begin{equation}
     p(\thetaVec\given\yobs)
  \propto p(\thetaVec)p(\yobs\given\thetaVec), 
  \label{eq:exact-posterior}
\end{equation}
and the intractability of the likelihood makes exact sampling from \eqref{eq:exact-posterior} typically unattainable, with some exceptions such as when a non-negative and unbiased approximation of the likelihood is available \citep{andrieu2009pseudo,andrieu2010particle}. For the case where simulation of data is feasible, but not the evaluation of the likelihood,  a large amount of literature has been produced within the simulation-based inference (SBI) context; see \cite{cranmer} and \cite{pesonen2023abc} for  reviews. Initially, much of the attention was directed to  approximate Bayesian computation (ABC) \citep{marin,sisson2018handbook}, synthetic likelihoods \citep{wood2010statistical,price2018bayesian}, and specific instances of pseudomarginal methods, when simple forward simulation is possible (as when the bootstrap filter is used to unbiasedly approximate the likelihood). The mentioned approaches have also been denoted as ``statistical SBI'' in \cite{wang2024comprehensive}, to distinguish those from more recent methods exploiting neural networks (typically normalizing flows) to approximate conditional densities, sometimes denoted ``neural conditional density estimation'' (NCDE). NCDE has been used to approximate likelihoods (\citealp{papamakarios2019,chen2021neural}), posterior distributions \citep{papamakarios2016,greenberg2019,durkan2020contrastive,chen2021neural,miller2021truncated,delaunoy2022towards}, or the likelihood and the posterior simultaneously \citep{wiqvist2021sequential,radev2023jana}.

The present work concerns statistical SBI and makes two main contributions.
The first is an approach for learning both likelihood and posterior
approximations from amortized surrogate models expressed as mixtures of
experts. Starting from a broadly trained surrogate, the procedure
sequentially localizes the approximation to the observed data $\yobs$,
yielding increasingly specialized surrogate likelihoods and posteriors.
The final posterior surrogate is then corrected by importance sampling
using a \textit{locally amortized} estimate of a likelihood-to-evidence ratio.
Likelihood-to-evidence ratio estimation by classification has previously
been considered in SBI, for example in
\cite{cranmer2015approximating,gutmann2018likelihood,thomas2022likelihood}. However our construction differs in that a single classifier is fitted jointly over a local sampler
$\thetaVec\sim g(\thetaVec\mid\yobs)$ and data-derived features. Because the
ratio approximation is locally amortized, it can
subsequently be evaluated at arbitrary proposal draws from that local
sampler. This contrasts with the linear LFIRE method of
\cite{thomas2022likelihood}, where a separate classifier is fitted at each
parameter value $\thetaVec$ using repeated model simulations conditional on
that value. Once our classifier has been fitted, the estimated ratio can be
evaluated for additional proposal draws without refitting the classifier or
performing additional model simulations.

The second contribution is a novel method for simulation-based calibration (SBC), which we call \textit{localized SBC}. This method can probe whether calibration is preserved
not only near the bulk of the reported posterior, but also over a wider region
of parameter space. 
This is done by checking calibration with parameters generated by a local design $\thetaVec\sim g_\lambda(\thetaVec\mid \yobs)$, where $\lambda\geq 1$ is a parameter set by the statistician to induce pseudo-true parameters to be sampled from a broader version (when $\lambda>1$) of the available posterior, or from the posterior itself (when $\lambda=1$), and the latter case corresponds to the \textit{posterior} SBC recently introduced in \cite{sailynoja2026posterior}. This construction is general and is not limited to SBI. However, it is
particularly useful in SBI, where inference is typically based on approximate rather than exact posterior distributions. For example, it is well documented that some SBI methods can return overconfident (under-dispersed) posterior approximations; see \cite{hermans2022a}. Therefore, assessing whether calibration is preserved in a moderately wider neighborhood of the parameter space seems appropriate.
As hinted above, we also show that localized SBC generalizes the posterior SBC of \cite{sailynoja2026posterior}.  Moreover, localized SBC is computationally lightweight relative to rerunning the complete inference pipeline independently for every pseudo-observation,
because we develop a conditional surrogate that is locally amortized: it is fitted only once on
simulations generated under the local design and is then reused across all
pseudo-observations generated from that design. This  makes the calibration
check relatively inexpensive in the current setting.

The ratio-corrected importance sampling we present as our first contribution shares similarities with the SeMPLE algorithm in \cite{haggstrom2024fast}, however, SeMPLE uses an MCMC approach for sampling. Here we motivate why we decided to consider a different route. 
The methodology in \cite{haggstrom2024fast} considered a surrogate generative model defined as a Gaussian locally linear map (GLLiM, \cite{deleforge}) with $K$ Gaussian components, which is the route we follow also in this work. Their resulting SeMPLE (Sequential Mixture Posterior and Likelihood Estimation) used mixtures of experts fitted via expectation-maximization (EM) to first determine amortized surrogates of the likelihood $p(\yVec\given \thetaVec)$ and of the posterior $p(\thetaVec\given  \yVec)$ for a generic $\yVec$. Then, with a sequence of estimation rounds, the surrogates were ``localized'' to the observed $\yobs$. The surrogate learning used training data given by parameter-data pairs, where the parameters were generated via Metropolis-Hastings (MH). Due to the accept/reject nature of MH, SeMPLE requires a sufficient number of posterior draws to preserve variety in the training data. Poor variety in the training data would affect the surrogate learning at the next round. This was also crucial in terms of exploration of all possible $K$ modes in the target.

In this work, we completely avoid MCMC sampling, and replace it with two rounds involving an importance-sampling correction step, with an estimation of the likelihood-to-evidence ratio, which we construct to be locally amortized as previously mentioned. The ratio-correction approach also involved the construction of a logistic regression classifier, and the identification of suitable data-summaries to be used as covariates in the classifier. The advantage of using an importance-sampling based approach, rather than MCMC, is threefold: 1) the resulting simulations contain independent proposal draws, rather than correlated samples affected by rejection, these reducing the effective diversity of the training data; 2) the strategy can be implemented in an embarrassingly parallel fashion; 3) it makes the computation of the novel localized SBC strategy straightforward, thanks to the locally amortized ratio estimation, for which we would otherwise need a separate MCMC chain for every simulated pseudo-true parameter.  

The paper is organized as follows: Section \ref{sec:gllim} summarizes inference for Gaussian locally linear maps (GLLiM) and the initial amortized surrogates for the likelihood and posterior; Section \ref{sec:tworound} constructs a sequential two-rounds learning for the surrogate posterior and surrogate likelihood; Section \ref{sec:ratio-correction} constructs a locally amortized ratio-correction for importance sampling; Section \ref{sec:local-sbc} builds the localized SBC approach and compares it to posterior SBC; Section \ref{sec:numerical-studies} presents results on three experiments: SIR model, Lotka-Volterra model and a real-data epidemiological model. Further methodological and experimental results are in Supplementary Material. Code to reproduce results is available at \url{https://github.com/umbertopicchini/ratioIS-locSBC}.

\section{GLLiM surrogates}\label{sec:gllim}

In the context of ``inverse regression'', Gaussian locally linear maps (GLLiM, \citealp{deleforge}) model the joint behavior of a low-dimensional variable, which in our context is represented by \(\thetaVec\), and a possibly high-dimensional simulator output, \(\yVec\) in our context, by introducing a latent mixture label \(Z\in\{1,\ldots,K\}\). More specifically, GLLiM is a mixture-of-experts (MoE) with $K$ Gaussian components. MoE are generalizations of neural network architectures proposed by \cite{jacobs1991adaptive}, and  their flexibility comes
from the fact that they allow the mixture weights (also called ``gating functions'') to depend on explanatory variables, together with the component densities (or the
experts). This is unlike classic mixture models, where mixture weights are constant. In the context of regression, MoE models with Gaussian experts and soft-max or normalized Gaussian gating functions (as in GLLiM) are the most popular choices. MoE are flexible conditional density approximators: with sufficiently many components, they can approximate a broad class of conditional distributions. Another useful feature of MoE is that they can be fitted to data using expectation-maximization (EM, \citealp{jordan1994hierarchical,xu1994alternative}), and in the particular MoE we consider, GLLiM, the complete-data model has a conjugate/algebraic structure allowing for closed-form EM updates, that is no optimization is required, only iterating between the E and the M steps.

 Background on the GLLiM construction for SBI can be found in \cite{forbes} and \cite{haggstrom2024fast} and, for ease of reading, some details are in Supplementary Material for the interested reader. Here it suffices to know that a GLLiM  defines a joint  Gaussian mixture $q_{\tilde{\phiVec}}(\yVec, \thetaVec)$ on the pair data-parameter $(\yVec,\thetaVec)$, where $\yVec$ is a simulation from the DGM $\mathcal{M}(\thetaVec)$ as $\yVec\sim p(\yVec|\thetaVec)$ (recall in our context $p(\yVec|\thetaVec)$ is intractable). Specifically, we have
\begin{equation*}
     q_{\tilde{\phiVec}}(\yVec, \thetaVec) = \sum_{k=1}^K  q_{\tilde{\phiVec}}(\yVec \!\given\! \thetaVec, z\!=\!k)  q_{\tilde{\phiVec}}(\thetaVec \!\given\! z\!=\!k)  q_{\tilde{\phiVec}}(z\!=\!k),
\end{equation*}
where 
$
    \tilde{\phiVec} = \{\pi_k, \tilde{\boldsymbol{\nu}}_k, \tilde{\boldsymbol \Gamma}_k, \tilde{\boldsymbol A}_k, \tilde{\boldsymbol b}_k, \tilde{\boldsymbol \Sigma}_k\}_{k=1}^K
$ is the full vector of mixture model parameters.
Given this joint distribution, we can easily deduce the conditional distributions $q_{\tilde \phiVec}(\yVec \given \thetaVec)$ and $q_{\phiVec}(\thetaVec \given \yVec)$ in closed-form. First, we have the surrogate likelihood
\begin{equation}
    \label{eq:surr_lik}
    q_{\tilde{\phiVec}}(\yVec \given \thetaVec) = \sum_{k=1}^{K} \tilde{\omega}_k(\thetaVec) \mathcal{N}(\yVec; \tilde{\boldsymbol{A_k}} \thetaVec + \tilde{\boldsymbol{b}}_k, \tilde{\boldsymbol{\Sigma}}_k),
\end{equation}
with
$
    \tilde{\omega}_k(\thetaVec) = \pi_k \mathcal{N}_l(\thetaVec; \tilde{\boldsymbol{\nu}}_k, \tilde{\boldsymbol{\Gamma}}_k)/(\sum_{j=1}^{K} \pi_j \mathcal{N}_l(\thetaVec; \tilde{\boldsymbol{\nu}}_j, \tilde{\boldsymbol{\Gamma}}_j))$, where we used $\Normal(\bm{\mu},\bm{\Sigma})$ to denote the multivariate Gaussian distribution with mean $\bm{\mu}$ and covariance matrix $\bm{\Sigma}$.
The parameters $\tilde{\phiVec}$ of the surrogate likelihood can be estimated via EM using training data (details are described later), and an important feature of GLLiM is that once an estimate for $\tilde{\phiVec}$ is obtained, it is possible to obtain immediately the parameters $\phiVec = \{\pi_k, \boldsymbol \nu_k, \boldsymbol \Gamma_k, \boldsymbol A_k, \boldsymbol b_k, \boldsymbol \Sigma_k\}_{k=1}^K$ of the surrogate posterior $q_{{\phiVec}}(\thetaVec \given \yVec)$, in closed form following algebraic relationships from \cite{deleforge}, also given in Supplementary Material. Therefore, the same training data gives estimates for both surrogate conditionals. 
The surrogate posterior is given by 
\begin{equation}
    \label{eq:surr_post}
    q_{{\phiVec}}(\thetaVec \given \yVec) = \sum_{k=1}^{K} \omega_k(\yVec) \mathcal{N}_l(\thetaVec; \boldsymbol A_k \yVec + \boldsymbol b_k, \boldsymbol \Sigma_k),
\end{equation}
with
%\begin{equation}
%    \label{eq:post_prob}
%    \omega_k(\yVec) = \frac{\pi_k \mathcal{N}(\yVec; \boldsymbol \nu_k, \boldsymbol \Gamma_k)}{\sum_{j=1}^{K} \pi_j \mathcal{N}(\yVec; \boldsymbol \nu_j, \boldsymbol \Gamma_j)}.
%\end{equation}
$
    \omega_k(\yVec) = \pi_k \mathcal{N}(\yVec; \boldsymbol \nu_k, \boldsymbol \Gamma_k)/(\sum_{j=1}^{K} \pi_j \mathcal{N}(\yVec; \boldsymbol \nu_j, \boldsymbol \Gamma_j)).
$
Notice that it is critical to first estimate $\tilde{\phiVec}$ and then deduce ${\phiVec}$, rather than the other way around (which is also possible): the reason is in terms of computational complexity. If the surrogate generative model was formulated in the opposite direction, with ``responses'' $\thetaVec$ and input/covariates $\yVec$, then the number of parameters to infer would be much larger, see Remark 2.1 in \cite{nguyen2024bayesian} for an example. 

Given this construction, in Section \ref{sec:tworound} we produce three levels of increasingly refined ``proposal functions'' for importance sampling, where the last proposal is finally ``corrected'' to provide an approximation to $p(\thetaVec|\yVec)$ (and hence to $p(\thetaVec|\yobs)$ once we plug $\yobs$ into $\yVec$). This is novel compared to the above mentioned references, and we detail the correction mechanism, as well as the defensive proposal sampling that we employ in our experiments. Then, in Section \ref{sec:local-sbc}, we introduce the novel \textit{localized} simulation-based calibration.

\section{Surrogates construction}
\label{sec:tworound}

In what follows, \(q_0\), \(q_1\), and \(q_2\) denote conditionals of the form \eqref{eq:surr_post}, fitted on  simulated training data that are progressively more compatible with $\yobs$. Optionally, should the user decide to provide a low-dimensional data representation for the simulated $\yVec$, as well as for the observed $\yobs$, that is when simulated/observed data are ``summarized'' via a function $S(\yVec)$, in the following we will still write $\yVec$ and $\yobs$  regardless of the application of a summary function.  Thus the method targets \(p(\thetaVec\given S(\yobs))\) when a summary map \(S\) is supplied, and \(p(\thetaVec\given\yobs)\) when \(S(\yVec)=\yVec\). We leave this specification to the user and will not refer to $S$ in the following, except when necessary to clarify.
The construction uses three simulated training data.  The first uses prior-predictive simulations, and produces \(q_{\hat{\phiVec}_0}(\thetaVec|\yVec)\).  The second one is sampled (with a defensive mechanism) from \(q_{\hat{\phiVec}_0}(\thetaVec\given\yobs)\) and produces \(q_{\hat{\phiVec}_1}(\thetaVec|\yVec)\).  The third one is sampled (again with a defensive mechanism) from \(q_{\hat{\phiVec}_1}(\thetaVec\given\yobs)\) and produces \(q_{\hat{\phiVec}_2}(\thetaVec|\yVec)\). Each of the estimates $\hat{\phiVec}_0$, $\hat{\phiVec}_1$ and $\hat{\phiVec}_2$ is algebraically obtained from the corresponding estimates of $\tilde{\phiVec}_0$, $\tilde{\phiVec}_1$ and $\tilde{\phiVec}_2$, as detailed in Section \ref{sec:gllim}. To obtain the three $\tilde{\phiVec}$ estimates we employed  the EM optimization routine provided in the R \texttt{xLLiM} package \citep{xllim}. 

In the following, for ease of notation we write \(\hat{q}_r(\thetaVec\given\yVec)\) instead of \(q_{\hat{\phiVec}_r}(\thetaVec\given\yVec)\), where $r=0,1,2$, and use $\bm{m}_{rk}=\bm{A}_{rk}\yVec+\bm{b}_{rk}$ to denote the theoretical mean of the $k$-th posterior component (see eq. \eqref{eq:surr_post}) for the posterior surrogate at the $r$-th round.
Finally, we denote the training data for $q_1$ and $q_2$ with \((\thetaVec_a,\yVec_a)\) and \((\thetaVec_b,\yVec_b)\) respectively.\\

\noindent
\textbf{Prior-predictive fit: construction of \(q_0\).}
We start by obtaining $N_0$ prior predictive samples as
\begin{equation}
  \thetaVec_0^{(i)}\sim p(\thetaVec),
  \qquad
  \yVec_0^{(i)}\sim p(\yVec\given\thetaVec_0^{(i)}),
  \qquad i=1,\ldots,N_0.
\end{equation}
Then we fit \eqref{eq:surr_lik} to
$
  \mathcal D_0=\{(\thetaVec_0^{(i)},\yVec_0^{(i)})\}_{i=1}^{N_0}$ via EM, to obtain an estimate $\hat{\tilde{\phiVec}}_0$ of $\tilde{\phiVec}_0$, and consequently an estimate $\hat{\phiVec}_0$ of ${\phiVec}_0$, using the relations in Supplementary Material. The $\hat{\phiVec}_0$ are then used to define the initial posterior surrogate
$
  \hat{q}_0(\thetaVec\given\yVec)
  = \sum_{k=1}^K \hat{\omega}_{0k}(\yVec)
    \Normal\{\thetaVec;\hat{\bm{m}}_{0k}(\yVec),\hat{\bm{\Sigma}}_{0k}\}.
$
When the observation $\yobs$ is plugged in place of $\yVec$, then \(\hat{q}_0(\thetaVec\given\yobs)\) is an initial surrogate posterior approximation based on observed data, which should be further refined since it has been learned from simulations generated by a potentially diffuse prior, that is $p(\thetaVec)$ may be very different from the true $p(\thetaVec|\yobs)$. Notice that the optimized parameters $\hat{\tilde{\phiVec}}_0$ are used to hot-start the optimization of  ${\tilde{\phiVec}}_1$ in the next round.\\

\noindent
\textbf{First local fit: construction of \(q_1\).} We now construct the first ``local'' fit, meaning that the next surrogate $q_1$ will be learned from training data containing information about $\yobs$, unlike the prior-predictive training data.
The first localized parameters are drawn ``defensively'' $N_a$ times from a mixture of the initial surrogate and the prior:
\begin{equation}
  \thetaVec_a^{(i)} \sim
  (1-\eps_1)\hat{q}_0(\thetaVec\given\yobs)
  + \eps_1 p(\thetaVec),
  \qquad i=1,\ldots,N_a.
  \label{eq:stage1-design}
\end{equation}
Simulation from \eqref{eq:stage1-design} is trivial: first simulate $u^*\sim U(0,1)$ from a uniform distribution, and if $u^*<\varepsilon_1$ then $\thetaVec_a^{(i)}$ is sampled from the prior $p(\thetaVec)$, and otherwise it is sampled from  $\hat{q}_0(\thetaVec\given\yobs)$.
The constant $\varepsilon_1>0$ is a small defensive term that is not intended to dominate the generation of the localized training data, but is meant to protect against the premature collapse of \(\hat{q}_0\) (for example if $\hat{q}_0(\thetaVec|\yobs)$ is too narrowly located around a posterior mode) by introducing a small number of draws from the prior to help preserve support in regions that may still be relevant for the posterior.
For each \(\thetaVec_a^{(i)}\), we simulate
$\yVec_a^{(i)}\sim p(\yVec\given\thetaVec_a^{(i)})$, then
fit the  surrogate \eqref{eq:surr_lik} anew to
$
  \mathcal D_a=\{(\thetaVec_a^{(i)},\yVec_a^{(i)})\}_{i=1}^{N_a}
$ by using $\hat{\tilde{\phiVec}}_0$ to hot-start the optimization of ${\tilde{\phiVec}}_1$, and then obtain $\hat{\tilde{\phiVec}}_1$ and consequently $\hat{{\phiVec}}_1$ (again algebraically). The latter defines the further refined surrogate
\[
  \hat{q}_1(\thetaVec\given\yVec)
  = \sum_{k=1}^K \hat{\omega}_{1k}(\yVec)
    \Normal\{\thetaVec;\hat{\bm{m}}_{1k}(\yVec),\hat{\bm{\Sigma}}_{1k}\}.
\]
\\

\noindent
\textbf{Second local fit: construction of \(q_2\).} The final training set is generated from \(\hat{q}_1(\thetaVec|\yobs)\), again with a prior-defensive component using a small constant $\varepsilon_2>0$:
\begin{equation}
  \thetaVec_b^{(i)} \sim
  (1-\eps_2)\hat{q}_1(\thetaVec\given\yobs)
  + \eps_2 p(\thetaVec),
  \qquad i=1,\ldots,N_b.
  \label{eq:stage2-design}
\end{equation}
Then as usual,
$
  \yVec_b^{(i)}\sim p(\yVec\given\thetaVec_b^{(i)})$,
$i=1,\ldots,N_b,
$
and the surrogate likelihood is fitted to
$
  \mathcal D_b=\{(\thetaVec_b^{(i)},\yVec_b^{(i)})\}_{i=1}^{N_b}
$. For this second local fit, the GLLiM optimization is initialized
independently rather than hot-started from $\hat{\tilde{\phiVec}}_1$, as for this surrogate we obtained better results with a cold-start.
The resulting estimate is $\hat{\tilde{\phiVec}}_2$, from which
$\hat{\phiVec}_2$ is obtained algebraically, yielding
\[
  \hat{q}_2(\thetaVec\given\yVec)
  = \sum_{k=1}^K \hat{\omega}_{2k}(\yVec)
    \Normal\{\thetaVec;\hat{\bm{m}}_{2k}(\yVec),\hat{\bm{\Sigma}}_{2k}\}.
\]
The resulting conditional density is not yet probabilistically valid to approximate $p(\thetaVec|\yobs)$, since it has been trained using parameters that have not been generated from the prior. A correction step is therefore necessary, which will modify $\hat{q}_2$, and this is described in Section \ref{sec:ratio-correction}. However, it is worth stating that further simulations from the DGM, that is simulations of the type $\yVec\sim p(\yVec|\thetaVec)$, will not be necessary towards obtaining posterior inference for $\thetaVec$: that is, the 
 total simulator budget is
$N_{\mathrm{sim}}=N_0+N_a+N_b$. Further model simulations may be produced for other purposes, such as running simulation-based calibration or posterior predictive checks.

Notice that, throughout this work, parameter values generated from
a surrogate proposal are required to lie within the support of the prior. More
precisely, whenever a proposed value $\thetaVec$ satisfies
$p(\thetaVec)=0$, it is discarded and a new value is generated; sampling
continues until the prescribed number  of admissible parameter values  has
been obtained. Thus, whenever we refer to draws from a surrogate, or from a
defensive mixture involving a surrogate, the corresponding sampling
distribution is implicitly restricted to the support of the prior.

\section{Locally amortized ratio correction}
\label{sec:ratio-correction}

As described in Section \ref{sec:tworound}, our method is sequential, and learns first an amortized conditional posterior $q_0(\thetaVec|\yVec)$ (altogether with an amortized likelihood), but then it learns two localized (observation dependent) approximations. The ratio estimator developed below is ``locally amortized'': this means that it is trained once
on parameter--simulation pairs generated from the final localized design, and
is then reused to evaluate the correction at arbitrary proposal draws from
that local region.
The surrogate \(\hat{q}_2(\thetaVec\given\yobs)\) is considered a proposal function towards obtaining final inference, and not as the object returning such inference without applying a correction to its output. For example, this has been discussed in \cite{papamakarios2016}, \cite{greenberg2019}, and in \cite{haggstrom2024fast}. In this work, for the correction we consider the ``ratio estimation'' approach, namely the estimation of the likelihood-to-evidence ratio $p(\yVec\given \thetaVec)/p(\yVec)$, which in SBI has been considered in e.g. \cite{cranmer2015approximating,gutmann2018likelihood,thomas2022likelihood}. Our goal is to estimate the ratio in an amortized way, though the amortization is restricted to training data depending on $\yobs$ ( hence ``locally amortized''), so that we can estimate it once and for all and reuse it to importance-correct all draws $\thetaVec^{(i)}\sim \hat{q}_2(\thetaVec\given \yobs)$. Resampled draws from an appropriately weighted set $\{\thetaVec^{(i)},w^{(i)}\}_i$ will then return the final posterior inference. The way we approach the ratio-estimation problem shares similarities with the
(linear) LFIRE method of \cite{thomas2022likelihood}, but also important
differences. In linear LFIRE, a separate classifier is fitted at each parameter
value, using repeated model simulations conditional on that value. In our
construction, a single classifier is instead trained jointly over parameters
from a localized design and data-derived features, and is subsequently reused
to evaluate the ratio at many proposal draws without refitting the classifier
or performing additional model simulations. These and other differences are
discussed in Section~\ref{sec:importance-reweight}.

Here we first give a summary of the ratio-estimation approach, before specializing it to our context.
The posterior conditioned on \(\yobs\) is expressed as
\begin{equation}
  p(\thetaVec\given \yobs)
  = p(\thetaVec) r(\thetaVec,\yobs),
  \quad \text{ with } \quad
  r(\thetaVec, \yVec)=\frac{p(\yVec\given\thetaVec)}{p(\yVec)},
  \label{eq:ratio-identity}
\end{equation}
where the ratio \(r\) can be estimated by binary classification that tries to distinguish between (i) simulated $\yVec^{(i)}\sim p(\yVec|\thetaVec)$ where the $(\thetaVec^{(i)},\yVec^{(i)})$ are generated from the \textit{joint pairs} model \(p(\thetaVec,\yVec)=p(\thetaVec)p(\yVec\given \thetaVec)\), and (ii) $\yVec^{(i)}\sim p(\yVec)$ where the $(\thetaVec^{(i)},\yVec^{(i)})$ are generated by the \textit{product-of-marginal pairs} model  \(p(\thetaVec,\yVec)=p(\thetaVec)p(\yVec)\). Say that a binary random variable $W\in\{0,1\}$ takes value 1 if and only if $(\thetaVec,\yVec)$ has been generated under the joint-pairs hypothesis (case (i)), and 0 otherwise. It is then possible to define the classifier's output as $D(\thetaVec,\yVec)=\mathbb{P}(W=1\given\thetaVec,\yVec)$, and hence $1-D(\thetaVec,\yVec)=\mathbb{P}(W=0\given\thetaVec,\yVec)$. Now, for simplicity we assume ``balanced classes'', that is $\mathbb{P}(W=1)=\mathbb{P}(W=0)$, then it is easy to show that  $r(\thetaVec,\yVec)=D(\thetaVec,\yVec)/(1-D(\thetaVec,\yVec))=p(\yVec\given \thetaVec)/p(\yVec)$, and such equality is derived in Supplementary Material. Therefore, working on the log-scale induces the familiar logit transformation 
\[\log r(\thetaVec,\yVec)=\log\biggl(\frac{D(\thetaVec,\yVec)}{1-D(\thetaVec,\yVec)}\biggr)=\log \frac{p(\yVec\given\thetaVec)}{p(\yVec)},\]
which can be learned through logistic regression using appropriate  training data and set of covariates, which will be specified precisely in Section \ref{sec:training_classifier}.
 For the sake of illustration, for the moment we denote with \(\mathcal {D}=\{(\thetaVec^{(i)},\yVec^{(i)})\}_{i=1}^{N_{\mathcal{D}}}\) a generic set of parameter-data pairs. Then we set the following
\begin{equation}
  \{(\thetaVec^{(i)},\yVec^{(i)},W^{(i)}=1)\}_{i=1}^{N_{\mathcal{D}}},\label{eq:joint_class}
\end{equation}
namely the entirety of \(\mathcal D\) is assigned to the joint class ($W=1$). To produce the product-of-marginals class, in $\mathcal{D}$ we permute the $\yVec^{(i)}$ (but not the $\thetaVec^{(i)}$), to obtain a new dataset
\begin{equation}
  \{(\thetaVec^{(i)},\yVec^{(\pi(i))},W^{(i)}=0)\}_{i=1}^{N_{\mathcal{D}}},\label{eq:product_class}
\end{equation}
where $\{\yVec^{(\pi(i))}\}_{i=1}^{N_\mathcal{D}}$ denotes a random permutation of $\{\yVec^{(i)}\}_{i=1}^{N_\mathcal{D}}$. Therefore, by permuting the $\yVec^{(i)}$'s while keeping the $\thetaVec^{(i)}$'s fixed to the original ordering, the association between  simulated data and the parameters generating them has been destroyed, making \eqref{eq:product_class} suitable for the class with $W=0$ (product class).
Ultimately, the classifier's training data is obtained by stacking \eqref{eq:joint_class} with \eqref{eq:product_class}, and has therefore $2N_\mathcal{D}$ rows.

We now go back to our specific scenario and adapt the ratio-estimation approach accordingly. What we are required to do is: (i) to precisely specify the training data for the classifier (the $\mathcal{D}$ above was a generic illustration);
(ii) to determine which covariates to use in the classifier.
Both points will be addressed later in Section \ref{sec:training_classifier}, because prior to perform classification, we need to perform data summarization, which will enter both the construction of the classifier's training data, and the covariates construction. 
We require a set of covariates for the classifier and we wish to learn $r(\thetaVec,\yVec)$ for arbitrary pairs of parameters and simulator outputs $\yVec$. To accomplish this, we first reduce each (possibly large-dimensional) $\yVec$ to a set of summaries $T(\yVec)$, which will be used as covariates for the logistic classifier altogether with the $\thetaVec$. For $T(\yVec)$ we use posterior summaries induced by the final localized surrogate $\hat{q}_2(\thetaVec\given \yVec)$, where $\yVec$ is a generic argument that can either accept simulated $\yVec^{(i)}$ values or the observed $\yobs$. Specifically, we use summaries that have been explored in \cite{forbes} within an ABC context, namely, posterior means, posterior variances and covariances, and mixture weights. In our work, we obtain these summaries from $\hat{q}_2(\thetaVec\given \yVec)$, that is
\begin{equation}
  T(\yVec)
  =
  \begin{pmatrix}
    \E_{\hat{q}_2}[\thetaVec\given \yVec]\\
    \mathrm{vech}\{\Var_{\hat{q}_2}(\thetaVec\given \yVec)\}\\
    \hat{\omega}_{21}(\yVec),\ldots,\hat{\omega}_{2K}(\yVec)
  \end{pmatrix}
  \label{eq:summaries}
\end{equation}
where $\mathrm{vech}(\mathbf{A})$ takes the lower-triangular part, including the diagonal, of a symmetric matrix $\mathbf{A}$, and $\hat{\omega}_{2k}$ represents the estimated $\omega_k(\yVec)$ when using $\hat{q}_2(\thetaVec\given \yVec)$.   As mentioned at the beginning of Section \ref{sec:tworound}, if a user-specified summary map \(S\) is supplied from the beginning of the inference procedure, the classifier can instead use \(S(\yVec)\) instead of $\yVec$, and therefore act on $T(S(\yVec))$. But again, to simplify the exposition we avoid referring to such $S(\cdot)$. Of course, other ways of constructing $T(\cdot)$ could be considered: data summarization in SBI is a heavily studied open problem, where summaries construction can be automated or be hand-crafted (e.g. \citealp{fearnhead2012constructing,blum2013comparative,wiqvist19a}). Here we picked the selection made in \cite{forbes} out of convenience, since these quantities are readily available. 
Also, notice that in practice the summaries are centered and scaled using the final local training data, before being used as covariates. Moreover, in \cite{forbes} functional summaries (hence infinite-dimensional objects) have also been explored, however for our logistic regression classifier we need finite dimensional summaries to be used as covariates, and hence we prefer to use  \eqref{eq:summaries}.
 
\subsection{Training data for the locally amortized classifier and choice of covariates}\label{sec:training_classifier}

We now specify the training data for the ratio-estimation classifier. In Section \ref{sec:ratio-correction}, when introducing the methodology in generic terms, we defined \(\mathcal {D}=\{(\thetaVec^{(i)},\yVec^{(i)})\}_{i=1}^{N_{\mathcal{D}}}\) as unspecified training data, however a possibility that we use in practice is taking  \(\mathcal {D}\equiv\mathcal {D}_2=\{(\thetaVec^{(i)}_b,\yVec^{(i)}_b)\}_{i=1}^{N_b}\), which was employed to train the second localized surrogate $\hat{q}_2$, while simultaneously employing the summaries $T(\yVec_b^{(i)})$ \eqref{eq:summaries}, instead of $\yVec^{(i)}_b$. Therefore, we train the classifier as in Section \ref{sec:ratio-correction}, but using
$
  \{(\thetaVec_b^{(i)},T(\yVec_b^{(i)}),W^{(i)}=1)\}_{i=1}^{N_b},
$
and the permuted data 
$
  \{(\thetaVec_b^{(i)},T(\yVec_b^{(\pi(i))}),W^{(i)}=0)\}_{i=1}^{N_b}.
$

\noindent
We now specify the covariates entering the binary classifier. For ease of notation, in the reminder of this section we stop writing $\thetaVec_b$ and $T(\yVec_b)$ and instead use $\thetaVec$ and $T(\yVec)$, respectively, but it should be understood that in practice training data use $(\thetaVec_b,T(\yVec_b))$. Let
$
\thetaVec=(\theta_1,\ldots,\theta_{d_\theta})^\top
$
and write
$
\mathbf{T}=T(\yVec)=(T_1,\ldots,T_{d_T})^\top
$
for the posterior-based summary vector defined in \eqref{eq:summaries}. Rather than
restricting the classifier to the concatenated vector
$(\thetaVec^\top,\mathbf{T}^\top)^\top$, we allow its linear predictor to
depend on a feature map $\psi(\thetaVec,\mathbf{T})$ containing selected
transformations and interactions of these two blocks, where the first ``block'' is $\thetaVec^\top$ and the second one is $\mathbf{T}^\top$.
The simplest covariates are the \emph{linear terms},
$
\psi_{\mathrm{lin}}(\thetaVec,\mathbf{T})
=
\left(
\theta_1,\ldots,\theta_{d_\theta},
T_1,\ldots,T_{d_T}
\right)^\top,
$
which contribute $d_\theta+d_T$ covariates.
We may additionally include \textit{within-block quadratic terms},
consisting of the componentwise squares
$
\psi_{\mathrm{quad}}(\thetaVec,\mathbf{T})
=
\left(
\theta_1^2,\ldots,\theta_{d_\theta}^2,
T_1^2,\ldots,T_{d_T}^2
\right)^\top,
$
giving a further $d_\theta+d_T$ covariates. 
A further potential enrichment consists of \textit{within-block interactions}: for the parameter block these are
$
\{\theta_j\theta_k:1\leq j<k\leq d_\theta\},
$
while for the summary block they are
$
\{T_\ell T_m:1\leq \ell<m\leq d_T\}.
$
The total number of such terms is therefore
$
{d_\theta\choose 2}+{d_T\choose 2}.
$
Finally, we consider \textit{between-block interactions}, namely all
products between a parameter coordinate and a summary coordinate,
$
\psi_{\mathrm{cross}}(\thetaVec,\mathbf{T})
=
\left\{
\theta_j T_\ell:
j=1,\ldots,d_\theta,\;
\ell=1,\ldots,d_T
\right\},
$
which contribute $d_\theta d_T$ additional covariates.
The between-block interactions are particularly important for the
permutation-based classifier. Covariates
that are functions of $\thetaVec$ alone or of $T(\yVec)$ alone cannot,
at the population level, distinguish the two classes. Indeed, the
joint and product-of-marginals classes have by construction exactly the same
empirical marginal distribution of the parameter block
$\thetaVec_b$, and also exactly the same empirical marginal
distribution of the summary block $T(\yVec_b)$.  Features coupling
the two blocks, such as the products $\theta_j T_\ell$, provide the
simplest way for a logistic classifier to learn this dependence.
For this reason, between-block interactions are the essential and our baseline specification nevertheless includes the linear terms together with all between-block interactions, that is we use
\begin{equation}
\psi(\thetaVec,\mathbf{T})
=
\left(
\psi_{\mathrm{lin}}(\thetaVec,\mathbf{T})^\top,
\psi_{\mathrm{cross}}(\thetaVec,\mathbf{T})^\top
\right)^\top,
\label{eq:covariates}
\end{equation}
whereas the quadratic and within-block interaction terms are regarded
as optional enrichments (these optional terms can substantially increase the
number of classifier covariates, especially when $d_T$ is moderately
large). 
In summary, when fitting training data we obtain the following estimated log density ratio 
\[
\log \hat r(\thetaVec,\mathbf T)
=
\log\left\{
\frac{\hat D(\thetaVec,\mathbf T)}
     {1-\hat D(\thetaVec,\mathbf T)}
\right\}
=
\hat\beta_0+
\hat{\boldsymbol\beta}^{\top}
\psi(\thetaVec,\mathbf T),
\qquad
\mathbf T=T(\yVec).
\]
The resulting classifier is fitted by regularized logistic regression, which in our implementation uses the elastic-net family via the R \texttt{glmnet} package \citep{Friedman2010,Tay2023}, with the
regularization parameter selected by cross-validation. In practice, same as in the ``linear LFIRE'' of \cite{thomas2022likelihood}, in our experiments we set the $\alpha$ parameter for the elastic-net framework to use lasso ($\alpha=1$). 
All classifier covariates are standardized internally before fitting. In Section \ref{sec:importance-reweight} we discuss differences between our classifier and the one in linear LFIRE.

\subsection{Importance reweighting}\label{sec:importance-reweight}

Given the fitted locally amortized likelihood-to-evidence ratio developed in the previous section, we are now ready to apply importance correction to new draws obtained from $\hat{q}_2$, thereby returning the final inference. We simulate a fresh set of $M$ parameter draws from $\hat{q}_2(\thetaVec\given\yobs)$, which is cheap to do since $\hat{q}_2$ is a Gaussian mixture and can be directly sampled, therefore $M$ can be safely taken large: 
\begin{equation}
  \thetaVec^{(m)}\sim \hat{q}_2(\thetaVec\given\yobs),
  \qquad m=1,\ldots,M,\label{eq_q2_draws}
\end{equation}
and then we evaluate the alrready fitted log-ratio at parameters and summaries of the observed data $(\thetaVec^{(m)},T(\yobs))$ as 
$
\log \hat r(\thetaVec^{(m)},\mathbf T_{\mathrm{obs}} )
=
\hat\beta_0+
\hat{\boldsymbol\beta}^{\top}
\psi(\thetaVec^{(m)},\mathbf T_{\mathrm{obs}} )$,
with $
\mathbf T_{\mathrm{obs}}=T(\yobs).
$
We compute an unnormalized correction weight as
\begin{equation}
  \wt w^{(m)}
  =
  \frac{p(\thetaVec^{(m)})\exp(\log\hat r\{\thetaVec^{(m)},\mathbf T_{\mathrm{obs}}\})}
       {\hat{q}_2(\thetaVec^{(m)}\given\yobs)}, \qquad m=1,...,M.
  \label{eq:raw-weight}
\end{equation}
Normalized weights are then computed as
  $w^{(m)}=\tilde{w}^{(m)}/\sum_{j=1}^M \tilde{w}^{(j)}$, and the final corrected posterior draws can be obtained by resampling $M$ times with replacement from the weighted parameters $\{\thetaVec^{(m)},w^{(m)}\}_{m=1}^M$, and we may call these resampled parameters as $\{\thetaVec^{(m)}_{\mathrm{post}}\}_{m=1}^M$. Ideally, $\{\thetaVec^{(m)}_{\mathrm{post}}\}_{m=1}^M$ constitute the basis for the reported posterior inference.
However, raw weights of the form \eqref{eq:raw-weight} can be unstable when the ratio estimate is noisy or when \(\hat{q}_2(\thetaVec\given\yobs)\) is underdispersed, and this can reduce the effective sample size (ESS) of $\{\thetaVec^{(m)}_{\mathrm{post}}\}_{m=1}^M$, namely $\mathrm{ESS}=1/\sum_{m=1}^M (w^{(m)})^2$. To mitigate this problem, possibilities include first sampling the $\thetaVec^{(m)}\sim g_\lambda(\thetaVec)$, that is $\hat{q}_2$ in \eqref{eq_q2_draws} would be replaced by a slightly inflated sampler $g_\lambda$, where the latter is identical to $\hat{q}_2(\thetaVec\given\yobs)$ except that the covariance matrices are multiplied by some factor $\lambda>1$. Another possibility is to use a deterministic clipping rule for the normalized weights $w^{(m)}$, like the one suggested in \cite{martino2018comparison}: take
$
  M_c=\lfloor\sqrt M\rfloor
$, then 
let \(w^{[1]}\ge\cdots\ge w^{[M]}\) be the ordered normalized raw weights (notice the square brackets to denote the ordered weights), and define the cap
$
  c=w^{[M_c]}.
$
Finally, the ``clipped weights'' are
\begin{equation}
  w^{(m)}_{\mathrm{clip}}=\min(w^{(m)},c),
  \qquad
  \bar{w}^{(m)}_{\mathrm{clip}}
  =\frac{w^{(m)}_{\mathrm{clip}}}{\sum_{j=1}^M w^{(j)}_{\mathrm{clip}}},
  \label{eq:clip}
\end{equation}
where computing normalized $\bar{w}^{(m)}_{\mathrm{clip}}$ is necessary as the $w^{(m)}_{\mathrm{clip}}$ are no longer normalized after clipping.
Then, the final ``clipped'' corrected posterior draws are given by resampling $M$ times with replacement from $\{\thetaVec^{(m)},\bar{w}^{(m)}_\mathrm{clip}\}_{m=1}^M$.
Formally, the \textit{clipped ratio-corrected} posterior approximation is the weighted empirical
measure
\begin{equation}
    \widehat\Pi_{\mathrm{clip}}(d\thetaVec\mid \yobs)
    =
    \sum_{m=1}^M
     \bar{w}^{(m)}_{\mathrm{clip}} \, \delta_{\thetaVec_m}(d\thetaVec), \label{eq:corrected-measure}
\end{equation}
where \(\delta(\cdot)\) is the dirac-measure. The associated effective sample size is
$
    \mathrm{ESS}_{\mathrm{clip}}
    = 1/\sum_{m=1}^M ( \bar{w}^{(m)}_\mathrm{clip})^2$, which can turn much larger than the ``raw ESS'' based on the original normalized weights $w^{(m)}$. 
In our experiments, the final inference results are based on samples from \eqref{eq:corrected-measure}, which we will compare against other inference methods, such as pseudomarginal particle MCMC and the uncorrected surrogate $q_2(\thetaVec|\yVec_{\mathrm{obs}})$. As it will be illustrated in applications, occasionally the proposal sampler $\hat{q}_2(\thetaVec|\yVec_{\mathrm{obs}})$ tends to be slightly overconfident prior to ratio-correction, while the correction in \eqref{eq:corrected-measure} tends to slightly inflate the posterior approximation, and we believe a conservative posterior is a safer approach to approximate Bayesian inference \citep{hermans2022a}. It is worth to emphasize that, in our implementation, the ratio correction is always based on the learned statistic \(T(y)\). Hence, even with an exact ratio estimator, the corresponding target is \(p(\thetaVec\mid T(\yobs))\), which coincides with \(p(\thetaVec\mid \yobs)\) when \(T\) is sufficient.

We prefer to just use clipped weights for the final approximation, and not to introduce a further tuning parameter $\lambda>1$ determining a new sampler $g_\lambda$, although this is certainly
possible\footnote{However, we do use an inflated $g_\lambda$ in a different context, when we introduce local simulation-based calibration.}. Weight clipping \citep{koblents2015population} can be viewed as a nonlinear
importance-sampling transformation designed to reduce the dominance of a small
number of large importance weights and thereby mitigate weight degeneracy.
Our particular rule corresponds to the ``minimum'' clipping strategy considered
by \citet{martino2018comparison}. In particular, \citet{koblents2015population}
establish almost-sure consistency for clipped importance-sampling
approximations under suitable regularity conditions when the number of
transformed weights grows sufficiently slowly; see also \citet{martino2018comparison}, who consider the choice
$M_c \leq \sqrt{M}$. 
Conditional on the fitted ratio estimator and proposal, the same consistency
argument applies to the importance-sampling target induced by the learned
ratio. Thus, provided the corresponding regularity conditions hold, for any
bounded test function $f$,
\[
\sum_{m=1}^M
\bar w_{\mathrm{clip}}^{(m)}f(\thetaVec^{(m)})
\ \xrightarrow[M\to\infty]{\mathrm{a.s.}}\
\int f(\thetaVec)\,
\widehat p_{\mathrm{corr}}(\thetaVec\mid T(\yVec_{\mathrm{obs}}))
\,d\thetaVec,
\]
where in our problem
$
\widehat p_{\mathrm{corr}}(\thetaVec\mid T(\yVec_{\mathrm{obs}}))
\propto
p(\thetaVec)\widehat r\{\thetaVec,T_{\mathrm{obs}}\}.
$
Accordingly, clipping acts as a finite-sample regularization of the importance
weights while preserving consistency for the distribution induced by the
fixed learned ratio. It does not, however, remove approximation error in the
ratio estimator itself. Hence, even as $M\to\infty$, the limiting distribution
need not coincide with
$
p(\thetaVec\mid T_{\mathrm{obs}})
\propto
p(\thetaVec) r\{\thetaVec,T_{\mathrm{obs}}\}
$,
unless the learned ratio is exact. Indeed, the accuracy of the learned ratio is determined by the classifier training stage.\\

\noindent
\textbf{Notable differences with the ``linear LFIRE'' of \cite{thomas2022likelihood}:} while our ratio-estimation step builds on the LFIRE principle of estimating likelihood-to-evidence ratios by probabilistic classification, its role in our methodology is different. LFIRE uses ratio estimation as the primary mechanism for posterior construction and, in its linear implementation, fits a classifier separately \textit{at each parameter value}. Specifically, the classification mechanism is applied from the very beginning, where parameters could be simulated from the prior or generated from some different mechanism, and conditionally on each $\thetaVec$ a number $n_\theta$ of DGM simulations are performed: that is, in linear LFIRE a set of $n_\theta$ simulations $\{\yVec_\theta^{(i)}\}_{i=1}^{n_\theta}$ is generated at each proposed $\thetaVec$. Instead, the ``marginal'' simulations corresponding to the ratio denominator are generated once and for all by drawing parameters from the prior $p(\thetaVec)$, in the number of $n_m$, using different $\thetaVec$-values, by simulating $\{\yVec_m^{(i)}\}_{i=1}^{n_m}$ as $\yVec_m^{(i)}\sim p(\yVec|\thetaVec_m^{(i)})$ and disregarding the information about the employed $\thetaVec_m^{(i)}$. Therefore, if the posterior is evaluated at $P$ parameter values, then $P$ ratio classifiers are needed, for a total of $n_m + n_\theta P$ calls of the DGM simulator.  
If the prior is broad then the $P$ posterior draws should have a fairly large $P$.  On the other hand, each individual classification problem in linear LFIRE is simpler than ours, since in linear LFIRE each classification is performed at a fixed parameter value $\thetaVec$, and therefore the classifier only needs to model variation in the simulated data summaries, and $\thetaVec$-dependent covariates do not need to be included. 
In our approach we instead fit \textit{a
single classifier} jointly over both $\thetaVec$ and $T(\yVec)$,
using the already available second-round training data $\mathcal{D}_2$, but then it must learn the dependence on both $\thetaVec$ and $T(\yVec)$, which requires a richer feature representation. However, in our approach, once $q_2$
and the ratio classifier have been fitted, evaluating the ratio at additional proposal draws from $q_2$ requires neither new DGM simulations nor new classifier fits.

\section{Localized simulation-based calibration}
\label{sec:local-sbc}

Simulation-based calibration (SBC) assesses whether a Bayesian inference procedure is calibrated under repeated data generation from the assumed model \citep{talts2018validating}. Here, calibration means that, across repeated simulations, the generating parameter values behave as draws from their corresponding posterior distributions. SBC can therefore reveal systematic bias or misrepresentation of posterior uncertainty. 
Even when the inference algorithm is implemented correctly, it may produce significantly biased inference for some posteriors. 
In the standard form provided in \cite{talts2018validating}, one repeatedly draws parameters and simulated observations 
$
  \thetaVec^\star \sim p(\thetaVec)
 $,
  $\yVec^\star \sim p(\yVec\given\thetaVec^\star),
$
runs the inference algorithm of choice on each pseudo-observation \(\yVec^\star\), thus producing posterior draws based on \(\yVec^\star\), and checks whether the rank of \(\thetaVec^\star\) among those posterior draws is uniform.   This is a global prior-predictive check of the algorithm (and is denoted ``prior SBC'' in \citealp{sailynoja2026posterior}).  It is also expensive, because the full inference procedure must be repeated for each of the many simulated observations. 
For an applied analysis one may instead want a calibration check concentrated in the region of parameter and data space relevant to the observed $\yobs$: this approach was named ``posterior SBC'' in \cite{sailynoja2026posterior}, which we discuss further below. We now introduce an SBC methodology that is analogous to posterior SBC, but generalizes it by assessing whether the pseudo-posterior is calibrated for pseudo-observations generated from parameters drawn from a neighborhood of the observed-data posterior. We call our method \textit{localized SBC}, and we show that it is particularly appealing in SBI studies.
Moreover, differences with posterior SBC are discussed in Section \ref{sec:comparison-post-sbc}.  

Let \(g\) be a local design distribution on \(\Theta\), where ``local'' implies that $g$ is somehow depending on $\yobs$.  In practice, in our implementation, the local design $g_\lambda$ is constructed from the fitted $\hat q_2(\thetaVec\mid\yobs)$ by 
using only non-negligible (``active'') mixture components for this task, after removing components
with negligible posterior responsibility (this is detailed in Supplementary Material), and the within-component covariance
matrices are then inflated by a factor $\lambda^2$. We make such choice to illustrate the localized SBC, however, it should be understood that $g$ can be chosen in other ways.  Specifically, in our implementation, given the following surrogate
$
  \hat{q}_2(\thetaVec\given\yobs)
  =
  \sum_{k=1}^K
  \hat{\omega}_k(\yobs)
  \Normal(\thetaVec;\hat{\mathbf{m}}_k,\hat{\mathbf{\Sigma}}_k),
$
then a broadened local design can be defined as
\begin{equation}
  g_\lambda(\thetaVec|\yobs)
  =
  \sum_{k=1}^K
  \hat{\omega}_k(\yobs)
  \Normal(\thetaVec;\hat{\mathbf{m}}_k,\lambda^2\hat{\mathbf{\Sigma}}_k),
  \qquad \lambda\ge 1.
  \label{eq:local-design}
\end{equation}
The value \(\lambda=1\) recovers the \(\hat{q}_2\) mixture, whereas  multiplying the component standard
deviations by a factor $\lambda\geq 1$ is equivalent to replacing each
component covariance matrix $\hat{\Sigma}_k$ by $\lambda^2\hat{\Sigma}_k$.
Thus, for example, 
$\lambda=1.25$ corresponds to a 25\% inflation of component
standard deviations. 

Generally, when simulating
$
  \thetaVec^\star \sim g_\lambda(\thetaVec\given \yobs)$ and
  $\yVec^\star\sim p(\yVec\given\thetaVec^\star)$, 
the corresponding exact pseudo-posterior is not the original posterior under the prior \(p(\thetaVec)\), but the posterior under the following conditional
\begin{equation}
  p_{g_\lambda}(\thetaVec\given \yVec^\star; \yobs)
  =
  \frac{g_\lambda(\thetaVec\given \yobs)p(\yVec^\star\given\thetaVec)}
       {\int g_\lambda(\mathbf{u}\given \yobs)p(\yVec^\star\given \mathbf{u})\,d\mathbf{u}},
  \label{eq:local-target}
\end{equation}
where the notation for $p_{g_\lambda}(\thetaVec\given \yVec^\star; \yobs)$ means that this is the pseudo-posterior of $\thetaVec$ conditioned on  $\yVec
^\star$, with a dependence on $\yobs$ that is implicit through $g_\lambda(\thetaVec\given \yobs)$.  
Thus localized SBC is ordinary SBC for the artificial prior \(g_\lambda\). Its purpose is therefore not to assess global calibration under the original prior, but rather to assess whether the pseudo-posterior \eqref{eq:local-target} is calibrated for pseudo-observations generated from parameters drawn from a neighborhood of the observed-data posterior, and below we discuss why this can be of interest. In settings where the exact posterior is unavailable, as in SBI, this neighborhood is necessarily constructed from an approximation to the observed-data posterior, such as the fitted surrogate or the ratio-corrected approximation considered here.
Expression \eqref{eq:local-target} defines the general full-data local
target. In our SBI implementation, however, ratio estimation is performed
using a learned statistic constructed from the local GLLiM fit. Specifically,
after fitting the local GLLiM model, we obtain a statistic
$T_\lambda(\yVec)$ using the same posterior-summary construction as in the
main inference procedure (consisting of posterior means, variances/covariances, and mixture weights obtained from the locally fitted GLLiM approximation), whose construction is clarified below. The statistic therefore depends on the local
GLLiM fit and, implicitly, on $\lambda$. The corresponding summary-conditional target therefore turns into 
\begin{equation}
p_{g_\lambda}
\left(
\thetaVec\mid T_\lambda(\yVec^\star);\yobs
\right)
\propto
g_\lambda(\thetaVec\mid\yobs)\,
p_{T_\lambda}
\left(
T_\lambda(\yVec^\star)\mid\thetaVec
\right).
\label{eq:local-summary-target}
\end{equation}
Then, since an exact local ratio estimator can be written as
\begin{equation}
r_\lambda(\thetaVec,T)
=
\frac{p_{T_\lambda}(T\mid\thetaVec)}
     {m_{g_\lambda}(T)},
\qquad
m_{g_\lambda}(T)
=
\int
p_{T_\lambda}(T\mid\mathbf u)\,
g_\lambda(\mathbf u\mid\yobs)\,d\mathbf u,
\label{eq:local-ratio}
\end{equation}
and since the evidence $m_{g_\lambda}(T_\lambda(\yVec^\star))$ does not
depend on $\thetaVec$, we have
\begin{equation}
p_{g_\lambda}
\left\{
\thetaVec\mid T_\lambda(\yVec^\star);\yobs
\right\}
\propto
g_\lambda(\thetaVec\mid\yobs)\,
r_\lambda
\left\{
\thetaVec,T_\lambda(\yVec^\star)
\right\}.\label{eq:local-target-ratioIS}
\end{equation}
This target coincides with the full-data pseudo-posterior
\eqref{eq:local-target} when $T_\lambda$ is sufficient; otherwise the two
targets are distinct.

Now that we have established the target distribution for a pseudo-posterior associated to a local design $g_\lambda$, and specified it for our corrected-ratio importance sampling method in \eqref{eq:local-target-ratioIS}, we are ready to show the detailed construction of the localized SBC method, see also Algorithm \ref{alg:local-sbc}.  
For a fixed local design $g_\lambda$, we use two independent sets of
simulations.  First, we generate local training data
\begin{equation}
  \thetaVec_i^{\mathrm{tr}}
  \sim g_\lambda(\thetaVec\mid\yobs),
  \qquad
  \yVec_i^{\mathrm{tr}}
  \sim p(\yVec\mid\thetaVec_i^{\mathrm{tr}}),
  \qquad i=1,\ldots,N_{\mathrm{loc}},
  \label{eq:local-training-bank}
\end{equation}
which is used only to construct the local inference procedure.  We fit a
GLLiM model once to this training data, initializing the EM algorithm from
the active components of the fitted
$\hat q_2(\thetaVec\mid\yobs)$ solution.  The resulting local fit is
then used to construct a learned summary
$T_\lambda(\yVec)$.
Using the same local training data set, we also fit a ratio estimator
$\widehat r_\lambda\{\thetaVec,T_\lambda(\yVec)\}$ with the same classifier
architecture and tuning strategy used in the main  procedure. 
Both the local GLLiM and the ratio estimator are fitted only
once for each value of $\lambda$ and are then kept fixed.
Calibration is assessed on an independent checking data set,
\begin{equation}
  \thetaVec_i^\star
  \sim g_\lambda(\thetaVec\mid\yobs),
  \qquad
  \yVec_i^\star
  \sim p(\yVec\mid\thetaVec_i^\star),
  \qquad i=1,\ldots,N_{\mathrm{check}},
  \label{eq:local-checking-bank}
\end{equation}
which is not used in fitting either the local GLLiM model or the ratio
estimator.
For each checking pseudo-observation $\yVec_i^\star$, we compute
$T_\lambda(\yVec_i^\star)$ and draw $M_{\mathrm{IS}}$ inexpensive
importance-sampling particles
$
  \thetaVec_i^{(m)}
  \sim
  g_\lambda(\thetaVec\mid\yobs)$,
  ($
  m=1,\ldots,M_{\mathrm{IS}}$).
Since the importance proposal is the local design itself, the
$g_\lambda$ terms cancel from the importance ratio and the unnormalized
weights are
$
  \widetilde w_{im}
  \propto
  \widehat r_\lambda
  \left\{
    \thetaVec_i^{(m)},
    T_\lambda(\yVec_i^\star)
  \right\}.
$
In practice, same as in the main inference procedure,  we apply the same deterministic clipping rule, this time to
$\{\widetilde w_{im}\}_{m=1}^{M_{\mathrm{IS}}}$, with
$M_c=\lfloor\sqrt{M_{\mathrm{IS}}}\rfloor$, and the resulting normalized 
weights are denoted by
$\bar w_{im,\mathrm{clip}}$.  Following \cite{cook2006validation}, for the $j$th coordinate the ideal
localized PIT associated with the summary-conditional target is
$
U_{i,j}^{\mathrm{loc}}
=
F^{T_\lambda}_{g_\lambda,j}
\!\left(
\theta_{ij}^\star
\mid
T_\lambda(\yVec_i^\star)
\right),
$
which is uniform on $[0,1]$ under the exact summary-conditional
pseudo-posterior \eqref{eq:local-summary-target}, where $F$ denotes its marginal cdf.
This can be approximated with the weighted empirical cdf (ECDF) evaluated at the pseudo-true parameter,
\begin{equation}
  \widehat U_{i,j}^{\mathrm{loc}}
  =
  \sum_{m=1}^{M_{\mathrm{IS}}}
  \bar w_{im,\mathrm{clip}}\,
  \mathbb{I}
  \left\{
    \theta_{ij}^{(m)}
    \leq
    \theta_{ij}^\star
  \right\}.
  \label{eq:local-weighted-pit}
\end{equation}
Histograms of $\widehat U_{i,j}^{\mathrm{loc}}$, for each dimension $j=1,...,d_\theta$, admit the usual SBC
interpretation: U-shaped histograms indicate underdispersion, hump-shaped
histograms indicate overdispersion, and monotone trends indicate location
bias \citep{talts2018validating}.  Because histogram-based assessment
depends on the choice of bins and ignores dependence across bins, following \citet{sailynoja2022graphical} we
additionally use ECDF-based displays with simultaneous confidence bands. A departure from uniformity indicates that the posterior approximation induced under the local design \(g_\lambda\) is not calibrated under the corresponding local simulation experiment, but does not by itself identify the source of the discrepancy. Possible causes include approximation error in the learned summary or ratio estimator, the effect of weight clipping, insufficient flexibility of the local surrogate, finite simulation budgets, optimization or implementation issues, or an unfavorable parametrization.

\begin{algorithm}[t!]
\small
\caption{Localized simulation-based calibration}
\label{alg:local-sbc}
\begin{algorithmic}[1]

\Require Observed data $\yobs$; local design
$g_\lambda(\thetaVec\mid\yobs)$; local training size
$N_{\mathrm{loc}}$; checking size $N_{\mathrm{check}}$;
importance-sampling size $M_{\mathrm{IS}}$.

\State Generate a set of local training data
\[
\thetaVec_i^{\mathrm{tr}}\sim
g_\lambda(\thetaVec\mid\yobs),
\qquad
\yVec_i^{\mathrm{tr}}\sim
p(\yVec\mid\thetaVec_i^{\mathrm{tr}}),
\qquad
i=1,\ldots,N_{\mathrm{loc}}.
\]

\State Fit one local GLLiM model to $\{\thetaVec_i^{\mathrm{tr}},\yVec_i^{\mathrm{tr}}\}_{i=1}^{N_{\mathrm{loc}}}$, warm-starting
from the active components of the fitted $\hat q_2$ model.

\State Construct the local learned summary $T_\lambda(\yVec)$ from this
GLLiM fit.

\State Using the local training data, fit once the ratio estimator
$\widehat r_\lambda\{\thetaVec,T_\lambda(\yVec)\}$ with the same
classifier specification as in the main analysis.

\State Generate a set of fresh simulations
\[
\thetaVec_i^\star\sim
g_\lambda(\thetaVec\mid\yobs),
\qquad
\yVec_i^\star\sim
p(\yVec\mid\thetaVec_i^\star),
\qquad
i=1,\ldots,N_{\mathrm{check}}.
\]

\For{$i=1,\ldots,N_{\mathrm{check}}$}

  \State Evaluate $T_\lambda(\yVec_i^\star)$, and draw
  $\thetaVec_i^{(m)}\sim
  g_\lambda(\thetaVec\mid\yobs)$,
  $m=1,\ldots,M_{\mathrm{IS}}$.

  \State Compute
  \[
  \widetilde w_{im}
  =
  \widehat r_\lambda
  \left\{
    \thetaVec_i^{(m)},
    T_\lambda(\yVec_i^\star)
  \right\}, \quad m=1,\ldots,M_{\mathrm{IS}}.
  \]

  \State Apply the clipping rule and
  normalize the resulting weights to obtain
  $\bar w_{im,\mathrm{clip}}$.

  \For{$j=1,\ldots,d_\theta$}
      \State Compute
      \[
      \widehat U_{i,j}^{\mathrm{loc}}
      =
      \sum_{m=1}^{M_{\mathrm{IS}}}
      \bar w_{im,\mathrm{clip}}
      \mathbb{I}
      \{
      \theta_{ij}^{(m)}
      \leq
      \theta_{ij}^\star
      \}.
      \]
  \EndFor

\EndFor

\State For each $j$, assess whether
$\{\widehat U_{i,j}^{\mathrm{loc}}\}_{i=1}^{N_{\mathrm{check}}}$
is compatible with a $\operatorname{Uniform}(0,1)$ distribution.

\end{algorithmic}
\end{algorithm}

For a fixed \(\lambda\), the inference components are \textit{locally amortized} over the checking pseudo-observations, since they are learned once from simulations generated under the local design $g_\lambda(\thetaVec\given\yobs)$. In fact, the local GLLiM model and ratio estimator are each fitted only once using the local training data and are subsequently reused for all \(N_{\mathrm{check}}\) pseudo-observations. Thus, the procedure avoids rerunning the complete surrogate-fitting and ratio-estimation pipeline separately for every pseudo-observation. 

We have described one particular construction of localized SBC, but the
local design $g_\lambda$ can be constructed in other ways. More generally,
the idea is not intrinsically restricted to SBI: whenever an approximate
posterior is available, one may define a parameter-generating distribution
in a neighborhood of that posterior and assess calibration of a conditional
approximation under the resulting local simulation experiment.
The case $g_\lambda \neq p(\thetaVec\mid\yobs)$ requires a different
interpretation from the posterior SBC of \cite{sailynoja2026posterior} (see also Section \ref{sec:comparison-post-sbc}). Suppose that an approximate posterior
$\hat p(\thetaVec\mid\yobs)$ has been obtained. A diagnostic whose
parameter-generating distribution is concentrated on this approximation
primarily probes inference in the region already emphasized by
$\hat p(\thetaVec\mid\yobs)$. If the approximation is itself
under-dispersed, nearby parameter regions receiving too little probability
under $\hat p(\thetaVec\mid\yobs)$ may consequently be explored only
weakly. Localized SBC with $\lambda>1$ deliberately broadens the
parameter-generating distribution and asks a different question: whether the locally trained approximation is calibrated under pseudo-experiments generated from that broader design.

Thus, results obtained for $\lambda>1$ should not be interpreted as direct
calibration statements about the reported posterior itself (in our study the ``reported posterior'' would be the clipped ratio corrected one in \eqref {eq:corrected-measure}). Instead, they
provide a local robustness diagnostic for the conditional approximation
under a broadened simulation design. 
Although this construction is not specific to SBI, the motivation is
particularly relevant in SBI, where some methods are known to produce under-dispersed
posterior approximations; see, for example, \cite{hermans2022a}. In this
setting, broadening $g_\lambda$ provides a controlled way to probe the
behavior of the conditional approximation in nearby parameter regions
that may receive insufficient mass under the reported approximation.

Moreover, calibration results need not vary monotonically with the inflation
factor $\lambda$. That is, a satisfactory result for a larger $\lambda$ does not ``rescue'' a calibration failure detected at a smaller value.
Suppose an approximation is badly calibrated in a small central region but well calibrated in the surrounding region. A diagnostic concentrated on the central region may detect the problem, whereas a broader diagnostic may average the central failure together with much larger well-behaved regions and therefore look acceptable. In other words, a broader design $g_\lambda$ may assign less weight to the parameter and
pseudo-observation regions in which the smaller-$\lambda$ diagnostic
detects a discrepancy, and calibration errors of opposite sign may
partly average out in the pooled PIT distribution. Thus, failure at
$\lambda=1$ remains evidence of a calibration problem in the
posterior-local region even if a broader diagnostic appears satisfactory.
To emphasize the matter from a practical perspective: a failure already at $\lambda=1$ indicates a
problem in the posterior-local region itself. If calibration is restored
for larger $\lambda$,  this should not be interpreted as correcting the
failure at $\lambda=1$, but rather, the pattern may motivate further
investigation of posterior concentration, support, or other sources of
local approximation error. The particular cause of a calibration failure
cannot, however, be inferred from the localized SBC diagnostic alone.

\subsection{Relation to posterior SBC}\label{sec:comparison-post-sbc}
The localized calibration check above is a generalization of the posterior simulation-based
calibration recently proposed in \cite{sailynoja2026posterior}. A pseudo-algorithm for posterior SBC is given in Supplementary Material, for ease of comparison. Posterior SBC considers draws
$
\thetaVec^\star \sim p(\thetaVec \mid \yVec_{\rm obs})$,
$\yVec^\star \sim p(\yVec \mid \thetaVec^\star),
$
and then compares \(\thetaVec^\star\) with the posterior obtained after augmenting
the observed data with the newly simulated dataset,
$
p(\thetaVec \mid \yobs,\yVec^\star)
\propto
p(\yVec^\star \mid \thetaVec)\,p(\thetaVec \mid \yobs).
$
Once more, for each scalar coordinate \(\theta_j\), calibration can be assessed through the
rank or a corresponding PIT value that we denote $U_{j}^\mathrm{post}$ and is
given by 
$U_{ij}^\mathrm{post}
=
P\left(
\theta_{ij} \leq \theta^\star_j
\mid \yobs,\yVec^\star
\right).
$
Therefore, in posterior SBC, $\yobs$ remains explicitly part of every calibration experiment. Posterior SBC addresses the following question: \textit{under repeated pseudo-experiments with 
$\thetaVec^\star \sim p(\thetaVec\mid \yobs)
$,
$\yVec^\star \sim p(\yVec\mid \thetaVec^\star)$,
is the posterior approximation based on the augmented data
$(\yobs,\yVec^\star)$ calibrated with respect to
$
p(\thetaVec\mid \yobs,\yVec^\star)$?
}
Instead in localized SBC  \(\yobs\) enters  indirectly, through the construction of \(g_\lambda(\thetaVec\mid \yobs)\). The question addressed in localized SBC is: \textit{under repeated pseudo-experiments generated from the local design $\thetaVec^\star \sim g_\lambda(\thetaVec\mid \yobs)$, $\yVec^\star \sim p(\yVec\mid \thetaVec^\star)$, is the posterior approximation calibrated with respect  to the corresponding local pseudo-posterior\footnote{Recall $p_{g_\lambda}(\thetaVec\mid\yVec^\star;\yobs)$ \textit{does not} denote a pseudo-posterior over an augmented $(\yVec^\star,\yobs)$; the dependence on $\yobs$ is indirect.}
$p_{g_\lambda}(\thetaVec\mid\yVec^\star;\yobs)$?} 
%The distribution $g_\lambda$ defines a neighborhood around the posterior
%obtained from $\yobs$, which for $\lambda>1$ can deliberately be broader
%than the posterior itself. In contrast, posterior SBC uses the posterior
%induced by $\yobs$ itself as the parameter-generating distribution for
%the subsequent pseudo-experiment.

Interestingly, the two constructions, posterior SBC and localized SBC, coincide when
$
g_\lambda(\thetaVec\mid \yobs)=p(\thetaVec\mid \yobs)
$, as in that case
\[
p_{g_\lambda}(\thetaVec\mid \yVec^\star; \yobs)
\propto
p(\yVec^\star\mid \thetaVec)p(\thetaVec\mid \yobs)
=
p(\thetaVec\mid \yobs,\yVec^\star).
\]
Therefore posterior SBC is a special case of localized SBC, when the local design $g_\lambda$ is taken to be the posterior itself. In our implementation of localized SBC, however, the ratio correction operates through
the learned statistic $T(\yVec)$. Hence the practical target is
$p_{g_\lambda}(\thetaVec\mid T(\yVec^\star);\yobs)$, and in our implementation exact
equality with posterior SBC additionally requires that $T(\yVec)$
retain all information relevant for the posterior update, that is $T$ should be a sufficient statistic.\\

\noindent
\textbf{Computational complexity of localized SBC and posterior SBC:}\\
  Posterior SBC is a general procedure that is not specifically designed for the intractable likelihood problems encountered in SBI, although it can certainly be used within SBI since the only requirement is the ability to simulate from a model. However, when posterior SBC is used with a SBI procedure, the inference
algorithm would in general need to be rerun for each of the
simulated $\yVec^\star$, since in posterior SBC each replication requires
inference under a different augmented dataset
$(\yobs,\yVec^\star)$. Consequently, in our current simulation-based inference approach, posterior SBC would require $N_{\mathrm{post}}$ separate GLLiM refits and ratio-estimation components for each augmented dataset. On the opposite side, localized SBC requires only one local GLLiM refit and one local ratio-estimator fit for each attempted \(\lambda\) value, after which these components are reused across all checking pseudo-observations. This makes our localized SBC particularly appealing for simulation-based inference.

\section{Numerical studies}
\label{sec:numerical-studies}

Here follow case studies illustrating results using our method, and compared to reference posteriors (when available). Some settings are kept constant across case studies: in defensive sampling, we always consider a 5\% fraction of samples from the prior, that is we set $\varepsilon_1=\varepsilon_2=0.05$, and an initial number\footnote{We talk of ``initial number'' of components, even though we do not intentionally prune mixture components during the inference algorithm, because the internals of the \texttt{xLLiM} package can remove negligible components to avoid problems during EM.} of ten mixture components, i.e. $K_{\mathrm{start}}=10$. For localized SBC we use a total of $ N_{\mathrm{loc}}=2,000$ and $N_{\mathrm{check}}=2,000$, consisting of 2,000 simulations used as training data and 2,000 extra simulations for out-of-sample checking, and for each checking pseudo-observation we used
$M_{\mathrm{IS}}=5,000$ importance-sampling particles. The maximum number of EM iterations was set to 300 for every  fitting of the surrogate (mixture) models. During the inference, we do not prune mixture's components when a component's weight $\omega_k$ become negligible (unlike in \citealp{haggstrom2024fast}), but we prune negligible components when running localized SBC. For the GLLiM fitting procedure we employed  the expectation-maximization (EM) routine provided in the R \texttt{xLLiM} package \citep{xllim}. The code reproducing the results is available at \url{https://github.com/umbertopicchini/ratioIS-locSBC}.

\subsection{SIR model}\label{ex:sir}

We first consider a deterministic susceptible--infected--removed (SIR) epidemic model, with binomial observation noise, taken from the \texttt{sbibm} package \citep{sbibm} available at \url{https://github.com/sbi-benchmark/sbibm}. We use the observations $\yVec_\mathrm{obs}$ from \texttt{sbibm} and compare results against the reference posteriors provided in \texttt{sbibm}. We coded the case study in R, using the same model settings, including priors, as in \texttt{sbibm}.
For a population of size \(P=10^6\), the latent dynamics are governed by the system of ODEs
\begin{equation}
\frac{dS(t)}{dt} = -\beta \frac{S(t)I(t)}{P}, \qquad
\frac{dI(t)}{dt} = \beta \frac{S(t)I(t)}{P} - \gamma I(t), \qquad
\frac{dR(t)}{dt} = \gamma I(t) .
\label{eq:sir-ode}
\end{equation}
The system is solved numerically at a fixed grid of times. The observed data are
binomially distributed counts obtained from the infected fraction at ten selected
observation times:
\[
Y_j \mid \beta,\gamma
\sim
\operatorname{Binomial}
\left(
1000,
\frac{I(t_j;\beta,\gamma)}{P}
\right),
\qquad j=1,\ldots,10.
\]
The initial condition $(S(0),I(0),R(0))$ is fixed throughout the experiment to $I(0)=1$, $R(0)=0$ and by difference $S(0)=P-I(0)-R(0)$, and the parameters to infer are only two, $\thetaVec=(\beta,\gamma)$.
The prior distribution is independent across the two parameters,
$
    p(\thetaVec)
    =
    p(\beta)p(\gamma),
$
with support restricted to the biologically meaningful region \(\beta>0\), \(\gamma>0\). We consider the same prior as in the \texttt{sbibm} package (details are in Supplementary Material).
%For this first example we briefly remind how our ratio-estimation importance sampling method works. We first simulate $N_0$ prior-predictive samples
%and fit an initial GLLiM model. Conditioning this fit on \(\yobs\) it gives the first approximate posterior proposal, denoted by
%$
%    q_0(\thetaVec\mid \yobs).
%$
%The first localized simulation stage then samples $N_a$ parameters $\thetaVec_i^{(a)}$ defensively via \eqref{eq:stage1-design},
%followed by corresponding $N_a$ simulator evaluations $\yVec_i^{(a)}\mid\thetaVec_i^{(a)}$.
%A second surrogate \(q_1(\thetaVec\mid \yVec)\) is fitted to these localized simulations. We then repeat the localization once more, sampling
%$N_b$ times the pairs $(\thetaVec^{(i)}_b,\yVec^{(i)}_b)$.
%The final localized surrogate, denoted by
%$
%    q_2(\thetaVec\mid \yVec),
%$
%is fitted to the second-round of $N_b$ localized simulations and is used as the final proposal distribution. In practice, 
To fit the sequence of posterior surrogates $q_0$, $q_1$ and $q_2$ we used
$
    N_0=2\times 10^4$,
    $N_a=2\times 10^4$,
    $
    N_b=10^4$ model simulations, respectively,
implying a total of $N_0+N_a+N_b=5\times 10^4$ runs of the generative model.
The number of mixture components was initialized at $K_{\mathrm{start}}=10$.
After obtaining \(\hat{q}_2(\thetaVec\mid \yobs)\), we finally generate
$
    M=10^4
$
independent proposal draws (no model simulations  required here)
$
    \thetaVec^{(m)}\sim \hat{q}_2(\thetaVec\mid \yVec_{\mathrm{obs}})$
    ($m=1,\ldots,M$),
which were then ratio-corrected to provide the final inference, as described in Section \ref{sec:importance-reweight}. Since the \texttt{sbibm} package provides ten different datasets $\yobs$ and corresponding reference posteriors, we also provide results using our ratio-corrected importance sampling based on the same observed data.

For validation, we compared the surrogate conditional \(\hat{q}_2\), the clipped ratio-corrected posteriors \eqref{eq:corrected-measure}, and the reference posteriors from \texttt{sbibm}, based on the same $\yVec_{\mathrm{obs}}$. An example is given in Figure \ref{fig:sir_obs5} while posteriors for all the remaining observations are in Supplementary Material. As displayed in Figure \ref{fig:sir_obs5}, the uncorrected surrogate $\hat{q}_2(\thetaVec|\yVec_{\mathrm{obs}})$ manages to find the bulk of the true posterior but it displays overconfidence and hence underestimate the tails mass, instead the ratio-corrected posterior \eqref{eq:corrected-measure} (with weight clipping)  manages to cover the reference posterior. In other words, the ratio-corrected posterior is a more conservative approximation, and this is confirmed in further datasets from this same experiment, reported in Supplementary Material, where we also report a table showing the much larger ESS values obtained from clipped weights, compared to the ESS from raw weights.
\begin{figure}
    \centering
    \includegraphics[width=0.55\linewidth]{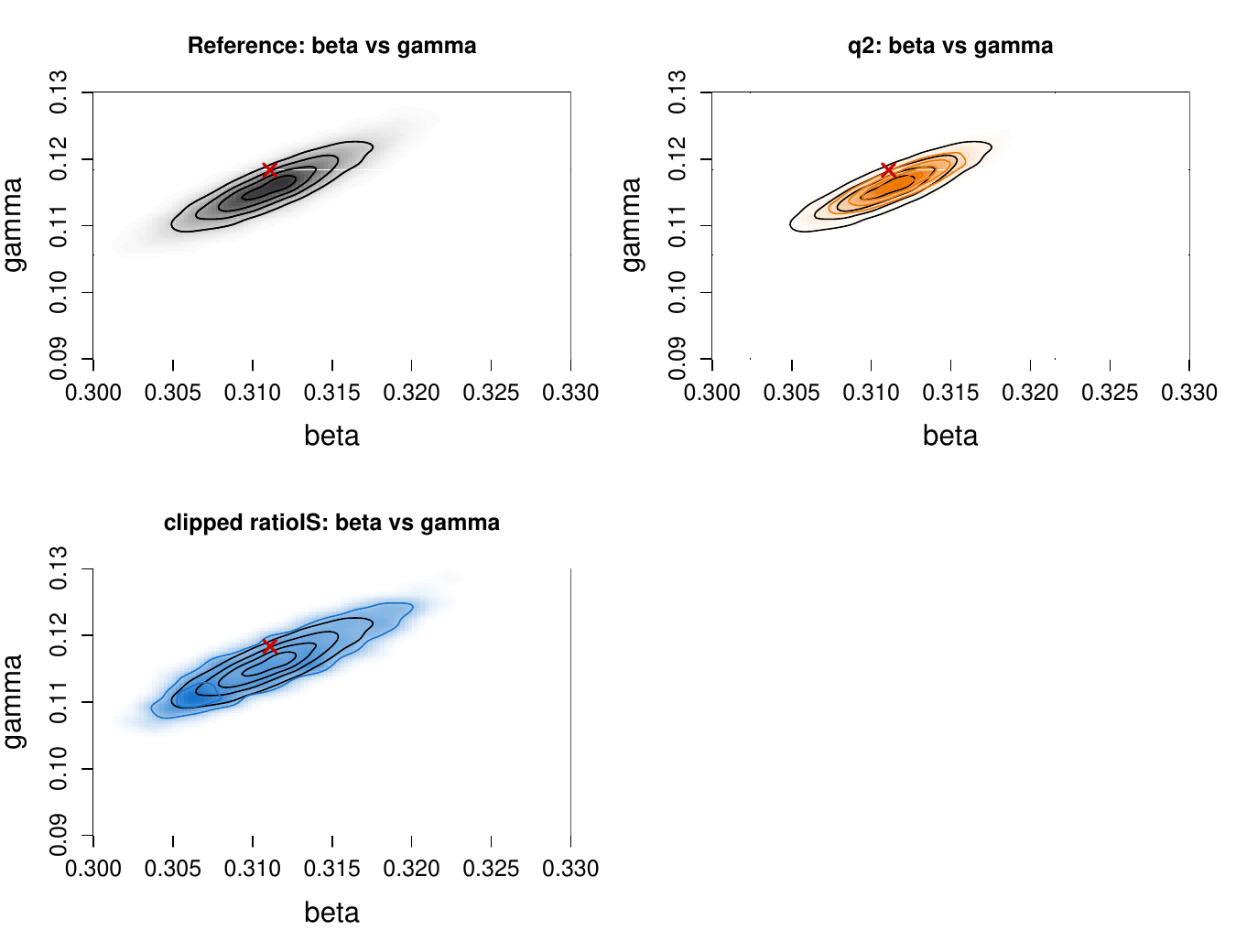}
    \caption{SIR: marginal posteriors based on observation 5. (Upper left) reference posterior; (upper right) uncorrected $\hat{q}_2(\thetaVec\mid \yobs)$ in yellow (and reference contour lines in black); (bottom left) clipped ratio corrected posterior in blue (and reference contour lines in black). Red crosses denote data-generating parameter values.}
    \label{fig:sir_obs5}
\end{figure}

\begin{figure}[!t]
    \centering
    \begin{subfigure}[t]{0.35\textwidth}
        \centering
        \includegraphics[width=\linewidth]{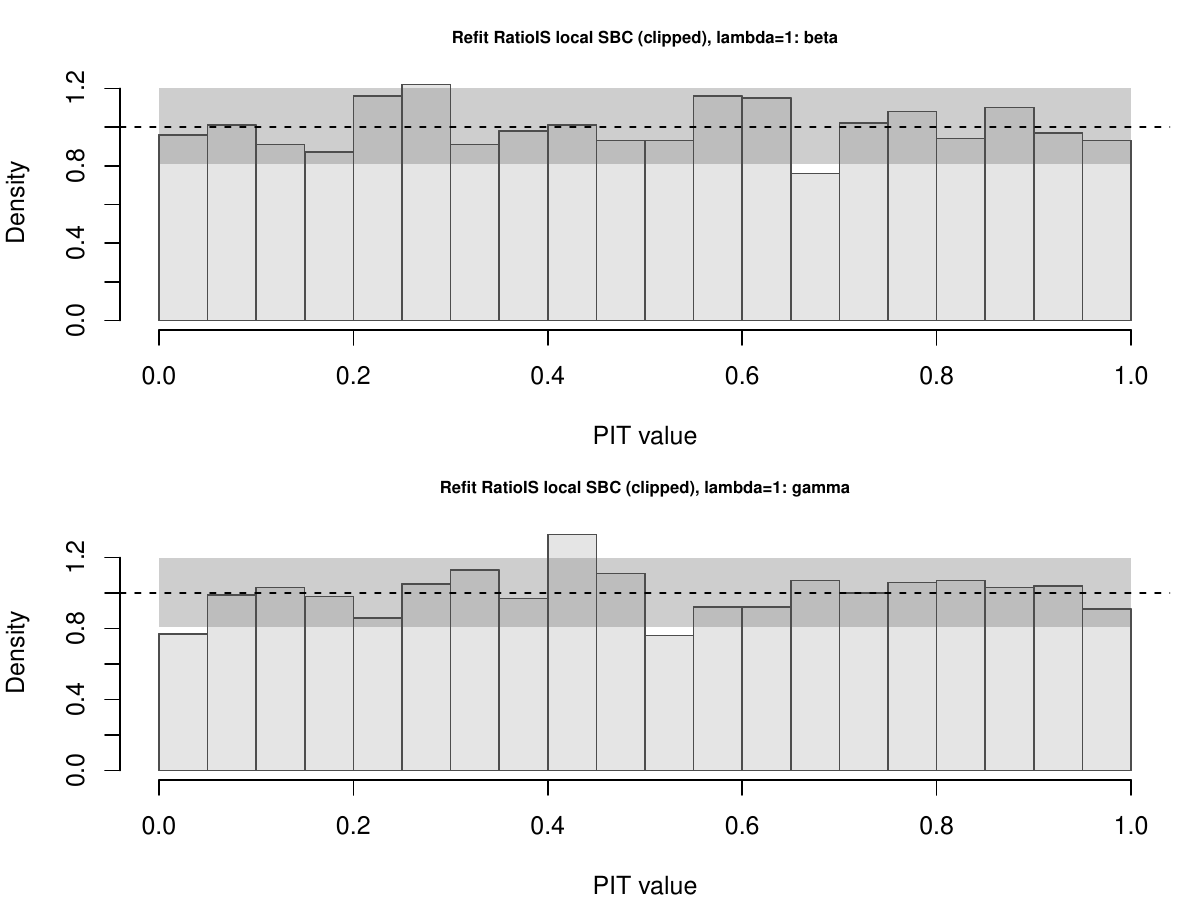}
        \caption{PIT histograms for $\beta$ (top) and $\gamma$ (bottom).}
    \end{subfigure}
\hspace{0.05\textwidth}
    \begin{subfigure}[t]{0.35\textwidth}
        \centering
        \includegraphics[width=\linewidth]{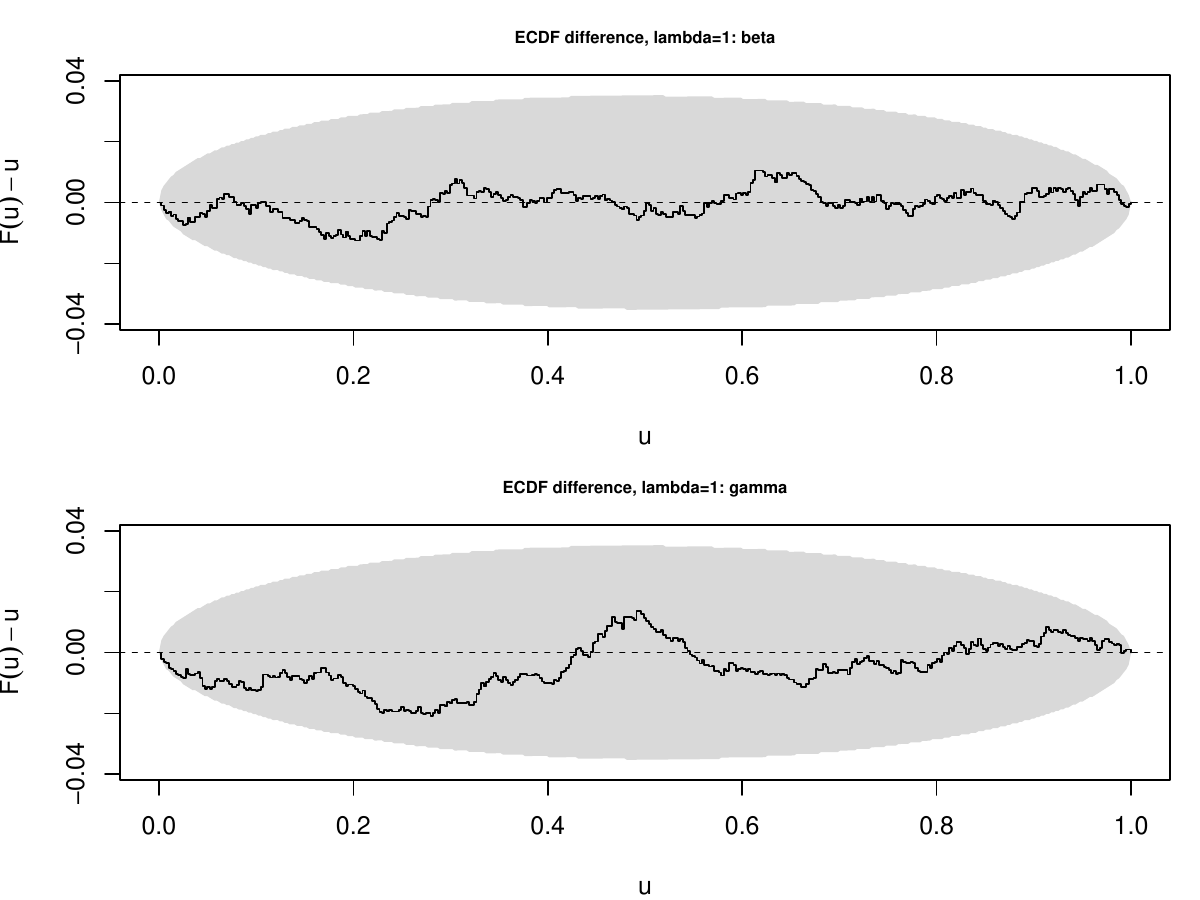}
        \caption{Difference ECDF of the PIT for $\beta$ (top) and $\gamma$ (bottom).}
    \end{subfigure}
    \caption{SIR: localized SBC when $\lambda=1$ for observation 5. }
    \label{fig:sir_obs5_sbc_lambda1}
\end{figure}

\begin{figure}[htbp]
    \centering
    \begin{subfigure}[t]{0.35\textwidth}
        \centering
        \includegraphics[width=\linewidth]{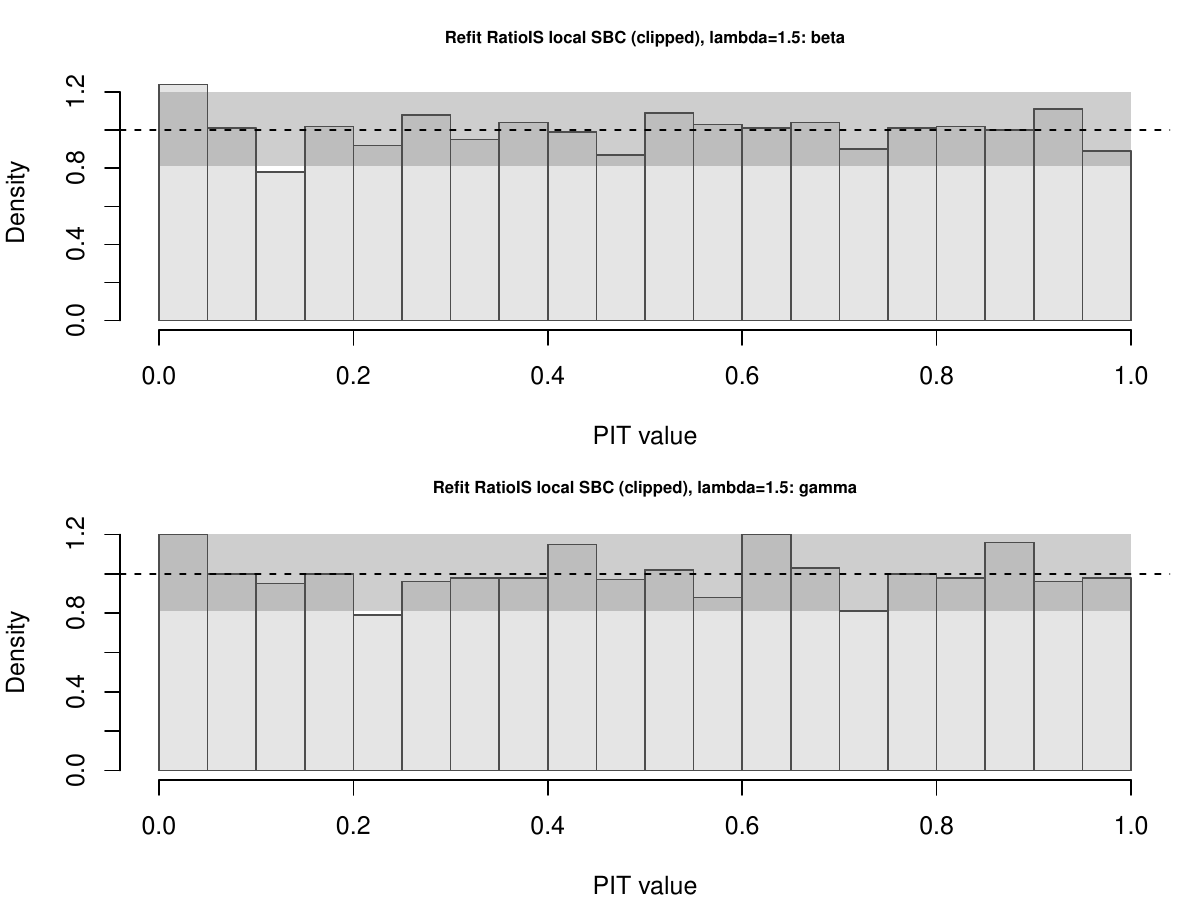}
        \caption{PIT histograms for $\beta$ (top) and $\gamma$ (bottom).}
    \end{subfigure}
\hspace{0.05\textwidth}
    \begin{subfigure}[t]{0.35\textwidth}
        \centering
        \includegraphics[width=\linewidth]{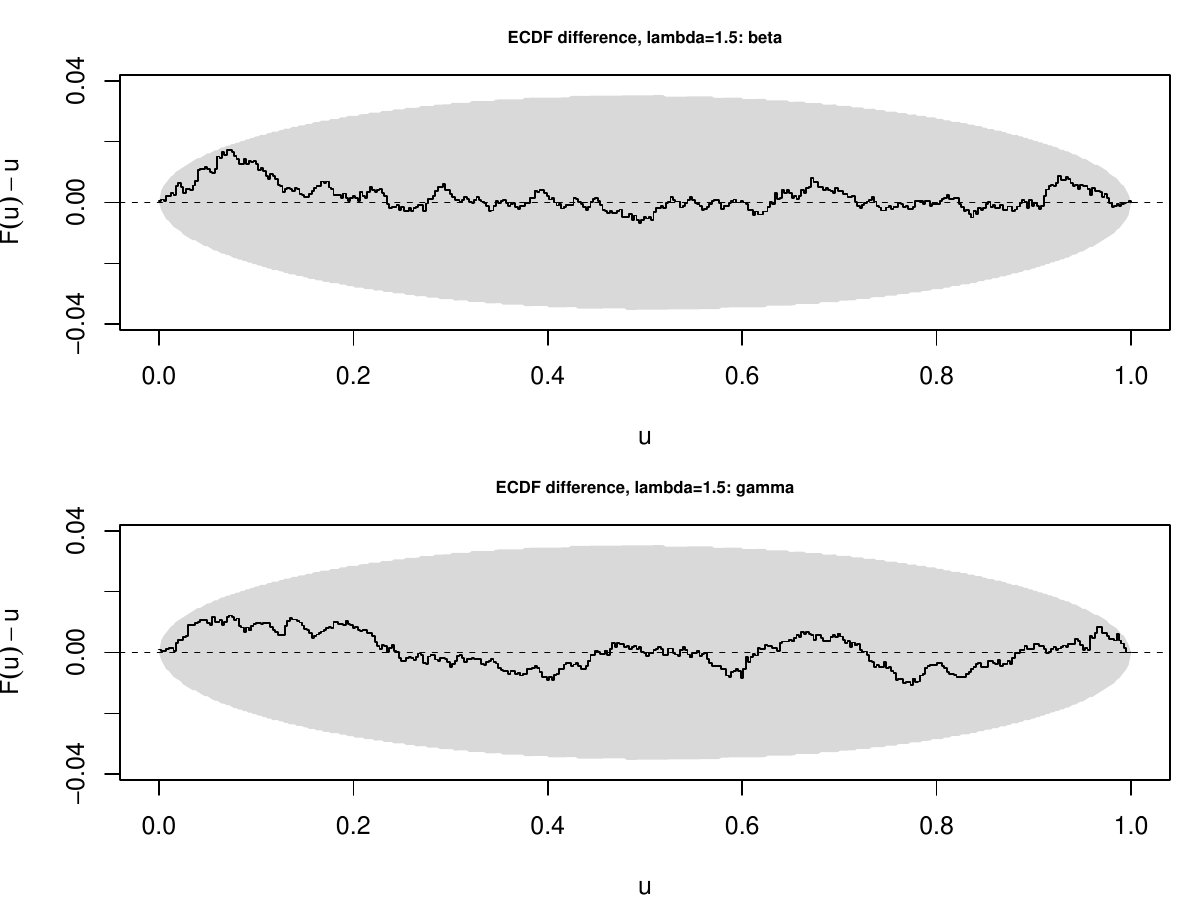}
        \caption{Difference ECDF of the PIT for $\beta$ (top) and $\gamma$ (bottom).}
    \end{subfigure}
    \caption{SIR: localized SBC when $\lambda=1.5$ for observation 5.}
    \label{fig:sir_obs5_sbc_lambda1p5}
\end{figure}

\begin{figure}
    \centering
    \includegraphics[width=0.55\linewidth]{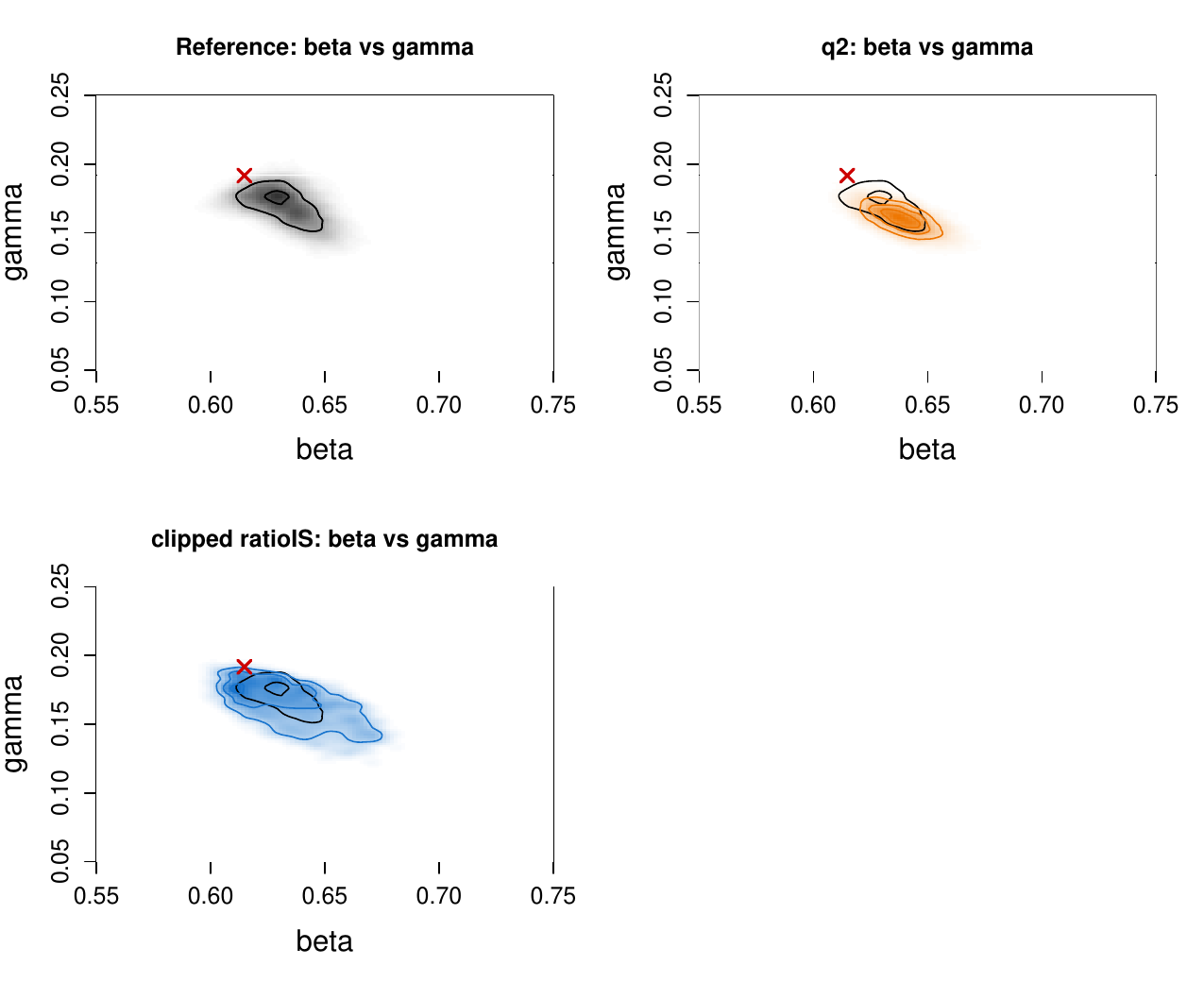}
    \caption{SIR: marginal posteriors based on observation 1. (Upper left) reference posterior; (upper right) uncorrected $\hat{q}_2(\thetaVec\mid \yobs)$ in yellow (and reference contour lines in black); (bottom left) clipped ratio corrected posterior in blue (and reference contour lines in black). Red crosses denote data-generating parameter values.}
    \label{fig:sir_obs1}
\end{figure}

\begin{figure}[t!]
    \centering
    \begin{subfigure}[t]{0.35\textwidth}
        \centering
        \includegraphics[width=\linewidth]{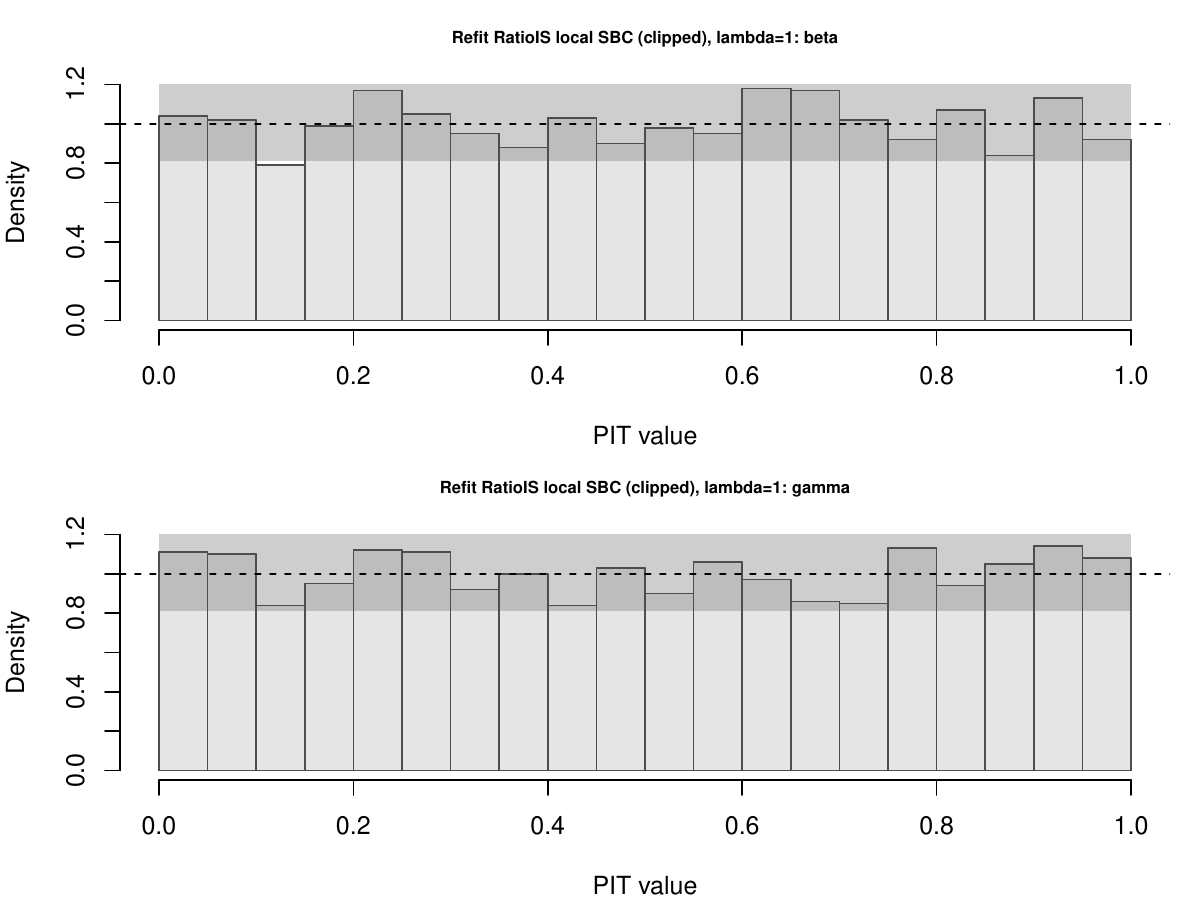}
        \caption{PIT histograms for $\beta$ (top) and $\gamma$ (bottom).}
    \end{subfigure}
\hspace{0.05\textwidth}
    \begin{subfigure}[t]{0.35\textwidth}
        \centering
        \includegraphics[width=\linewidth]{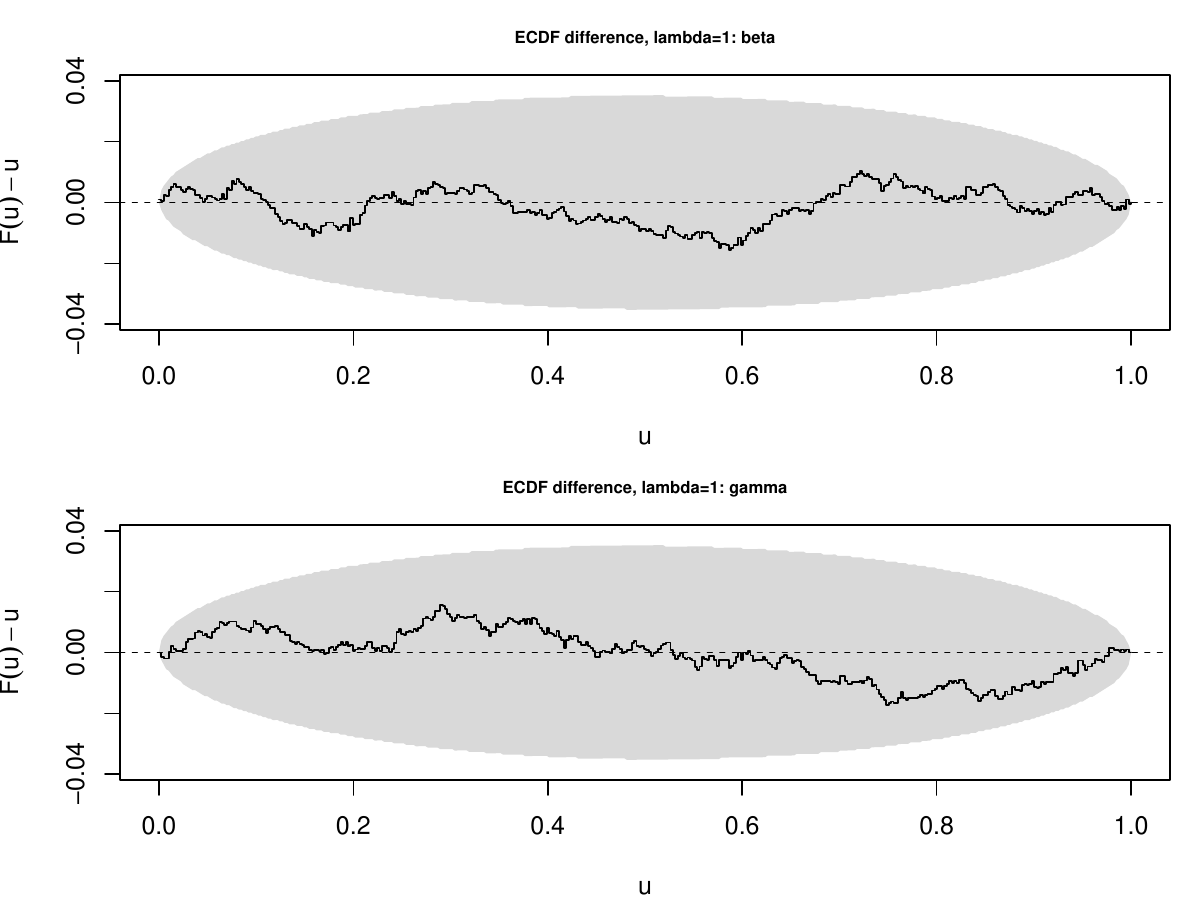}
        \caption{Difference ECDF of the PIT for $\beta$ (top) and $\gamma$ (bottom).}
    \end{subfigure}
    \caption{SIR: localized SBC when $\lambda=1$ for observation 1. }
    \label{fig:sir_obs1_sbc_lambda1}
\end{figure}

\begin{figure}[htbp]
    \centering

    \begin{subfigure}[t]{0.35\textwidth}
        \centering
        \includegraphics[width=\linewidth]{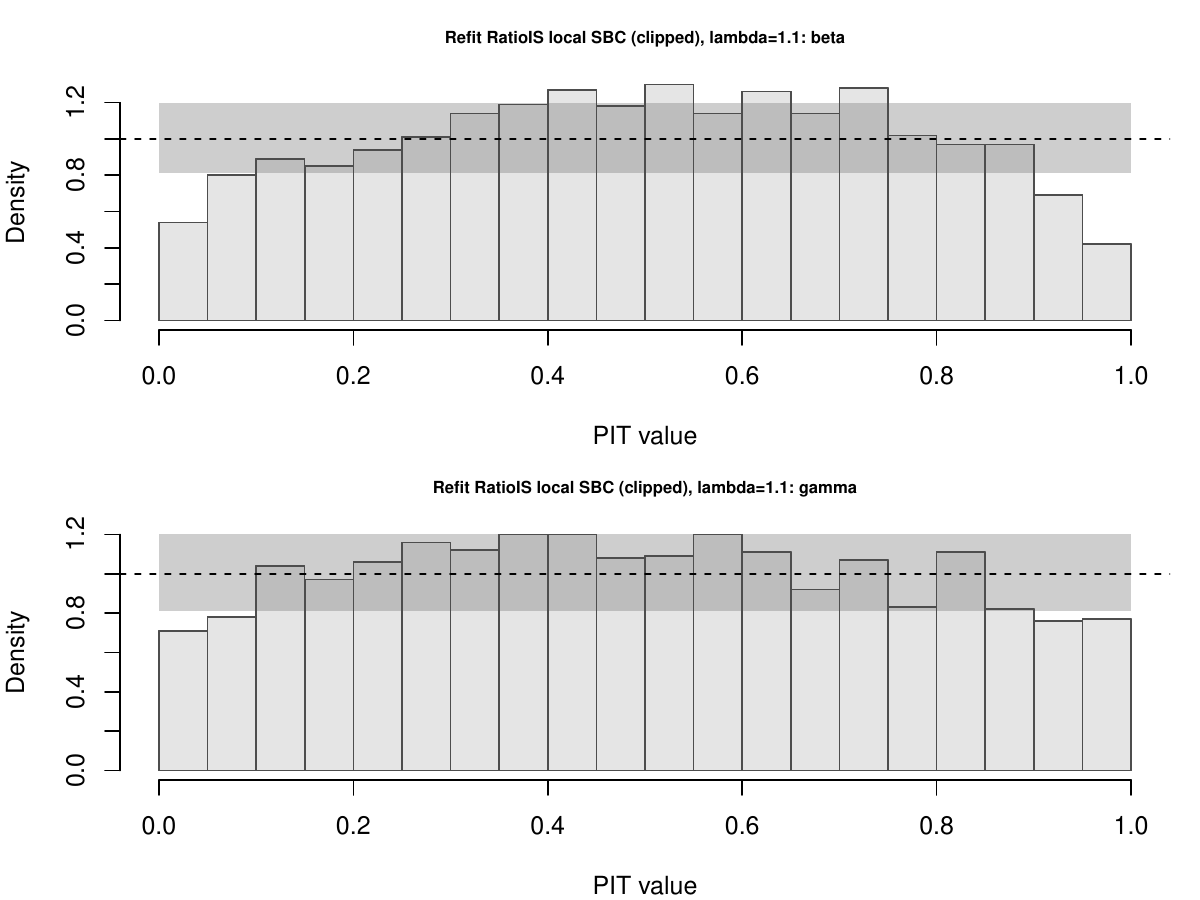}
        \caption{PIT histograms for $\beta$ (top) and $\gamma$ (bottom).}
    \end{subfigure}
\hspace{0.05\textwidth}
    \begin{subfigure}[t]{0.35\textwidth}
        \centering
        \includegraphics[width=\linewidth]{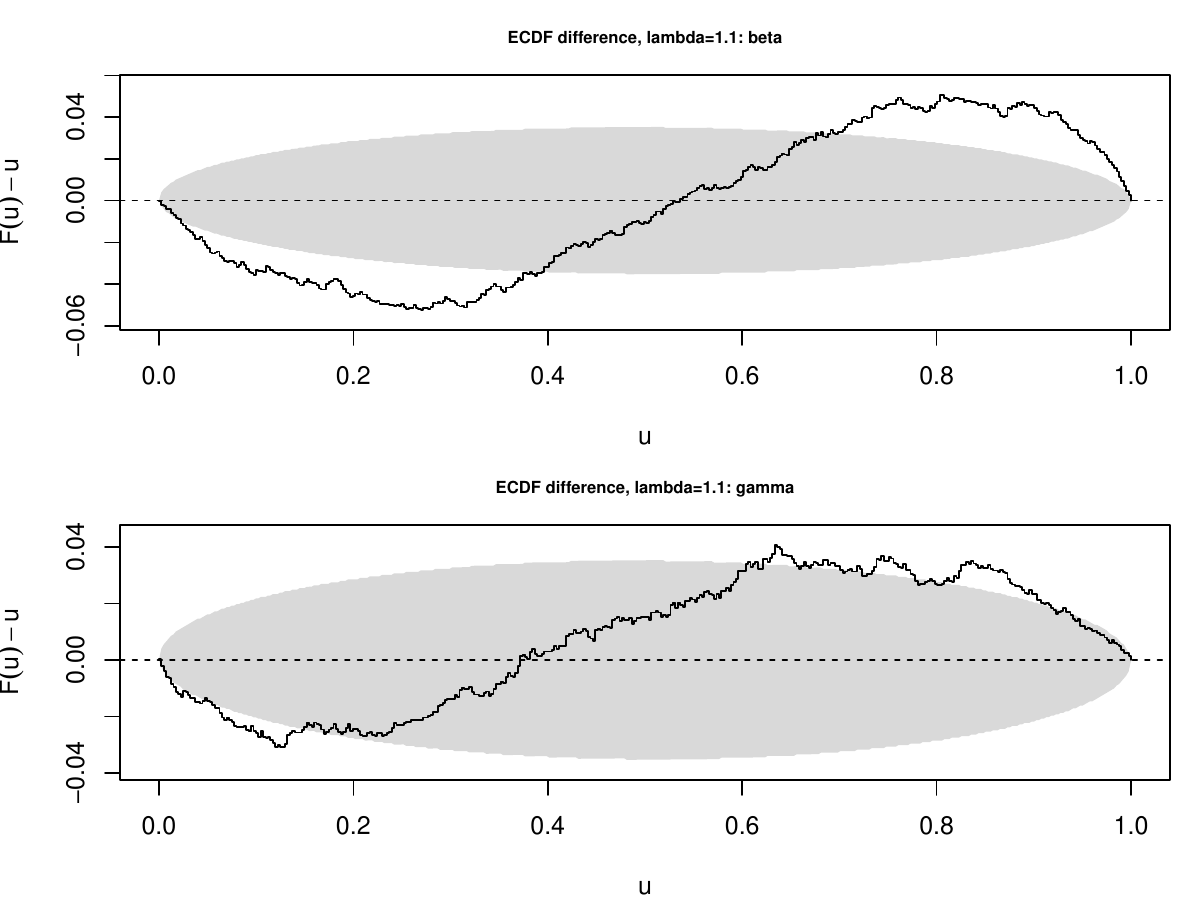}
        \caption{Difference ECDF of the PIT for $\beta$ (top) and $\gamma$ (bottom).}
    \end{subfigure}

    \caption{SIR: localized SBC when $\lambda=1.1$ for observation 1.}
    \label{fig:sir_obs1_sbc_lambda1p1}
\end{figure}

We applied localized simulation-based calibration to the corrected clipped
posterior, generating parameters from the local design
$g_\lambda(\thetaVec\mid\yobs)$ for several inflation factors
$\lambda\in\{1,1.05,1.1,1.25,1.5,2\}$ (there is no inflation when $\lambda=1$). 
 For each parameter $\theta_j$
($j=1,2$ here, since $\thetaVec=(\beta,\gamma)$), we primarily assess
calibration using plots of the difference between the empirical
cumulative distribution function (ECDF) of the
$\hat{U}^{\mathrm{loc}}_{i,j}$ and the CDF of a
$\operatorname{Uniform}(0,1)$ distribution, together with simultaneous
confidence bands, following \cite{sailynoja2022graphical}. For
completeness, we also report histograms of the
$\hat{U}^{\mathrm{loc}}_{i,j}$; although these provide the usual visual
interpretation of SBC, they are less informative than the ECDF-based
diagnostic because they depend on binning and do not account for the
dependence across bins, as discussed in Section~\ref{sec:local-sbc}.
In summary, out of the 10 tested observations, in 8 cases posteriors are calibrated at $\lambda=1$ and in six cases these are calibrated for $1\leq\lambda\leq 1.5$, and for no observation we observed calibration when $\lambda=2$. More information is available in Supplementary Material.
As an illustration, for observation $\yobs$ number \#5 in \texttt{sbibm}, for which posteriors are in Figure \ref{fig:sir_obs5}, localized SBC diagnostics using $\lambda=1$ are  in Figure \ref{fig:sir_obs5_sbc_lambda1}, and in Figure \ref{fig:sir_obs5_sbc_lambda1p5} using $\lambda=1.5$. These confirm calibration for the considered $\lambda$. For economy of space we do not report the figures based on the other $\lambda$ values, but we observed that, for the same observation, inference is calibrated also when $\lambda\in\{1.05,1.1,1.25\}$ but not for the more extreme case $\lambda=2$. 
For further illustration we now consider observation \#1: inference is accurate, see Figure \ref{fig:sir_obs1}, and calibrated according to our localized SBC when $\lambda=1$ but not for a slightly larger value $\lambda=1.1$ (and not calibrated also for even larger values), see Figure \ref{fig:sir_obs1_sbc_lambda1} and Figure \ref{fig:sir_obs1_sbc_lambda1p1} respectively. Indeed, the clipped ratio-corrected posterior for this observation is slightly more dispersed than that for observation 5, and when the local design is broadened by a factor \(\lambda>1\), the localized SBC reveals a loss of calibration, indicating that the inference procedure is less reliable over the corresponding broader neighborhood of the reported posterior.

\subsection{Lotka--Volterra model}
\label{sec:lotka_volterra}

\begin{figure}[t!]
    \centering

    \includegraphics[width=0.7\linewidth]{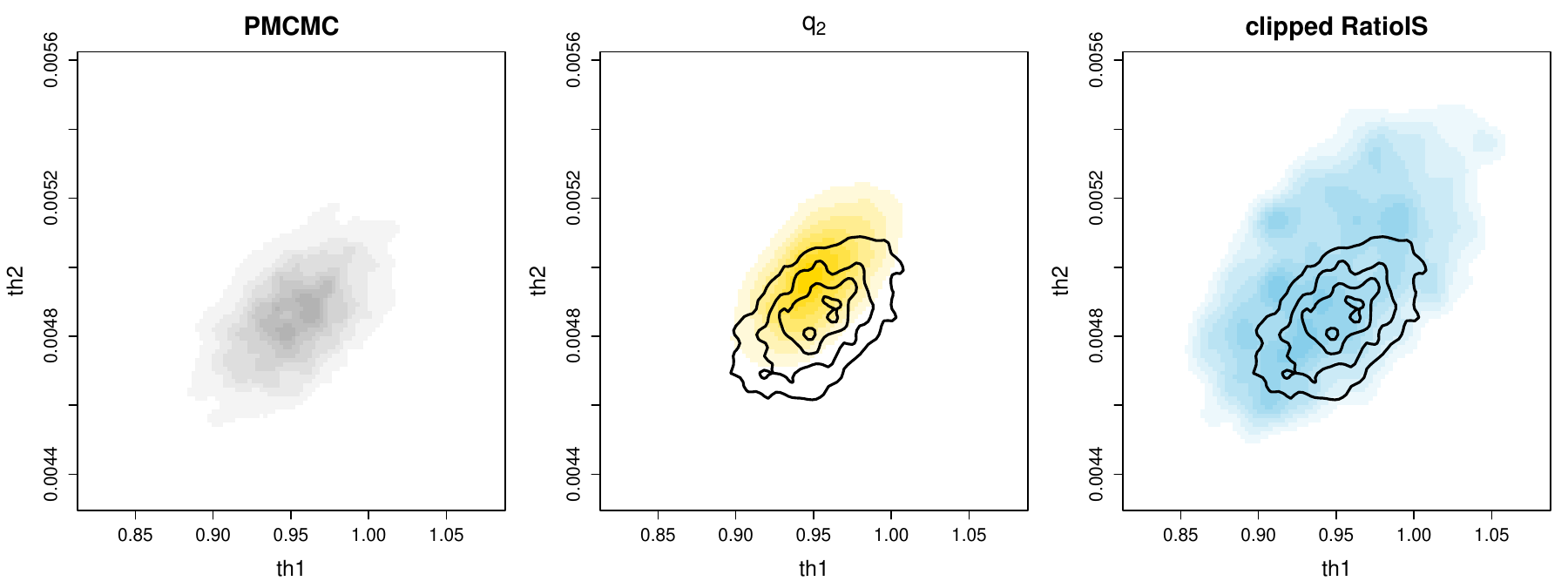}

    \caption{Lotka-Volterra: left plot shows the PMCMC reference posterior, the center plot shows in yellow the uncorrected \(q_2\) surrogate and in black the contour lines of the reference, the right plot shows in blue the clipped ratio-corrected approximation and in black the contour lines of the reference.}
    \label{fig:lv_contours}
\end{figure}

\begin{figure}
    \centering
    \includegraphics[width=0.5\linewidth]{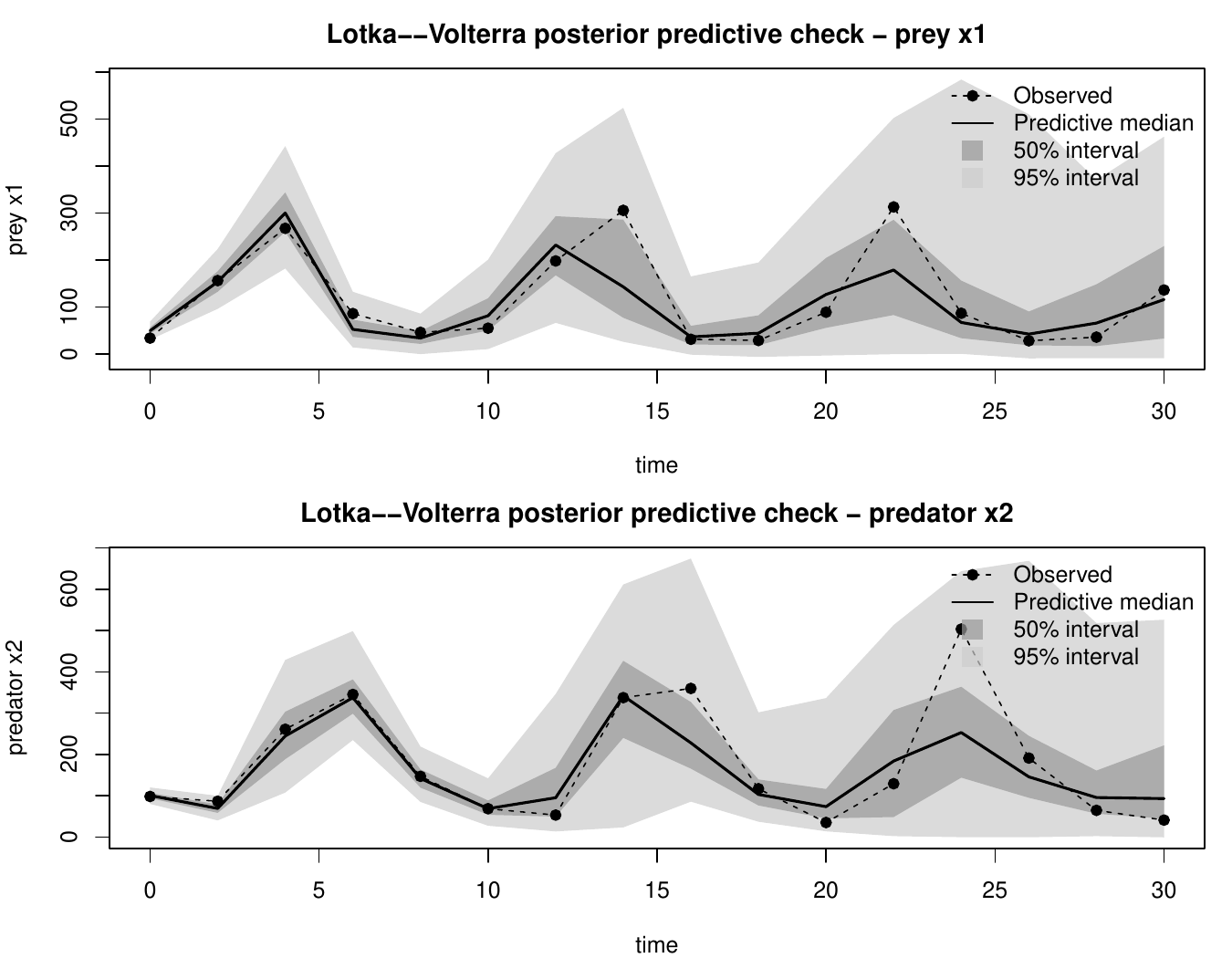}
    \caption{Lotka-Volterra: posterior predictive trajectories produced with the clipped ratio-corrected posterior draws. Circles are observed data, the bold solid line is the median of the simulated trajectories, the dark-gray region encloses 50\% posterior mass, and the light-gray region encloses 95\% mass.}
    \label{fig:lv_ppc_correctedclipped}
\end{figure}

As a second stochastic kinetic model, we consider the Lotka--Volterra predator--prey system implemented in the \texttt{smfsb} package \citep{smfsb}. The model describes the interaction between two populations, denoted by $\{X_1(t)\}_t$ and $\{X_2(t)\}_t$, representing the prey and predator populations, respectively. The model is formulated as a continuous-time Markov jump process, and we follow the three-reactions specification used in \texttt{smfsb} (reproduction of prey, predation, and predator's death). 
Parameters to infer are three reaction-rates
$\thetaVec=(\theta_1,\theta_2,\theta_3)$ and the stochastic dynamics can be simulated exactly using Gillespie's stochastic simulation algorithm.
The parameter $\theta_1$ controls prey reproduction, $\theta_2$ controls the frequency of predator--prey interactions, and $\theta_3$ determines the predator death rate. The interaction parameter $\theta_2$ couples the two populations and gives rise to the characteristic oscillatory predator--prey dynamics. We assume that the populations are observed with measurement error, and the observed data for each population ($j=1,2$) are given by
\[
Y_j(t_i)=X_j(t_i)+\varepsilon_{ij},
\qquad
\varepsilon_{ij}\overset{\mathrm{iid}}{\sim}\mathcal{N}(0,10^2),
\qquad j=1,2.
\]
This observational setup corresponds to the \texttt{LVnoise10} dataset shipped with \texttt{smfsb}, and in practice we load \texttt{LVnoise10} to represent the $\yVec_{\mathrm{obs}}$ in our experiments.
Ground-truth parameters are
$
(c_1,c_2,c_3)=(1,0.005,0.6),
$
and fix an initial state
$
X(0)=(50,100)^\top$. 
The number of mixture components was initialized at $K_{\mathrm{start}}=10$, then
we used $
    N_0=2\times 10^4$,
    $N_a=2\times 10^4$,
    $N_b=10^4$,
hence a total of $N_0+N_a+N_b=5\times 10^4$ runs of the generative model were performed, which is a relatively small simulation budget for this model. It is known that, for the given model configuration, the parameter region giving rise to the characteristic oscillatory behavior in the population is very narrow: Figure \ref{fig:lv_contours} gives the ``exact'' posterior (in black) based on $8\times 10^4$ post burn-in iterations of a pseudomarginal MCMC algorithm, which samples exactly from the true posterior as the number of iterations goes to infinity (to approximate the likelihood when obtaining exact posterior draws, we used a bootstrap filter with 200 particles, which produced excellent mixing). Despite the tight posterior region, as shown in Figure \ref{fig:lv_contours} our method manages to recover it, and once more we notice that the uncorrected  surrogate $\hat{q}_2$ covers a too narrow posterior region, while the clipped ratio-corrected posterior produces a more dispersed yet plausible approximation. In fact the  posterior predictive simulations in Figure \ref{fig:lv_ppc_correctedclipped} (conditionally on the ratio-corrected draws) are almost identical to those obtained with pseudomarginal samples, which are found in Supplementary Material. 
As for localized SBC, we consider a perturbation scheme $\lambda\in\{1,1.05,1.1,1.25,1.5,2\}$ but in this case already with $\lambda=1$ localized SBC displays no calibration for $\theta_1$, while this issue is not present for $\theta_2$ and $\theta_3$. For $\lambda>1$ miscalibration appears also for the other two parameters. Results are shown in Supplementary Material.

\subsection{Epidemiological application} 
\label{sec:ecoli}

We now consider a real-data epidemiological application based on the model and
observed data of \cite{ojala2025basic}, who studied the transmission dynamics
of pandemic \textit{Escherichia coli} ST131 clades. 
The model combines
a deterministic compartmental model for asymptomatic colonization with a
stochastic observation model relating the latent colonization process to
observed bloodstream infections (BSIs). The resulting likelihood is
intractable.
We consider inference for the clade denoted ST131-A and use their same annual BSI observations (with the same methodology we could also report results for clades ST131-C1 and ST131-C2, but these are left out). Unlike the ABC analysis in \cite{ojala2025basic}, which reduces the annual BSI
trajectory to three summary statistics, in our analysis the complete
vector of annual BSI observations is supplied to the
GLLiM-based inference procedure. 
The latent colonization dynamics are described by the same SIR model as in \eqref{eq:sir-ode}, with parameters $(\beta,\gamma)$, and the observational model is a Poisson distribution. Following \cite{ojala2025basic}, rather than working directly with $(\beta,\gamma)$, inference is
performed using the epidemiologically more interpretable parametrization
$
R_0=\beta/\gamma$,
$\tau=\beta-\gamma,
$
where $R_0$ is the basic reproduction number and $\tau$ is the net
transmission rate. Therefore,
$
\gamma=\tau/(R_0-1)$,
$
\beta=\tau R_0/(R_0-1)$. A third parameter, $D_t$, is a delay parameter specific to the observational model which, to ease the inference, we transform as $\widetilde D_t = 0.001D_t$ same as in \cite{ojala2025basic}. The parameter vector to infer is therefore
$
\thetaVec=(\tau,R_0,\widetilde D_t).
$ All the model-specific details are in Supplementary Material.
For the inference we considered a total of $10^5$ model simulations, distributed as $N_0=5\times 10^4$, $N_a=3\times 10^4$ and $N_b=2\times 10^4$, and use $K=10$ components.
The inference resulted in a posterior comparable to that reported in \cite{ojala2025basic} (see Supplementary Material), with informative posterior regions for $\tau$ and $R_0$ and a non-informative posterior for $D_t$ (after transforming $\tilde{D}_t$ to the original scale). Therefore, we recover the same broad scientific conclusion
about the delay parameter as in \cite{ojala2025basic}: the data contain too little information about $D_t$. 
Finally, we performed localized SBC for $\lambda\in\{1,1.05,1.1, 1.25, 1.5,2\}$ and we observed calibration for  all tested $\lambda<2$ values, hence for the more dispersed case using $\lambda=2$ we observe miscalibration. See the Supplementary Material section for results.

\section{Discussion}

We have introduced two main contributions. First, an importance sampling methodology for simulation-based inference (SBI), producing cheap-to-evaluate surrogate approximations of the intractable likelihood and posterior distribution -- both based on Gaussian mixtures-of-experts -- and employing a likelihood-to-evidence ratio correction implemented as a regression-based classifier. The ratio estimator is fitted only once for a given local design and is then locally amortized. It can subsequently be evaluated cheaply for arbitrary parameter--summary simulations within that local design. This is particularly attractive and marries well with the development of a lightweight simulation based calibration methodology, which is our second contribution.
We highlighted differences with the ``linear LFIRE'' method of \cite{thomas2022likelihood}: with linear LFIRE, a separate ratio classifier is fitted for each parameter value, whereas our classifier is fitted \textit{once} over the local parameter--summary space. 
The proposed workflow keeps the main computational attraction of SeMPLE \citep{haggstrom2024fast}: Gaussian locally linear maps initially provide amortized closed-form Gaussian-mixture inverse conditionals that can be evaluated and sampled cheaply. As in SeMPLE, when the amortized conditionals get specialized to learn conditionals for a specific observation, a correction term is required to produce (approximate) posterior draws. Here we avoid the MCMC-based correction step used in SeMPLE, in favor of ratio-correction importance sampling.  The success of the new strategy depends on the quality and support of the final proposal, the stability of the ratio estimator, and the appropriateness of the data representation used for fitting. The advantage of using an importance-sampling based approach rather than MCMC is that the resulting simulations contain independent proposal draws, rather than correlated samples affected by rejection, which reduces the effective diversity of the training data. Moreover our novel approach can be implemented in an embarrassingly parallel fashion, and the computational cost of importance sampling is more predictable, as opposed to the problem-dependent number of MCMC iterations required to obtain sufficiently mixed chains. However, the importance-sampling formulation introduces a different challenge: if the final proposal does not adequately cover the corrected target, the importance weights may become degenerate. This issue is alleviated via a weight-clipping strategy. Perhaps more importantly, the ratio-correction approach marries particularly well with localized SBC, as using the original SeMPLE with localized SBC would have resulted in a more computationally intensive tool, as we discuss below.

Our second contribution is a novel simulation-based calibration tool (SBC), named \textit{localized SBC}, based on a local (proposal) design $g_\lambda$, depending on a parameter $\lambda\geq 1$ set by the experimenter.
A satisfactory localized SBC at \(\lambda=1\) indicates calibration in the region emphasized by the reported posterior approximation. Under progressively broader local designs $g_\lambda$, corresponding to an increasing \(\lambda\), it probes whether this calibration persists when pseudo-true parameters are drawn from progressively broader neighborhoods of that approximation. In our experiments, observations for which the ratio-corrected posterior is close to the reference posterior tend to retain calibration over a wider range of \(\lambda\), whereas broader local designs can reveal calibration problems when the posterior approximation is less accurate. While localized SBC is not restricted to SBI, the method is particularly appealing for SBI studies. In our implementation, where localized SBC is combined with the locally amortized ratio correction, the procedure is computationally lightweight in that the full inference workflow does not need to be repeated for every simulated
pseudo-experiment $(\thetaVec^\star,\yVec^\star)$. In this implementation, localized SBC remains dependent on the quality of the learned data representation and the ratio estimator. It is possible to implement localized SBC with other inference methods, including the MCMC-based
SeMPLE correction. However, in that case, each of the
$N_{\mathrm{check}}$ simulated pseudo-experiments
$(\thetaVec^\star,\yVec^\star)$ would require a separate MCMC chain. If each chain requires $N_{\mathrm{MCMC}}$ iterations, the total number of MCMC iterations for localized SBC would be $N_{\mathrm{MCMC}}\times N_{\mathrm{check}}$, together with the non-negligible overhead of evaluating Gaussian-mixture densities at each
iteration. With localized SBC implemented using ratio-corrected importance sampling, the local correction is instead performed through independent proposal draws and evaluations of the fitted ratio estimator. This avoids
repeated Gaussian-mixture density evaluations for the $M_{\mathrm{IS}}\times N_{\mathrm{check}}$ particles, which can also be processed in parallel across the $N_{\mathrm{check}}$ pseudo-experiments.

\bibliographystyle{abbrvnat}
\bibliography{biblio}

\begin{thebibliography}{44}
\providecommand{\natexlab}[1]{#1}
\providecommand{\url}[1]{\texttt{#1}}
\expandafter\ifx\csname urlstyle\endcsname\relax
  \providecommand{\doi}[1]{doi: #1}\else
  \providecommand{\doi}{doi: \begingroup \urlstyle{rm}\Url}\fi

\bibitem[Andrieu and Roberts(2009)]{andrieu2009pseudo}
C.~Andrieu and G.~O. Roberts.
\newblock The pseudo-marginal approach for efficient {Monte Carlo}
  computations.
\newblock \emph{The Annals of Statistics}, 2009.

\bibitem[Andrieu et~al.(2010)Andrieu, Doucet, and
  Holenstein]{andrieu2010particle}
C.~Andrieu, A.~Doucet, and R.~Holenstein.
\newblock Particle {M}arkov chain {M}onte {C}arlo methods.
\newblock \emph{Journal of the Royal Statistical Society Series B: Statistical
  Methodology}, 72\penalty0 (3):\penalty0 269--342, 2010.

\bibitem[Blum et~al.(2013)Blum, Nunes, Prangle, and
  Sisson]{blum2013comparative}
M.~G. Blum, M.~A. Nunes, D.~Prangle, and S.~A. Sisson.
\newblock A comparative review of dimension reduction methods in approximate
  {B}ayesian computation.
\newblock \emph{Statistical Science}, pages 189--208, 2013.

\bibitem[Chen et~al.(2021)Chen, Zhang, Gutmann, Courville, and
  Zhu]{chen2021neural}
Y.~Chen, D.~Zhang, M.~U. Gutmann, A.~Courville, and Z.~Zhu.
\newblock Neural approximate sufficient statistics for implicit models.
\newblock In \emph{International Conference on Learning Representations}, 2021.

\bibitem[Cook et~al.(2006)Cook, Gelman, and Rubin]{cook2006validation}
S.~R. Cook, A.~Gelman, and D.~B. Rubin.
\newblock Validation of software for {B}ayesian models using posterior
  quantiles.
\newblock \emph{Journal of Computational and Graphical Statistics}, 15\penalty0
  (3):\penalty0 675--692, 2006.

\bibitem[Cranmer et~al.(2015)Cranmer, Pavez, and
  Louppe]{cranmer2015approximating}
K.~Cranmer, J.~Pavez, and G.~Louppe.
\newblock Approximating likelihood ratios with calibrated discriminative
  classifiers.
\newblock \emph{arXiv preprint arXiv:1506.02169}, 2015.

\bibitem[Cranmer et~al.(2020)Cranmer, Brehmer, and Louppe]{cranmer}
K.~Cranmer, J.~Brehmer, and G.~Louppe.
\newblock The frontier of simulation-based inference.
\newblock \emph{Proceedings of the National Academy of Sciences}, 117:\penalty0
  201912789, 05 2020.
\newblock \doi{10.1073/pnas.1912789117}.

\bibitem[Delaunoy et~al.(2022)Delaunoy, Hermans, Rozet, Wehenkel, and
  Louppe]{delaunoy2022towards}
A.~Delaunoy, J.~Hermans, F.~Rozet, A.~Wehenkel, and G.~Louppe.
\newblock Towards reliable simulation-based inference with balanced neural
  ratio estimation.
\newblock \emph{Advances in Neural Information Processing Systems},
  35:\penalty0 20025--20037, 2022.

\bibitem[Deleforge et~al.(2014)Deleforge, Forbes, and Horaud]{deleforge}
A.~Deleforge, F.~Forbes, and R.~Horaud.
\newblock {High-dimensional regression with Gaussian mixtures and
  partially-latent response variables}.
\newblock \emph{Statistics and Computing}, 25\penalty0 (5):\penalty0 893--911,
  mar 2014.

\bibitem[Durkan et~al.(2020)Durkan, Murray, and
  Papamakarios]{durkan2020contrastive}
C.~Durkan, I.~Murray, and G.~Papamakarios.
\newblock On contrastive learning for likelihood-free inference.
\newblock In \emph{International conference on machine learning}, pages
  2771--2781. PMLR, 2020.

\bibitem[Fearnhead and Prangle(2012)]{fearnhead2012constructing}
P.~Fearnhead and D.~Prangle.
\newblock Constructing summary statistics for approximate {B}ayesian
  computation: semi-automatic approximate {B}ayesian computation.
\newblock \emph{Journal of the Royal Statistical Society Series B: Statistical
  Methodology}, 74\penalty0 (3):\penalty0 419--474, 2012.

\bibitem[Forbes et~al.(2022)Forbes, Nguyen, Nguyen, and Arbel]{forbes}
F.~Forbes, H.~D. Nguyen, T.~Nguyen, and J.~Arbel.
\newblock {Summary Statistics and Discrepancy Measures for Approximate Bayesian
  Computation via Surrogate Posteriors}.
\newblock \emph{Statistics and Computing}, 32\penalty0 (5), oct 2022.

\bibitem[Friedman et~al.(2010)Friedman, Hastie, and Tibshirani]{Friedman2010}
J.~Friedman, T.~Hastie, and R.~Tibshirani.
\newblock Regularization paths for generalized linear models via coordinate
  descent.
\newblock \emph{Journal of Statistical Software}, 33\penalty0 (1), 2010.
\newblock ISSN 1548-7660.
\newblock \doi{10.18637/jss.v033.i01}.

\bibitem[Greenberg et~al.(2019)Greenberg, Nonnenmacher, and
  Macke]{greenberg2019}
D.~Greenberg, M.~Nonnenmacher, and J.~Macke.
\newblock Automatic posterior transformation for likelihood-free inference.
\newblock In \emph{International Conference on Machine Learning}, pages
  2404--2414. PMLR, 2019.

\bibitem[Gutmann et~al.(2018)Gutmann, Dutta, Kaski, and
  Corander]{gutmann2018likelihood}
M.~U. Gutmann, R.~Dutta, S.~Kaski, and J.~Corander.
\newblock Likelihood-free inference via classification.
\newblock \emph{Statistics and Computing}, 28\penalty0 (2):\penalty0 411--425,
  2018.

\bibitem[H{\"a}ggstr{\"o}m et~al.(2024)H{\"a}ggstr{\"o}m, Rodrigues,
  Oudoumanessah, Forbes, and Picchini]{haggstrom2024fast}
H.~H{\"a}ggstr{\"o}m, P.~L. Rodrigues, G.~Oudoumanessah, F.~Forbes, and
  U.~Picchini.
\newblock Fast, accurate and lightweight sequential simulation-based inference
  using gaussian locally linear mappings.
\newblock \emph{Transactions on Machine Learning Research}, 2024.
\newblock ISSN 2835-8856.
\newblock URL \url{https://openreview.net/forum?id=Q0nzpRcwWn}.

\bibitem[Hermans et~al.(2022)Hermans, Delaunoy, Rozet, Wehenkel, Begy, and
  Louppe]{hermans2022a}
J.~Hermans, A.~Delaunoy, F.~Rozet, A.~Wehenkel, V.~Begy, and G.~Louppe.
\newblock A crisis in simulation-based inference? beware, your posterior
  approximations can be unfaithful.
\newblock \emph{Transactions on Machine Learning Research}, 2022.
\newblock ISSN 2835-8856.
\newblock URL \url{https://openreview.net/forum?id=LHAbHkt6Aq}.

\bibitem[Jacobs et~al.(1991)Jacobs, Jordan, Nowlan, and
  Hinton]{jacobs1991adaptive}
R.~A. Jacobs, M.~I. Jordan, S.~J. Nowlan, and G.~E. Hinton.
\newblock Adaptive mixtures of local experts.
\newblock \emph{Neural computation}, 3\penalty0 (1):\penalty0 79--87, 1991.

\bibitem[Jordan and Jacobs(1994)]{jordan1994hierarchical}
M.~I. Jordan and R.~A. Jacobs.
\newblock Hierarchical mixtures of experts and the em algorithm.
\newblock \emph{Neural computation}, 6\penalty0 (2):\penalty0 181--214, 1994.

\bibitem[Koblents and M{\'\i}guez(2015)]{koblents2015population}
E.~Koblents and J.~M{\'\i}guez.
\newblock A population {Monte Carlo} scheme with transformed weights and its
  application to stochastic kinetic models.
\newblock \emph{Statistics and Computing}, 25\penalty0 (2):\penalty0 407--425,
  2015.

\bibitem[Lueckmann et~al.(2021)Lueckmann, Boelts, Greenberg, Goncalves, and
  Macke]{sbibm}
J.-M. Lueckmann, J.~Boelts, D.~Greenberg, P.~Goncalves, and J.~Macke.
\newblock Benchmarking simulation-based inference.
\newblock In \emph{International conference on artificial intelligence and
  statistics}, pages 343--351. PMLR, 2021.

\bibitem[Marin et~al.(2012)Marin, Pudlo, Robert, and Ryder]{marin}
J.-M. Marin, P.~Pudlo, C.~P. Robert, and R.~J. Ryder.
\newblock {Approximate Bayesian computational methods}.
\newblock \emph{Statistics and Computing}, 22\penalty0 (6):\penalty0
  1167--1180, 2012.

\bibitem[Martino et~al.(2018)Martino, Elvira, Míguez, Artés-Rodríguez, and
  Djuri{\'c}]{martino2018comparison}
L.~Martino, V.~Elvira, J.~Míguez, A.~Artés-Rodríguez, and P.~Djuri{\'c}.
\newblock A comparison of clipping strategies for importance sampling.
\newblock In \emph{2018 IEEE Statistical Signal Processing Workshop (SSP)},
  pages 558--562. IEEE, 2018.

\bibitem[Miller et~al.(2021)Miller, Cole, Forr{\'e}, Louppe, and
  Weniger]{miller2021truncated}
B.~K. Miller, A.~Cole, P.~Forr{\'e}, G.~Louppe, and C.~Weniger.
\newblock Truncated marginal neural ratio estimation.
\newblock \emph{Advances in Neural Information Processing Systems},
  34:\penalty0 129--143, 2021.

\bibitem[Nguyen et~al.(2024)Nguyen, Forbes, Arbel, and
  Duy~Nguyen]{nguyen2024bayesian}
T.~Nguyen, F.~Forbes, J.~Arbel, and H.~Duy~Nguyen.
\newblock Bayesian nonparametric mixture of experts for inverse problems.
\newblock \emph{Journal of Nonparametric Statistics}, pages 1--60, 2024.

\bibitem[Ojala et~al.(2025)Ojala, Pesonen, Gladstone, M{\"a}klin, Tonkin-Hill,
  Marttinen, and Corander]{ojala2025basic}
F.~Ojala, H.~Pesonen, R.~A. Gladstone, T.~M{\"a}klin, G.~Tonkin-Hill,
  P.~Marttinen, and J.~Corander.
\newblock Basic reproduction number varies markedly between closely related
  pandemic {E}scherichia coli clones.
\newblock \emph{Nature Communications}, 16\penalty0 (1):\penalty0 9490, 2025.

\bibitem[Papamakarios and Murray(2016)]{papamakarios2016}
G.~Papamakarios and I.~Murray.
\newblock {Fast $\varepsilon$-free inference of simulation models with Bayesian
  conditional density estimation}.
\newblock \emph{Advances in Neural Information Processing Systems}, 29, 2016.

\bibitem[Papamakarios et~al.(2019)Papamakarios, Sterratt, and
  Murray]{papamakarios2019}
G.~Papamakarios, D.~Sterratt, and I.~Murray.
\newblock {Sequential Neural Likelihood: Fast Likelihood-free Inference with
  Autoregressive Flows}.
\newblock In \emph{Proceedings of the Twenty-Second International Conference on
  Artificial Intelligence and Statistics}, volume~89, pages 837--848. PMLR,
  16--18 Apr 2019.

\bibitem[Perthame et~al.(2026)Perthame, Forbes, Deleforge, Devijver, and
  Gallopin]{xllim}
E.~Perthame, F.~Forbes, A.~Deleforge, E.~Devijver, and M.~Gallopin.
\newblock \emph{xLLiM: High Dimensional Locally-Linear Mapping}, 2026.
\newblock URL \url{https://github.com/epertham/xLLiM}.
\newblock R package version 2.3, commit
  dba703acda2955c9462c0afab8b2b9e6ce45827f.

\bibitem[Pesonen et~al.(2023)Pesonen, Simola, K{\"o}hn-Luque, Vuollekoski, Lai,
  Frigessi, Kaski, Frazier, Maneesoonthorn, Martin, and
  Corander]{pesonen2023abc}
H.~Pesonen, U.~Simola, A.~K{\"o}hn-Luque, H.~Vuollekoski, X.~Lai, A.~Frigessi,
  S.~Kaski, D.~T. Frazier, W.~Maneesoonthorn, G.~M. Martin, and J.~Corander.
\newblock {ABC} of the future.
\newblock \emph{International Statistical Review}, 91\penalty0 (2):\penalty0
  243--268, 2023.

\bibitem[Price et~al.(2018)Price, Drovandi, Lee, and Nott]{price2018bayesian}
L.~F. Price, C.~C. Drovandi, A.~Lee, and D.~J. Nott.
\newblock Bayesian synthetic likelihood.
\newblock \emph{Journal of Computational and Graphical Statistics}, 27\penalty0
  (1):\penalty0 1--11, 2018.

\bibitem[Radev et~al.(2023)Radev, Schmitt, Pratz, Picchini, Koethe, and
  Buerkner]{radev2023jana}
S.~T. Radev, M.~Schmitt, V.~Pratz, U.~Picchini, U.~Koethe, and P.~Buerkner.
\newblock {JANA}: Jointly amortized neural approximation of complex {B}ayesian
  models.
\newblock In \emph{Proceedings of the Thirty-Ninth Conference on Uncertainty in
  Artificial Intelligence}, pages 1695--1706. PMLR, 2023.

\bibitem[S{\"a}ilynoja et~al.(2022)S{\"a}ilynoja, B{\"u}rkner, and
  Vehtari]{sailynoja2022graphical}
T.~S{\"a}ilynoja, P.-C. B{\"u}rkner, and A.~Vehtari.
\newblock Graphical test for discrete uniformity and its applications in
  goodness-of-fit evaluation and multiple sample comparison.
\newblock \emph{Statistics and Computing}, 32\penalty0 (2):\penalty0 32, 2022.

\bibitem[S{\"a}ilynoja et~al.(2026)S{\"a}ilynoja, Schmitt, B{\"u}rkner, and
  Vehtari]{sailynoja2026posterior}
T.~S{\"a}ilynoja, M.~Schmitt, P.-C. B{\"u}rkner, and A.~Vehtari.
\newblock Posterior {SBC}: simulation-based calibration checking conditional on
  data.
\newblock \emph{Statistics and Computing}, 36\penalty0 (2):\penalty0 78, 2026.

\bibitem[Sisson et~al.(2018)Sisson, Fan, and Beaumont]{sisson2018handbook}
S.~A. Sisson, Y.~Fan, and M.~Beaumont, editors.
\newblock \emph{Handbook of approximate {B}ayesian computation}.
\newblock CRC Press, 2018.

\bibitem[Talts et~al.(2018)Talts, Betancourt, Simpson, Vehtari, and
  Gelman]{talts2018validating}
S.~Talts, M.~Betancourt, D.~Simpson, A.~Vehtari, and A.~Gelman.
\newblock Validating bayesian inference algorithms with simulation-based
  calibration.
\newblock \emph{arXiv preprint arXiv:1804.06788}, 2018.

\bibitem[Tay et~al.(2023)Tay, Narasimhan, and Hastie]{Tay2023}
J.~K. Tay, B.~Narasimhan, and T.~Hastie.
\newblock Elastic net regularization paths for all generalized linear models.
\newblock \emph{Journal of Statistical Software}, 106\penalty0 (1), 2023.
\newblock ISSN 1548-7660.
\newblock \doi{10.18637/jss.v106.i01}.

\bibitem[Thomas et~al.(2022)Thomas, Dutta, Corander, Kaski, and
  Gutmann]{thomas2022likelihood}
O.~Thomas, R.~Dutta, J.~Corander, S.~Kaski, and M.~U. Gutmann.
\newblock Likelihood-free inference by ratio estimation.
\newblock \emph{Bayesian Analysis}, 17\penalty0 (1):\penalty0 1--31, 2022.

\bibitem[Wang et~al.(2024)Wang, Kelly, Jenner, Warne, and
  Drovandi]{wang2024comprehensive}
X.~Wang, R.~P. Kelly, A.~L. Jenner, D.~J. Warne, and C.~Drovandi.
\newblock A comprehensive guide to simulation-based inference in computational
  biology.
\newblock \emph{arXiv preprint arXiv:2409.19675}, 2024.

\bibitem[Wilkinson(2024)]{smfsb}
D.~Wilkinson.
\newblock \emph{smfsb: Stochastic Modelling for Systems Biology}, 2024.
\newblock URL \url{https://CRAN.R-project.org/package=smfsb}.
\newblock R package version 1.5.

\bibitem[Wiqvist et~al.(2019)Wiqvist, Mattei, Picchini, and
  Frellsen]{wiqvist19a}
S.~Wiqvist, P.-A. Mattei, U.~Picchini, and J.~Frellsen.
\newblock Partially exchangeable networks and architectures for learning
  summary statistics in approximate {B}ayesian computation.
\newblock In \emph{Proceedings of the 36th International Conference on Machine
  Learning}, volume~97, pages 6798--6807, 09--15 Jun 2019.

\bibitem[Wiqvist et~al.(2021)Wiqvist, Frellsen, and
  Picchini]{wiqvist2021sequential}
S.~Wiqvist, J.~Frellsen, and U.~Picchini.
\newblock Sequential neural posterior and likelihood approximation.
\newblock \emph{arXiv preprint arXiv:2102.06522}, 2021.

\bibitem[Wood(2010)]{wood2010statistical}
S.~N. Wood.
\newblock Statistical inference for noisy nonlinear ecological dynamic systems.
\newblock \emph{Nature}, 466\penalty0 (7310):\penalty0 1102--1104, 2010.

\bibitem[Xu et~al.(1994)Xu, Jordan, and Hinton]{xu1994alternative}
L.~Xu, M.~Jordan, and G.~E. Hinton.
\newblock An alternative model for mixtures of experts.
\newblock \emph{Advances in neural information processing systems}, 7, 1994.

\end{thebibliography}

\subsection*{Acknowledgement of AI use}

The author used OpenAI's ChatGPT to improve spelling, grammar, clarity, and readability during manuscript preparation. The author carefully reviewed and edited the text and takes full responsibility for the final content. ChatGPT was also used to assist with writing, refining, and debugging the computational code accompanying this manuscript. The scientific methodology, algorithmic design, and interpretation of the results were developed by the author. All AI-assisted code was reviewed and tested by the author, who takes full responsibility for its implementation and the reported results.

\clearpage
\newpage

\section*{Supplementary Material}

\subsection{GLLiM models}

We give some background on the GLLiM surrogate posterior and likelihood construction that is not present in the main paper, but that can be found in \cite{forbes,haggstrom2024fast}. 
In the ``forward'' direction $\thetaVec\rightarrow\yVec$, where $\thetaVec$ affects $\yVec$, the locally affine representation in GLLiM is written as
\begin{equation}
    \yVec = \sum_{k=1}^K \mathbb{I}_{\{z = k\}} (\tilde{\boldsymbol A}_k \thetaVec + \tilde{\boldsymbol b}_k + \tilde{\boldsymbol \epsilon}_k),\label{eq:affine}
\end{equation}
where it is clear that $\tilde{\boldsymbol A}_k \thetaVec+\tilde{\boldsymbol b}_k$ induces an affine representation, while the ``locality'' is due to the fact that each component $k$ has its own specific affine representation\footnote{The term ``local'' used here is unrelated to ``local design'' or to the localized simulation-based calibration that is developed in the paper.}. We can consider \eqref{eq:affine} as a surrogate generative model for $\yVec$, that is \eqref{eq:affine} is \textit{not} the assumed DGM model for observations $\yobs$. The DGM that is proposed by the researcher (and which may of course be itself mispecified) is, for example, a probabilistic or a deterministic model $\mathcal{M}(\thetaVec)$ (e.g. a differential equation model, a point process model, a Markov jump process, etc.), with an associated likelihood $p(\yVec|\thetaVec)$, and we write a simulated $\yVec$ from the DGM $\mathcal{M}(\thetaVec)$ as $\yVec\sim p(\yVec|\thetaVec)$ (recall in our context $p(\yVec|\thetaVec)$ is intractable). The researcher  assumes that $\mathcal{M}(\thetaVec)$ is  suitable to represent/model an experiment that gave rise to data observations $\yobs$. That is, for some ``pseudo-true'' parameter value $\thetaVec_\mathrm{o}$, we assume that $\mathcal{M}(\thetaVec_o)\rightarrow \yobs$, that is the assumed DGM has generated the observations $\yobs$, but this is of course up to DGM model's mispecification (hence why we denoted $\thetaVec_\mathrm{o}$ as pseudo-true).
Instead in GLLiM it is assumed that the density of $\yVec\given \thetaVec$ induced by \eqref{eq:affine} when conditionally on component $z=k$, is given by
\begin{equation}
    q_{\tilde{\phiVec}}(\yVec \given \thetaVec, z=k) = \mathcal{N}(\yVec; \tilde{\boldsymbol A}_k \thetaVec + \tilde{\boldsymbol b}_k, \tilde{\boldsymbol \Sigma}_k),
\end{equation}
where we used $\Normal(\bm{\mu},\bm{\Sigma})$ to denote the multivariate Gaussian distribution with mean $\bm{\mu}$ and covariance matrix $\bm{\Sigma}$.
To complete the hierarchical model, $\thetaVec$ is assumed to follow a mixture of Gaussian distributions specified by
\begin{equation}
%\begin{alignat}{2}
     q_{\tilde{\phiVec}}(\thetaVec \given z=k) = \mathcal{N}(\thetaVec; \tilde{\boldsymbol \nu}_k, \tilde{\boldsymbol \Gamma}_k), \quad
     q_{\tilde{\phiVec}}(z=k) = \pi_k. \label{eq:pik}
%\end{alignat}
\end{equation}
The rest of the construction can be found in the main paper.

\subsection{From \texorpdfstring{$\tilde{\phiVec}$}{tilde phi} 
to \texorpdfstring{$\phiVec$}{phi}}

Let $
    \tilde{\phiVec} = \{\pi_k, \tilde{\boldsymbol{\nu}}_k, \tilde{\boldsymbol \Gamma}_k, \tilde{\boldsymbol A}_k, \tilde{\boldsymbol b}_k, \tilde{\boldsymbol \Sigma}_k\}_{k=1}^K.
$ denote parameters of the surrogate GLLiM likelihood $q_{\tilde{\phiVec}}(\yVec|\thetaVec)$, then the surrogate GLLiM posterior  $q_{{\phiVec}}(\thetaVec|\yVec)$ has parameters $\phiVec = \{\pi_k, \boldsymbol \nu_k, \boldsymbol \Gamma_k, \boldsymbol A_k, \boldsymbol b_k, \boldsymbol \Sigma_k\}_{k=1}^K$ given by \cite{deleforge}: 
\begin{align} 
    \boldsymbol{\nu}_k &= \tilde{\boldsymbol A_k} \tilde{\boldsymbol \nu}_k + \tilde{\boldsymbol b}_k, \nonumber \\
    \boldsymbol \Gamma_k &= \tilde{\boldsymbol \Sigma}_k + \tilde{\boldsymbol A_k} \tilde{\boldsymbol \Gamma}_k \tilde{\boldsymbol A_k}^\top, \label{eq:forwardParam_line1} \nonumber\\
    \boldsymbol \Sigma_k &= \big(\tilde{\boldsymbol \Gamma}_k^{-1} + \tilde{\boldsymbol A_k}^\top \tilde{\boldsymbol \Sigma}_k^{-1} \tilde{\boldsymbol A_k} \big)^{-1},\nonumber \\
    \boldsymbol A_k &= \boldsymbol \Sigma_k \tilde{\boldsymbol A_k}^\top \tilde{\boldsymbol \Sigma}_k^{-1}, \nonumber\\
    \boldsymbol b_k &= \boldsymbol \Sigma_k \big(\tilde{\boldsymbol \Gamma}_k^{-1} \tilde{\boldsymbol{\nu}}_k - \tilde{\boldsymbol A_k}^\top \tilde{\boldsymbol \Sigma}_k^{-1} \tilde{\boldsymbol b}_k \big).\nonumber
\end{align}

\subsection{Background on ratio estimation via classification}
We show that $D(\thetaVec,\yVec)/(1-D(\thetaVec,\yVec))=r(\thetaVec,\yVec)$, which is the key equality for the ratio-estimation approach via classification. 
Let $W\in\{0,1\}$ denote the class label, where $W=1$ corresponds
to samples from the joint density $p(\thetaVec,\yVec)$ and $W=0$ to samples
from the product of marginals $p(\thetaVec)p(\yVec)$. Define
$
D(\thetaVec,\yVec)
=
\Pr(W=1\mid \thetaVec,\yVec),
$
then
\[
D(\thetaVec,\yVec)=\Pr(W=1\mid \thetaVec,\yVec)
=
\frac{
p(\thetaVec,\yVec\mid W=1)\Pr(W=1)
}{
p(\thetaVec,\yVec)
},
\]
where
\[
p(\thetaVec,\yVec)
=
p(\thetaVec,\yVec\mid W=1)\Pr(W=1)
+
p(\thetaVec,\yVec\mid W=0)\Pr(W=0).
\]
Since
$
p(\thetaVec,\yVec\mid W=1)=p(\thetaVec,\yVec)
$
for the joint class, and
$
p(\thetaVec,\yVec\mid W=0)=p(\thetaVec)p(\yVec)
$
for the product-of-marginals class, we obtain
\[
D(\thetaVec,\yVec)
=
\frac{
\Pr(W=1)\,p(\thetaVec,\yVec)
}{
\Pr(W=1)\,p(\thetaVec,\yVec)
+
\Pr(W=0)\,p(\thetaVec)p(\yVec)
}.
\]
Similarly,
\[
1-D(\thetaVec,\yVec)
=
\Pr(W=0\mid\thetaVec,\yVec)
=
\frac{
\Pr(W=0)\,p(\thetaVec)p(\yVec)
}{
\Pr(W=1)\,p(\thetaVec,\yVec)
+
\Pr(W=0)\,p(\thetaVec)p(\yVec)
}.
\]
Finally,
\[
\frac{D(\thetaVec,\yVec)}
     {1-D(\thetaVec,\yVec)}
=
\frac{
\Pr(W=1)\,p(\thetaVec,\yVec)
}{
\Pr(W=0)\,p(\thetaVec)p(\yVec)
},
\]
and under balanced classes (i.e. assuming $P(W=0)=P(W=1)$),
\[
\frac{D(\thetaVec,\yVec)}
     {1-D(\thetaVec,\yVec)}
=
\frac{p(\thetaVec,\yVec)}
     {p(\thetaVec)p(\yVec)}=
\frac{p(\yVec\mid\thetaVec)}
     {p(\yVec)}
\equiv r(\thetaVec,\yVec).
\]

\subsection{Implementation of the localized SBC with active components}

In our implementation of localized SBC, the local GLLiM refit used only those components of the final \(\widehat q_2\) having posterior responsibility at \(\yobs\) larger than a small threshold \(\epsilon_{\rm post}\). This restriction was used only for the localized SBC refit, in order to avoid carrying essentially unsupported mixture components into the local GLLiM refit based on the comparatively small simulation bank of size $N_{\mathrm{loc}}$. Instead, no component pruning was performed during the main inference procedure. In all our examples we used \(\epsilon_{\rm post}=10^{-6}\), so the active-component restriction removes only components having essentially negligible posterior responsibility at $\yobs$.

More precisely, when running localized SBC we define the set of non-negligible mixture components (``active components'') induced by a small threshold $\epsilon_{\mathrm{post}}$:  \[
\mathcal K_{\mathrm{act}}
=
\left\{
k\in\{1,\ldots,K\}:
\widehat\omega_{2k}(\yobs)>\epsilon_{\mathrm{post}}
\right\}.
\]
Then we define the surrogate $\widehat q_2^{\mathrm{act}}$ with ``active components'' \[
\widehat q_2^{\mathrm{act}}
(\thetaVec\mid\yobs)
=
\sum_{k\in\mathcal K_{\mathrm{act}}}
\widetilde\omega_{2k}(\yobs)\,
\mathcal N\!\left\{
\thetaVec;
\widehat{\mathbf{m}}_{2k}(\yobs),
\widehat{\mathbf{\Sigma}}_{2k}
\right\},
\]
with renormalized weights
\[
\widetilde\omega_{2k}(\yobs)
=
\frac{
\widehat\omega_{2k}(\yobs)
}{
\displaystyle
\sum_{\ell\in\mathcal K_{\mathrm{act}}}
\widehat\omega_{2\ell}(\yobs)
},
\qquad
k\in\mathcal K_{\mathrm{act}}.
\]
The local design distribution is then obtained by inflating the within-component covariance
matrices of the active mixture,
\[
g_\lambda(\thetaVec)
=
\sum_{k\in\mathcal K_{\mathrm{act}}}
\widetilde\omega_{2k}(\yobs)\,
\mathcal N\!\left\{
\thetaVec;
\widehat{\mathbf{m}}_{2k}(\yobs),
\lambda^2\widehat{\mathbf{\Sigma}}_{2k}
\right\},
\qquad \lambda\ge1.
\]
Thus $\lambda=1$ reproduces the active posterior mixture, whereas
$\lambda>1$ defines a broader local neighborhood for the calibration check. 

\subsection{Implementation settings}

Unless stated otherwise for a specific experiment, the following implementation
settings were used throughout. The learned statistic $T(y)$ entering the ratio
classifier consists of the posterior mean, the lower-triangular entries of the
posterior covariance matrix (including its diagonal), and the posterior mixture
weights obtained from the final localized GLLiM fit, as defined in \eqref{eq:summaries}.
Each coordinate of $T(y)$ is centered and scaled using the corresponding
second-round training bank before classifier fitting; no further
dimension-reduction transformation is applied. The logistic ratio classifier
uses the baseline feature map in~(19), namely the linear terms in
$(\theta,T)$ together with all parameter--summary interactions
$\{\theta_j T_\ell\}$, while componentwise quadratic terms and within-block
interactions are omitted. We fit the classifier using \texttt{glmnet} with
lasso regularization ($\alpha=1$), with the penalty selected by
cross-validation using the \texttt{lambda.1se} rule. We use 10-fold
cross-validation unless stated otherwise. Classifier covariates are standardized internally by
\texttt{glmnet}.

For the GLLiM fits we used version 2.3 of the development version of the \texttt{xLLiM} R package \citep{xllim}, installed via 

\begin{verbatim}
 devtools::install_github("epertham/xLLiM",
  ref = "dba703acda2955c9462c0afab8b2b9e6ce45827f")   
\end{verbatim}
Moreover, we modified the EM stopping criterion in the original function \texttt{gllim}, as follows
\small
\begin{verbatim}
 if (iter >= min_iter && iter > patience) {  # "patience" means that several consecutive 
 # small improvements are required, instead of just one. It avoids 
 # stopping because of a single flat step.
    recent_delta <- diff(LL[(iter - patience):iter])
    recent_rel <- recent_delta / (abs(LL[(iter - patience):(iter - 1)]) + 1)
      
      converged <- all(is.finite(recent_rel)) &&
        all(recent_delta >= -1e-8) &&
        all(recent_rel < em_tol)
    }
  # In the original gllim function the stopping criterion followed the approach below   
  #  deltaLL_total <- max(LL[1:iter]) - min(LL[1:iter])
  #  deltaLL <- LL[iter] - LL[iter - 1]
  #  converged <- deltaLL <= 0.001 * deltaLL_total
\end{verbatim}
\normalsize
Inference is always initialized with $K_{\mathrm{start}}=10$ mixture components,
and we allowed up to 300 EM iterations, and did not remove components according to
small global mixture probabilities or posterior responsibilities. The first localized GLLiM fit is initialized from the
prior-predictive fit, whereas the final $q_2$ fit is initialized independently.
At both localization stages, 5\% of the training parameters are drawn
defensively from the prior. The final importance-sampling correction uses
$M=10^4$ fresh draws from $\widehat q_2(\theta\mid y_{\mathrm{obs}})$ and
reports the clipped-weight approximation in~(23).

For localized SBC, for each value of $\lambda$ we use
$N_{\mathrm{loc}}=2000$ simulations to fit the local GLLiM and ratio
classifier and an independent $N_{\mathrm{check}}=2000$ simulations for
calibration checking. The local GLLiM is warm-started from the active
components of the fitted $q_2$, where a component is retained when its
posterior responsibility at $\yobs$ exceeds
$\epsilon_{\mathrm{post}}=10^{-6}$. The local ratio classifier uses the same
summary construction, feature map, lasso penalty and cross-validation rule as
the corresponding main inference analysis. For each checking
pseudo-observation we draw $M_{\mathrm{IS}}=5000$ particles from
$g_\lambda$ and apply the same deterministic clipping rule,
$M_c=\lfloor\sqrt{M_{\mathrm{IS}}}\rfloor$, before computing the localized
PIT values.

\subsection{Posterior SBC of \cite{sailynoja2026posterior}}

For ease of comparison with localized SBC, algorithm \ref{alg:posterior-sbc} gives a pseudo-algorithm of the posterior SBC method of \cite{sailynoja2026posterior}. Following the notation in \cite{sailynoja2026posterior}, $\mathcal{A}$ denotes an algorithm of choice for posterior inference.

\begin{algorithm}[t]
\small
\caption{Posterior SBC checking (\citealp{sailynoja2026posterior})}
\label{alg:posterior-sbc}
\begin{algorithmic}[1]

\Require Observed data $\yobs$, inference algorithm $\mathcal{A}$,
model implementing $p(\yVec,\thetaVec)$, number of replications
$N_\mathrm{post}$, number of posterior draws $S$.

\State Apply $\mathcal{A}$ to $\yobs$ to obtain
\[
    \thetaVec^\star_1,\ldots,\thetaVec^\star_{N_\mathrm{post}}
    \sim p(\thetaVec \mid \yobs).
\]

\For{$i=1,\ldots,N_\mathrm{post}$}

    \State Simulate
    \[
        \yVec_i^\star
        \sim p(\yVec\mid\thetaVec^\star_i).
    \]

    \State Apply $\mathcal{A}$ to the augmented data
    $(\yobs,\yVec_i^\star)$ to obtain
    \[
        \thetaVec^{\star\star}_{i,1},\ldots,
        \thetaVec^{\star\star}_{i,S}
        \sim
        p(\thetaVec\mid\yVec_i^\star,\yobs).
    \]

    \State For $j=1,\ldots,d_\theta$, compute the PIT values
    \[
        U_{i,j}^{\mathrm{post}}
        =
        \Pr\!\left(
            \theta^{\star\star}_{j}
            \leq
            \theta^\star_{i,j}
            \mid \yVec_i^\star,\yobs
        \right),
    \]
    or their Monte Carlo approximations
    \[
        \widehat U_{i,j}^{\mathrm{post}}
        =
        \frac{1}{S}
        \sum_{s=1}^{S}
        \mathbb{I}
        \left\{
            \theta^{\star\star}_{i,s,j}
            \leq
            \theta^\star_{i,j}
        \right\}.
    \]

\EndFor

\State Separately for each $j=1,\ldots,d_\theta$, assess whether
\[
    U_{1,j}^{\mathrm{post}},\ldots,
    U_{N_\mathrm{post},j}^{\mathrm{post}}
    \sim \operatorname{Uniform}(0,1),
\]
and similarly for the Monte Carlo approximations.

\end{algorithmic}
\end{algorithm}

\clearpage
\newpage

\subsection{SIR model: further details and experimental results}

The unknown parameter is
$
\thetaVec=(\beta,\gamma),
$
where \(\beta\) and \(\gamma\) denote the infection and recovery rates,
respectively. Independent lognormal priors are assigned to the two
parameters,
\[
\beta \sim
\operatorname{LogNormal}\!\left(\log(0.4),\,0.5^2\right),
\qquad
\gamma \sim
\operatorname{LogNormal}\!\left(\log(0.125),\,0.2^2\right).
\]
The ODE system is numerically integrated from day \(0\) to day \(160\)
on a daily grid using the \texttt{lsoda} solver implemented in the
\texttt{deSolve} R package. Observations are retained at
$
t_j = 17(j-1)$,
$j=1,\ldots,10$,
that is, at days
$
0,17,34,51,68,85,102,119,136,153
$. Figures \ref{fig:obs1-2}-\ref{fig:obs9-10} below show the reference posteriors\footnote{These are provided in \texttt{sbibm} via a rejection sampler, with a proposal function learned via a neural spline flow.}, the non-corrected conditionals $\hat{q}_2(\thetaVec\mid \yobs)$ and the clipped ratio-corrected posterior for all the 10 datasets provided in the \texttt{sbibm} package \citep{sbibm}. Inference appears to be very satisfactory for all observations except observation \#10. On the other hand, for observation \#10 the generating parameter lies outside the main high-posterior-density region of the reference posterior, making this dataset somewhat atypical. In fact, this is the only observation that is not calibrated for $\lambda=1$, see below. 

\noindent
For each observation, Table \ref{tab:sir_ess} gives the raw and clipped effective sample sizes (ESS). The raw ESS is computed from the importance weights associated with the 10,000 proposal draws from
$\hat q_2(\theta\mid y_{\mathrm{obs}})$, before weight clipping, whereas the clipped ESS is computed after applying the deterministic clipping strategy to the ratio-corrected importance weights, and corresponds to the posterior approximation reported in the analysis. The raw importance weights can be highly concentrated for some observations, resulting in very small raw ESS values. Deterministic weight clipping substantially increases the effective sample size of the corrected particle approximation, which is the posterior approximation used for inference.\\

\begin{table}[ht]
\centering
\caption{Effective sample sizes of the importance-sampling correction for the ten SIR observations. These values are based on 10,000 posterior draws.
The raw ESS is computed from the unmodified importance weights, while the clipped ESS is computed after deterministic weight clipping.}
\begin{tabular}{rcc}
\toprule
Observation & ESS (raw) & ESS (clipped) \\
\midrule
1  & 176.7 & 1614.5 \\
2  & 79.1  & 1630.2 \\
3  & 1.2   & 1374.4 \\
4  & 4.2   & 1700.1 \\
5  & 15.7  & 1321.4 \\
6  & 7.0   & 1547.2 \\
7  & 194.4 & 1454.9 \\
8  & 3.6   & 1666.3 \\
9  & 68.6  & 1719.9 \\
10 & 3.7   & 899.9 \\
\bottomrule
\end{tabular}
\label{tab:sir_ess}
\end{table}

\noindent
Regarding calibration via localized SBC, we considered $\lambda\in \{1,1.05,1.1,1.25,1.5,2\}$. Here is a summary of the main findings:
\begin{itemize}
    \item observation 1: calibrated for $\lambda=1$ but not for larger values of $\lambda$;
    \item observation 2: calibrated for $\lambda=1$ but not for larger values of $\lambda$;
    \item observation 3: calibrated for $\lambda=1$ and $\lambda=1.25$ but not for other values of $\lambda$;
    \item observation 4: calibrated for $1\leq\lambda\leq 1.5$ but not for $\lambda=2$;
    \item observation 5: calibrated for $1\leq\lambda\leq 1.5$ but not for $\lambda=2$;
    \item observation 6: calibrated for $1\leq\lambda\leq 1.5$ but not for $\lambda=2$;
    \item observation 7: calibrated for $1\leq\lambda\leq 1.1$ but not for $\lambda\geq 1.25$;
    \item observation 8: calibrated for $1\leq\lambda\leq 1.5$ but not for $\lambda=2$;
    \item observation 9: calibrated for $1\leq\lambda\leq 1.5$ but not for $\lambda=2$;
    \item observation 10: not calibrated for any $\lambda$;
\end{itemize}
For illustration, we add extra diagnostic plots for the case $\lambda=1$ but only for observations 2 and 3 (notice the corresponding plots for observations 1 and 5 are in the main paper), since these diagnostic plots tend to be rather similar one to the other. Figure \ref{fig:sir_obs2_sbc_lambda1} is for observation 2 and Figure \ref{fig:sir_obs3_sbc_lambda1} is for observation 3.

\begin{figure}[htbp]
    \centering

    \begin{subfigure}[t]{0.6\textwidth}
        \centering
        \includegraphics[width=\linewidth]{figures/sir/contours/obs1.pdf}
        \caption{observation 1}
        \label{fig:sub1}
    \end{subfigure}
    \\
    \begin{subfigure}[t]{0.6\textwidth}
        \centering
        \includegraphics[width=\linewidth]{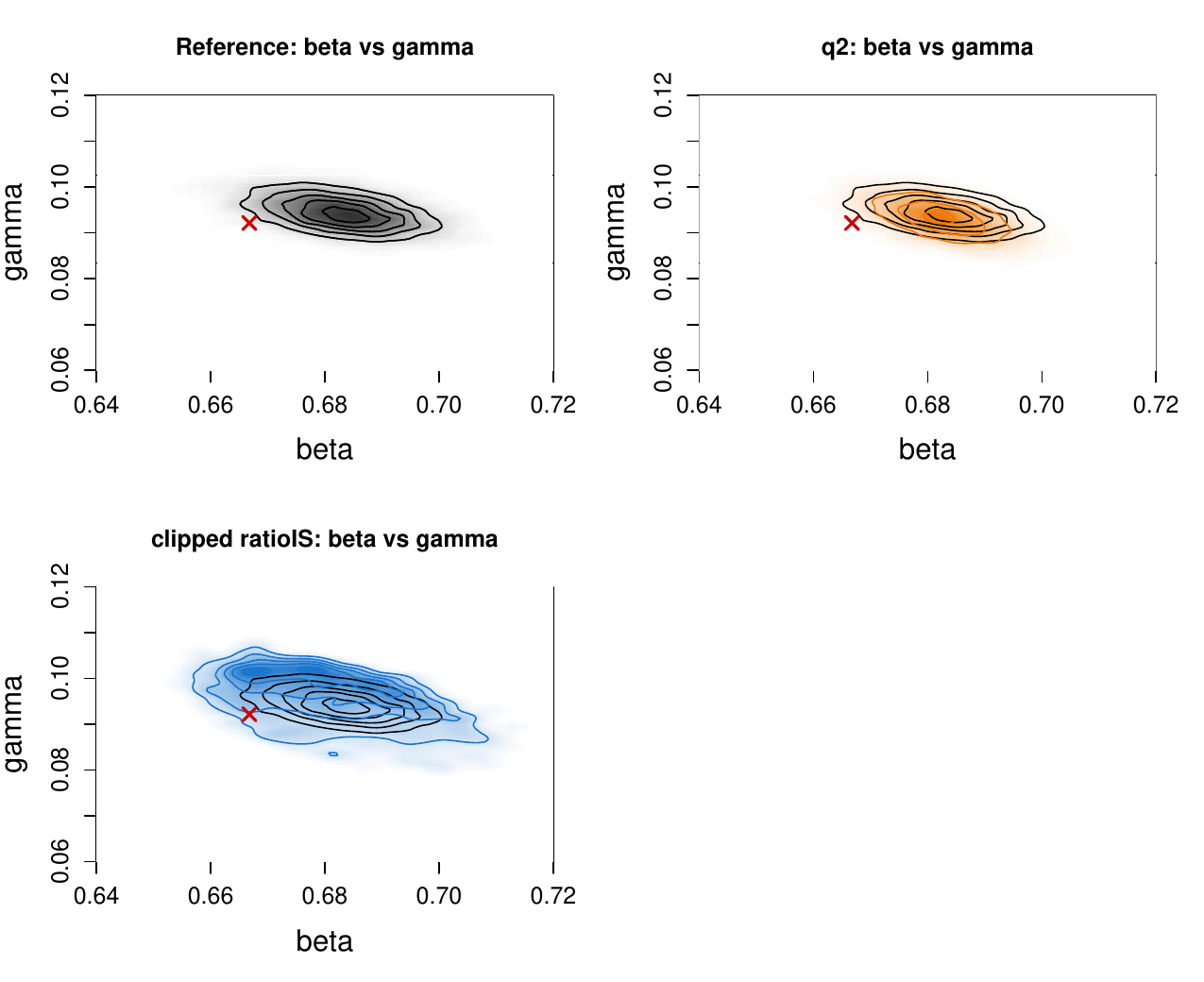}
        \caption{observation 2}
        \label{fig:sub2}
    \end{subfigure}

    \caption{SIR: marginal posteriors for observations 1 and 2. The reference posterior is in black, $\hat{q}_2(\thetaVec\mid \yobs)$ in yellow, and the clipped ratio-corrected posterior is in blue.}
    \label{fig:obs1-2}
\end{figure}

\begin{figure}[htbp]
    \centering

    \begin{subfigure}[t]{0.6\textwidth}
        \centering
        \includegraphics[width=\linewidth]{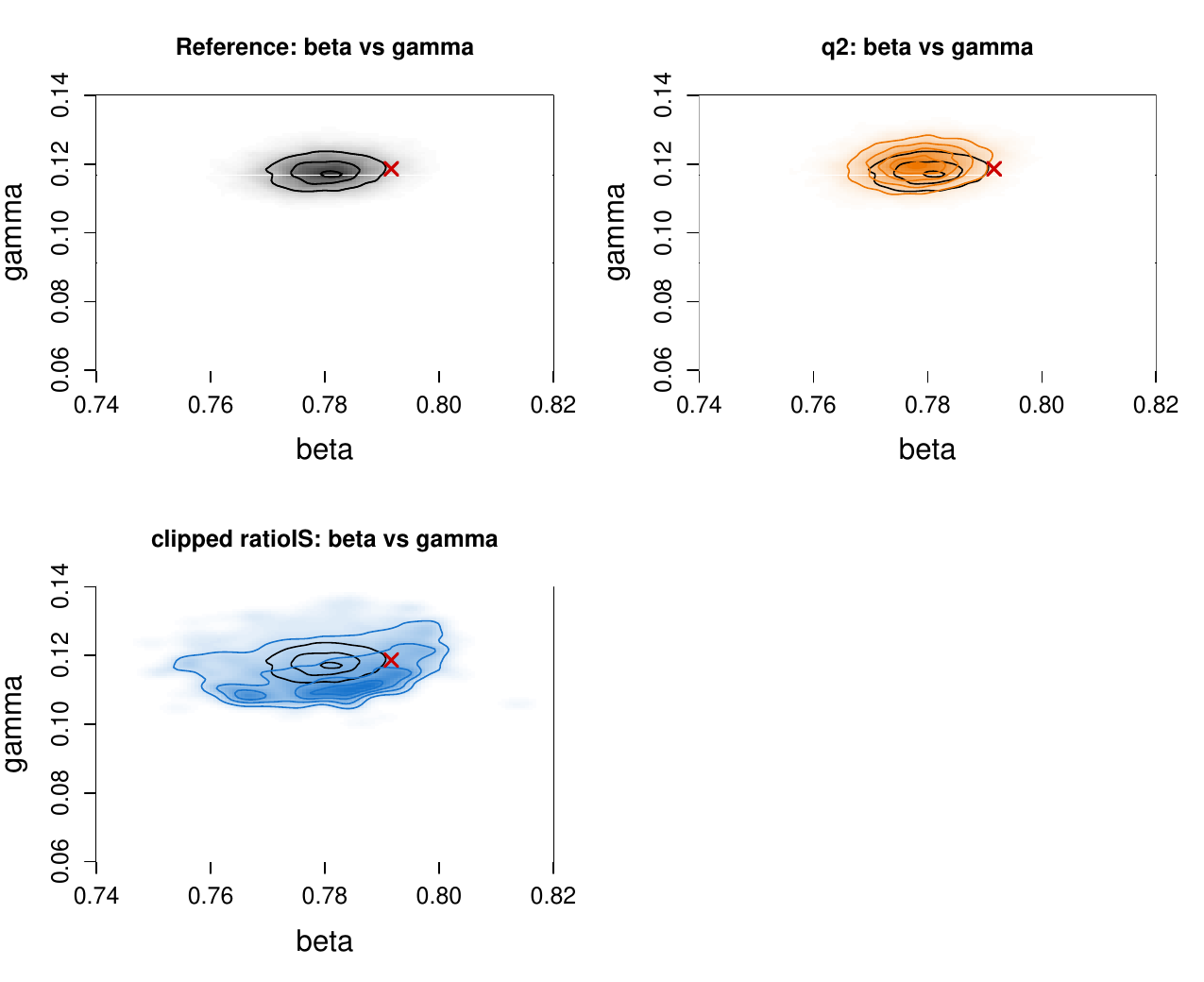}
        \caption{observation 3}
        \label{fig:sub3}
    \end{subfigure}
    \\
    \begin{subfigure}[t]{0.6\textwidth}
        \centering
        \includegraphics[width=\linewidth]{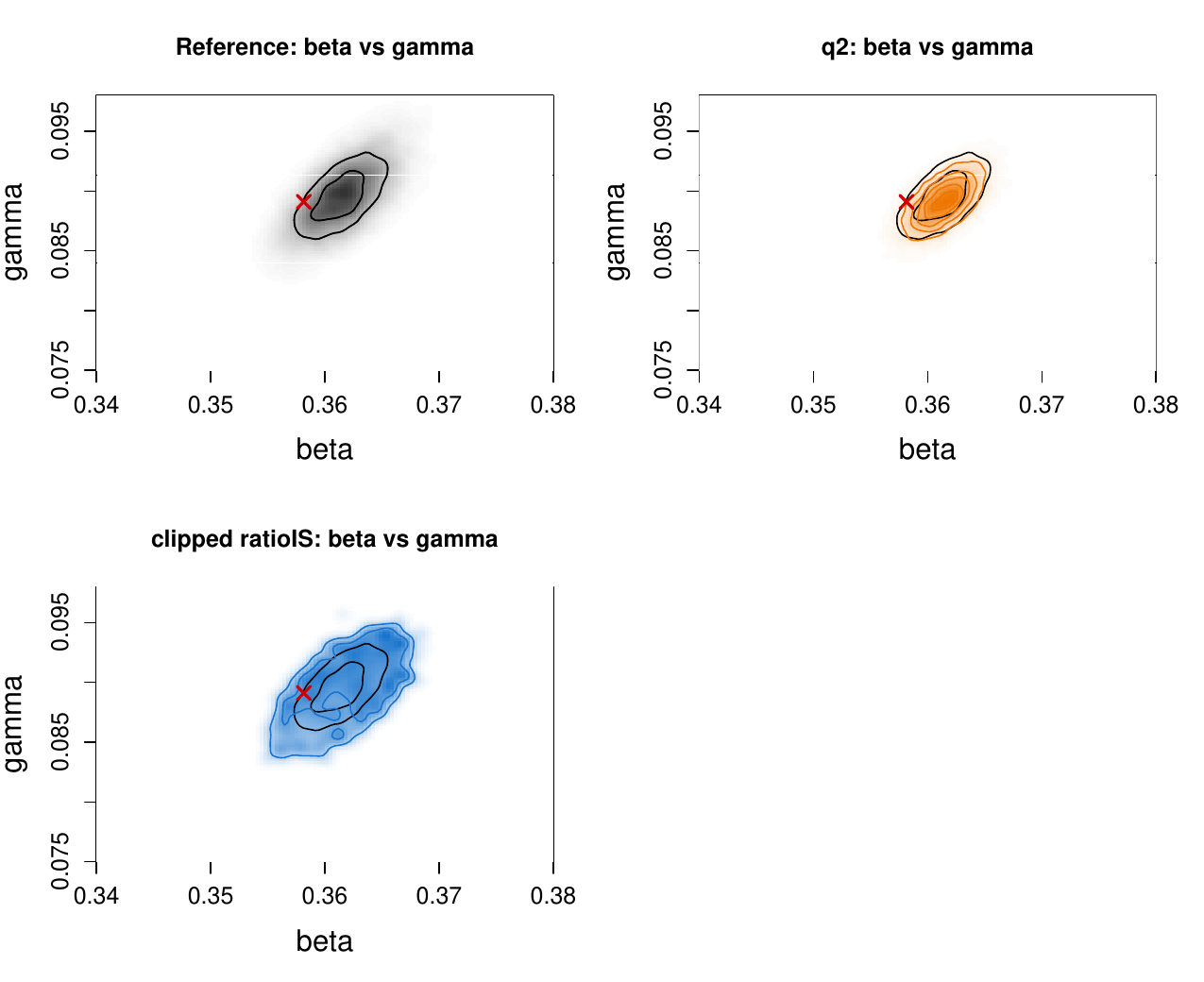}
        \caption{observation 4}
        \label{fig:sub4}
    \end{subfigure}

    \caption{SIR: marginal posteriors for observations 3 and 4. The reference posterior is in black, $\hat{q}_2(\thetaVec\mid \yobs)$ in yellow, and the clipped ratio-corrected posterior is in blue.}
    \label{fig:obs3-4}
\end{figure}

\begin{figure}[htbp]
    \centering

    \begin{subfigure}[t]{0.6\textwidth}
        \centering
        \includegraphics[width=\linewidth]{figures/sir/contours/obs5.pdf}
        \caption{observation 5}
        \label{fig:sub5}
    \end{subfigure}
    \\
    \begin{subfigure}[t]{0.6\textwidth}
        \centering
        \includegraphics[width=\linewidth]{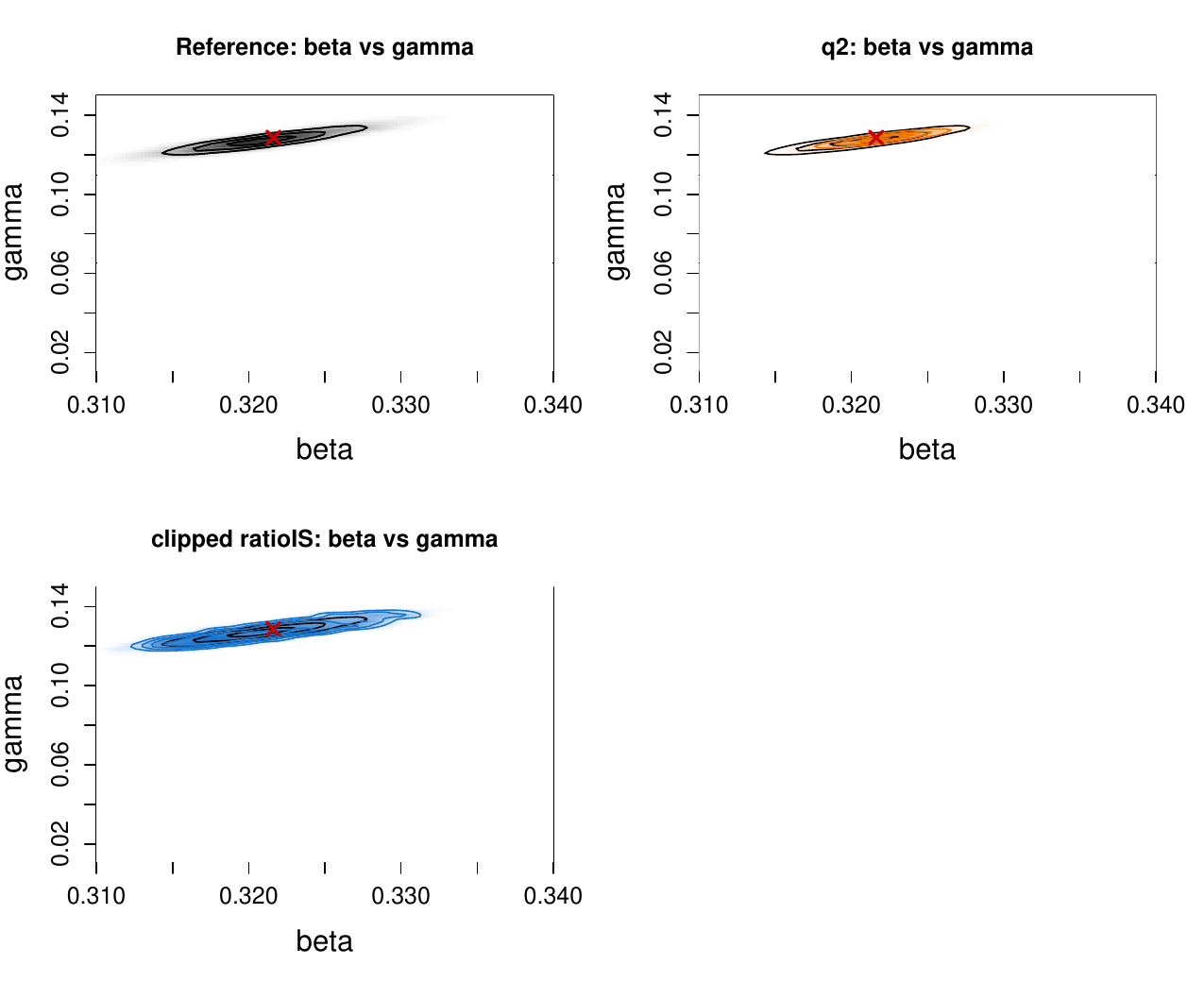}
        \caption{observation 6}
        \label{fig:sub6}
    \end{subfigure}

    \caption{SIR: marginal posteriors for observations 5 and 6. The reference posterior is in black, $\hat{q}_2(\thetaVec\mid \yobs)$ in yellow, and the clipped ratio-corrected posterior is in blue.}
    \label{fig:obs5-6}
\end{figure}

\begin{figure}[htbp]
    \centering

    \begin{subfigure}[t]{0.6\textwidth}
        \centering
        \includegraphics[width=\linewidth]{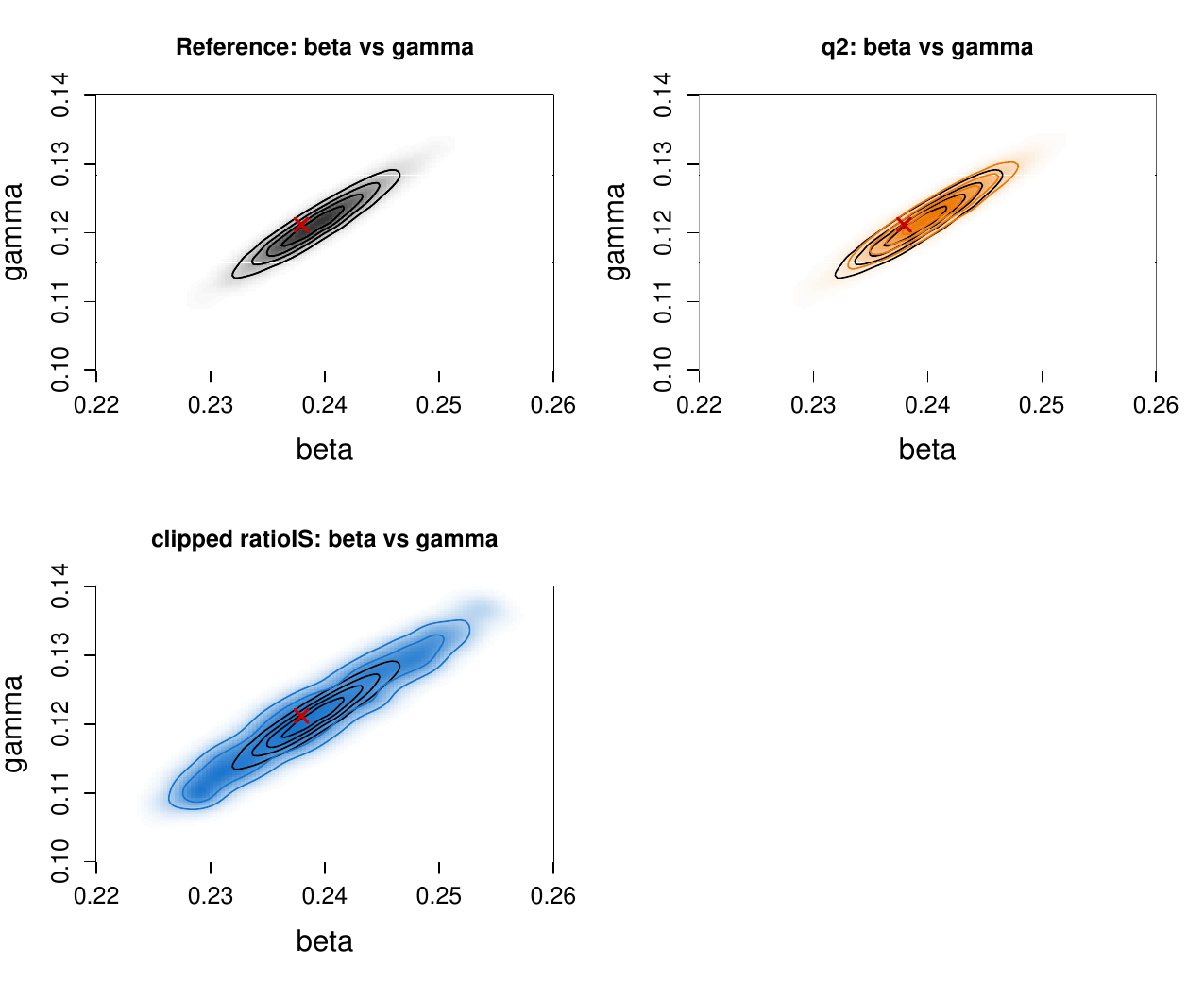}
        \caption{observation 7}
        \label{fig:sub7}
    \end{subfigure}
    \hfill
    \begin{subfigure}[t]{0.6\textwidth}
        \centering
        \includegraphics[width=\linewidth]{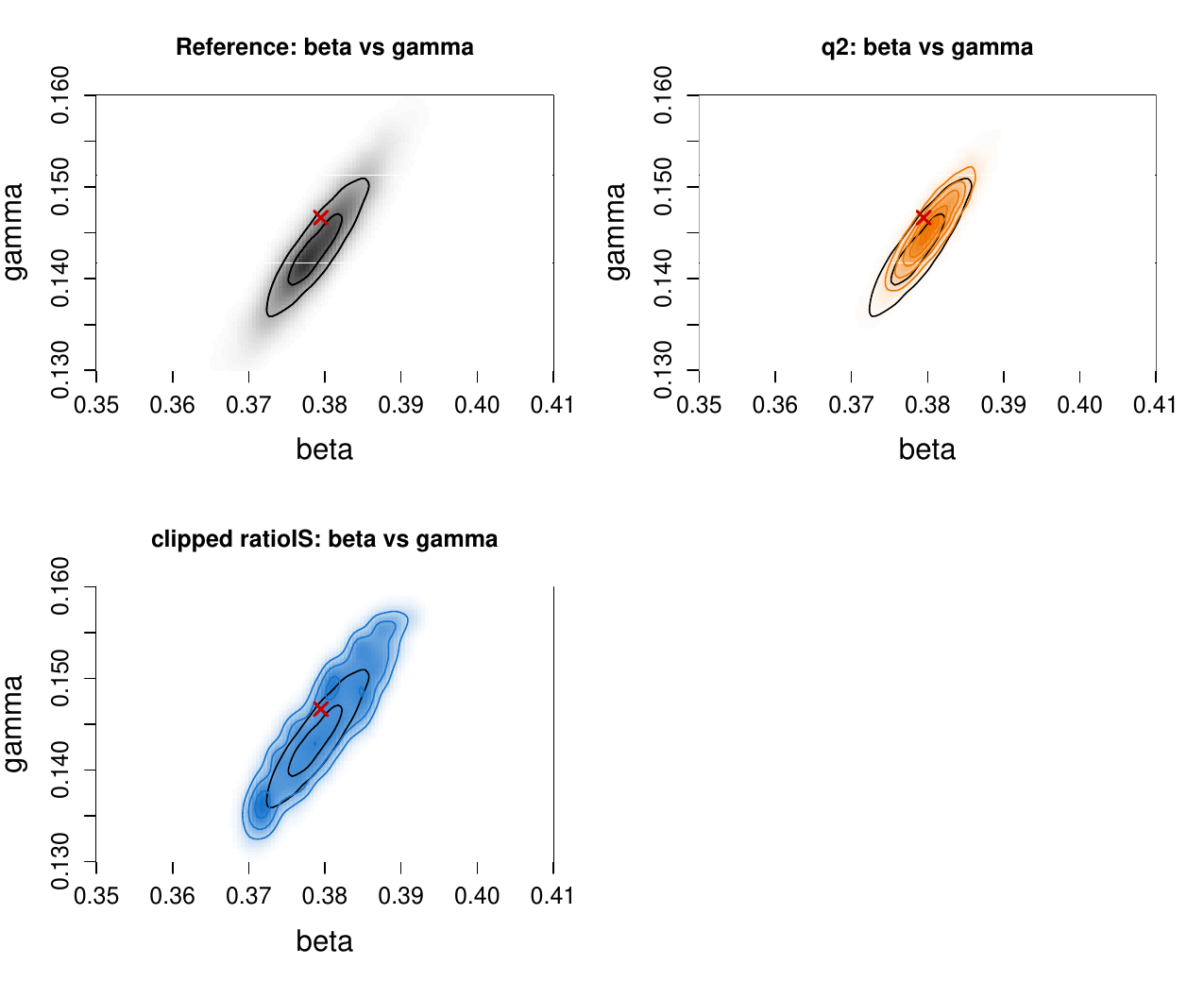}
        \caption{observation 8}
        \label{fig:sub8}
    \end{subfigure}

    \caption{SIR: marginal posteriors for observations 7 and 8. The reference posterior is in black, $\hat{q}_2(\thetaVec\mid \yobs)$ in yellow, and the clipped ratio-corrected posterior is in blue.}
    \label{fig:obs7-8}
\end{figure}

\begin{figure}[htbp]
    \centering

    \begin{subfigure}[t]{0.6\textwidth}
        \centering
        \includegraphics[width=\linewidth]{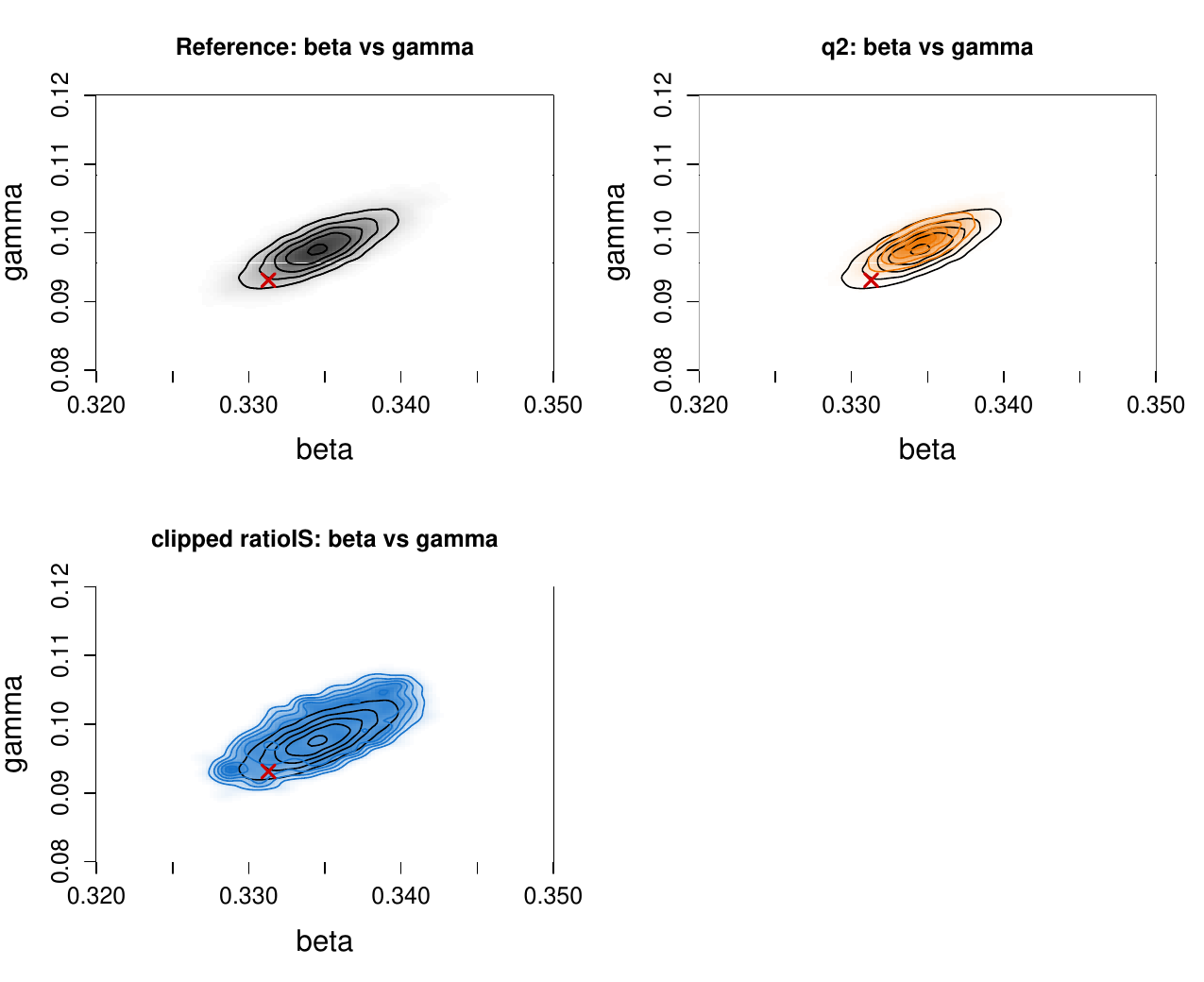}
        \caption{observation 9}
        \label{fig:sub9}
    \end{subfigure}
    \hfill
    \begin{subfigure}[t]{0.6\textwidth}
        \centering
        \includegraphics[width=\linewidth]{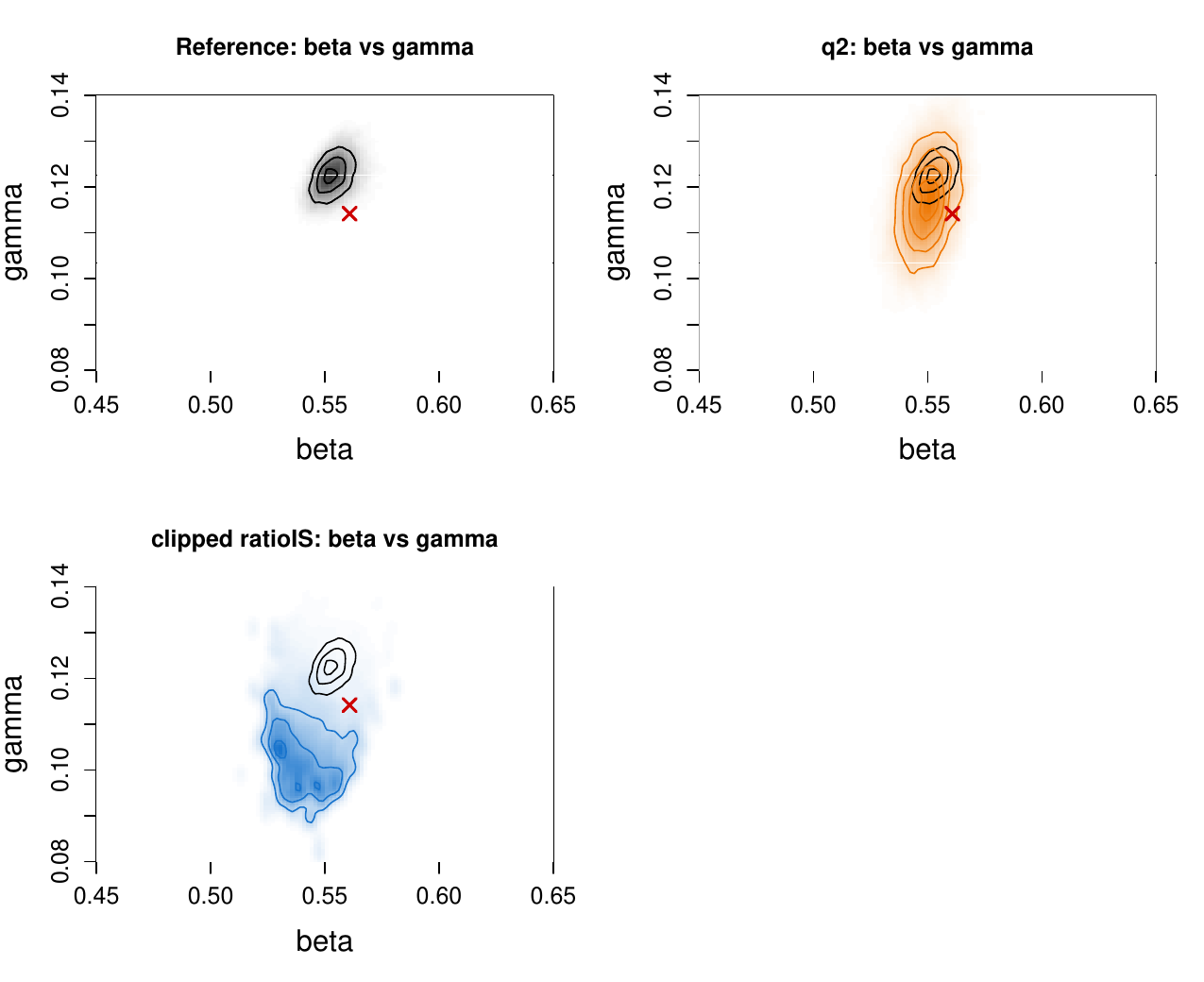}
        \caption{observation 10}
        \label{fig:sub10}
    \end{subfigure}

    \caption{SIR: marginal posteriors for observations 9 and 10. The reference posterior is in black, $\hat{q}_2(\thetaVec\mid \yobs)$ in yellow, and the clipped ratio-corrected posterior is in blue.}
    \label{fig:obs9-10}
\end{figure}

\begin{figure}[htbp]
    \centering

    \begin{subfigure}[t]{0.35\textwidth}
        \centering
        \includegraphics[width=\linewidth]{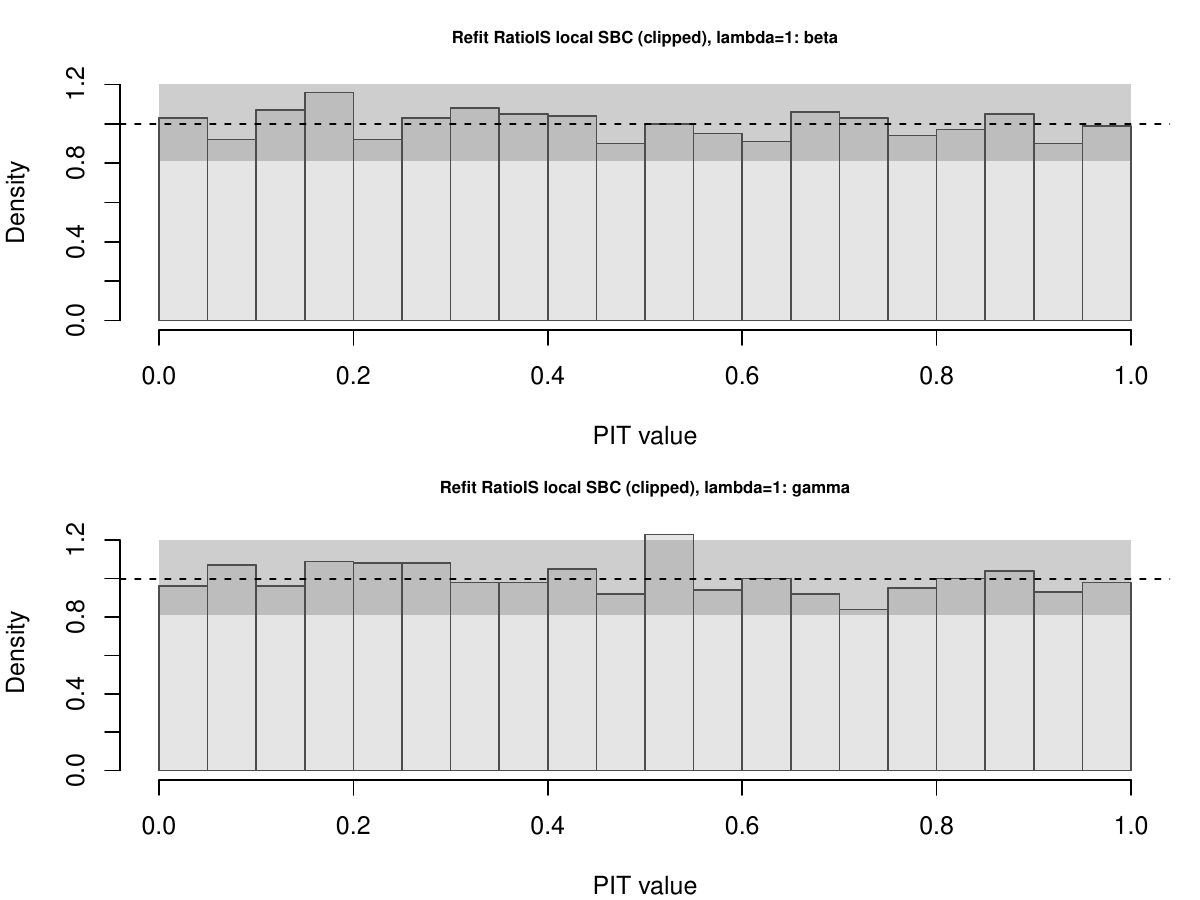}
        \caption{PIT histograms for $\beta$ (top) and $\gamma$ (bottom).}
    \end{subfigure}
\hspace{0.05\textwidth}
    \begin{subfigure}[t]{0.35\textwidth}
        \centering
        \includegraphics[width=\linewidth]{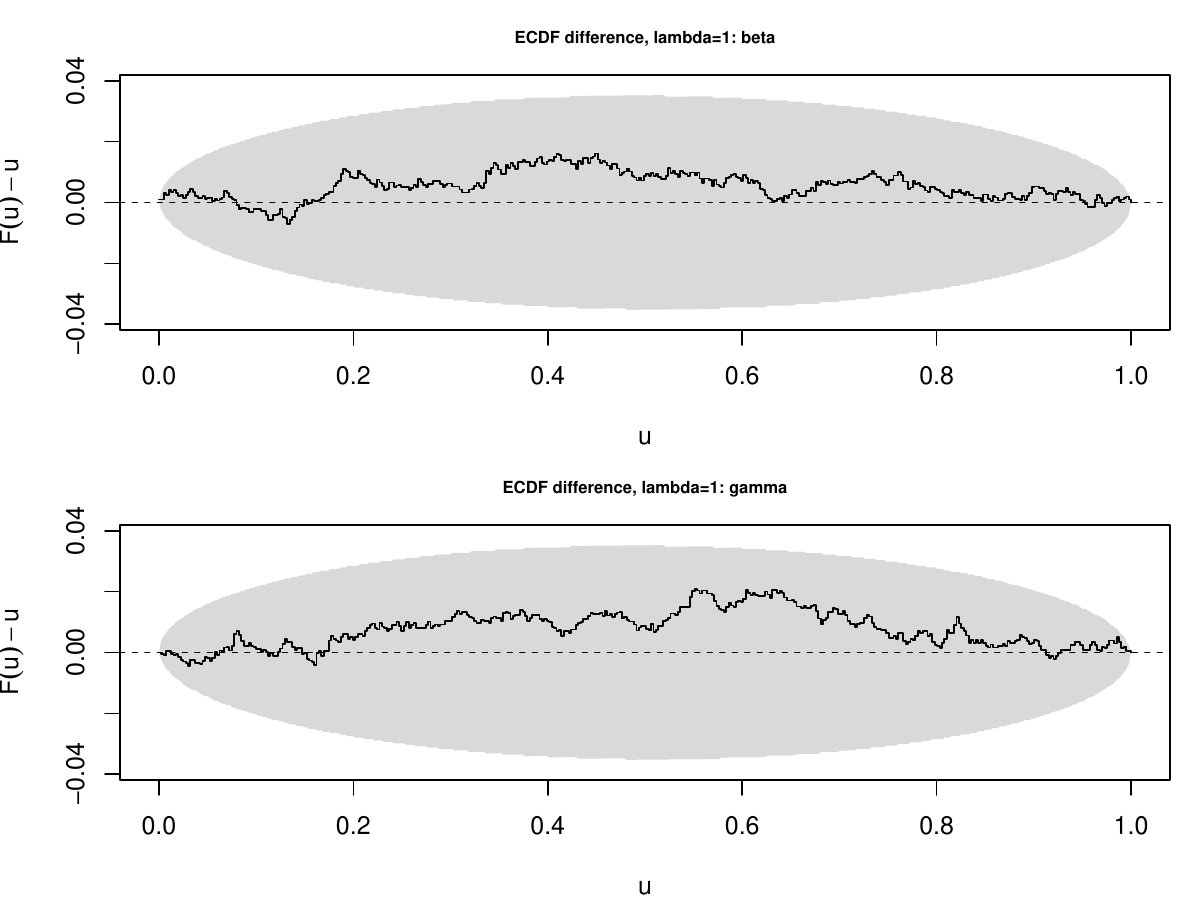}
        \caption{Difference ECDF of the PIT for $\beta$ (top) and $\gamma$ (bottom).}
    \end{subfigure}

    \caption{SIR: localized SBC when $\lambda=1$ for observation 2.}
    \label{fig:sir_obs2_sbc_lambda1}
\end{figure}

\begin{figure}[htbp]
    \centering

    \begin{subfigure}[t]{0.35\textwidth}
        \centering
        \includegraphics[width=\linewidth]{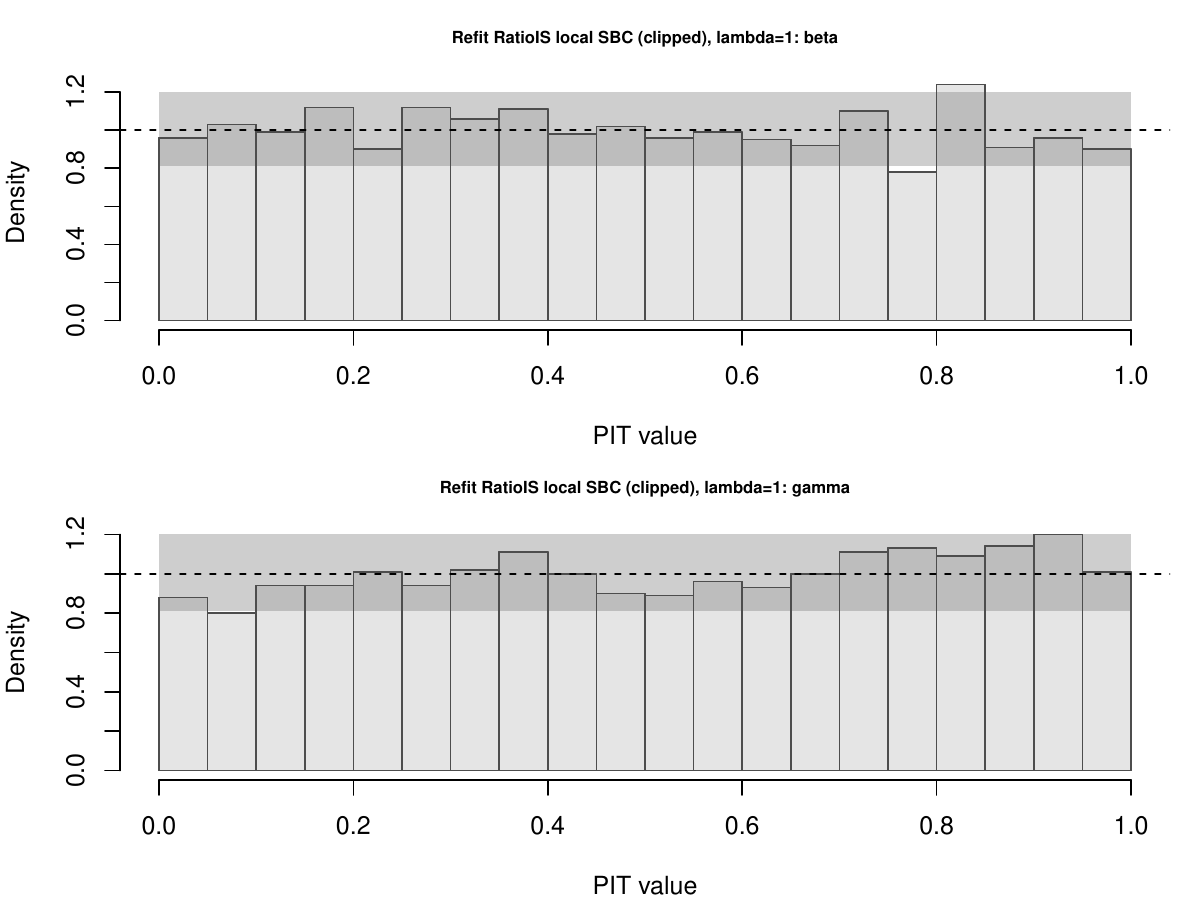}
        \caption{PIT histograms for $\beta$ (top) and $\gamma$ (bottom).}
    \end{subfigure}
\hspace{0.05\textwidth}
    \begin{subfigure}[t]{0.35\textwidth}
        \centering
        \includegraphics[width=\linewidth]{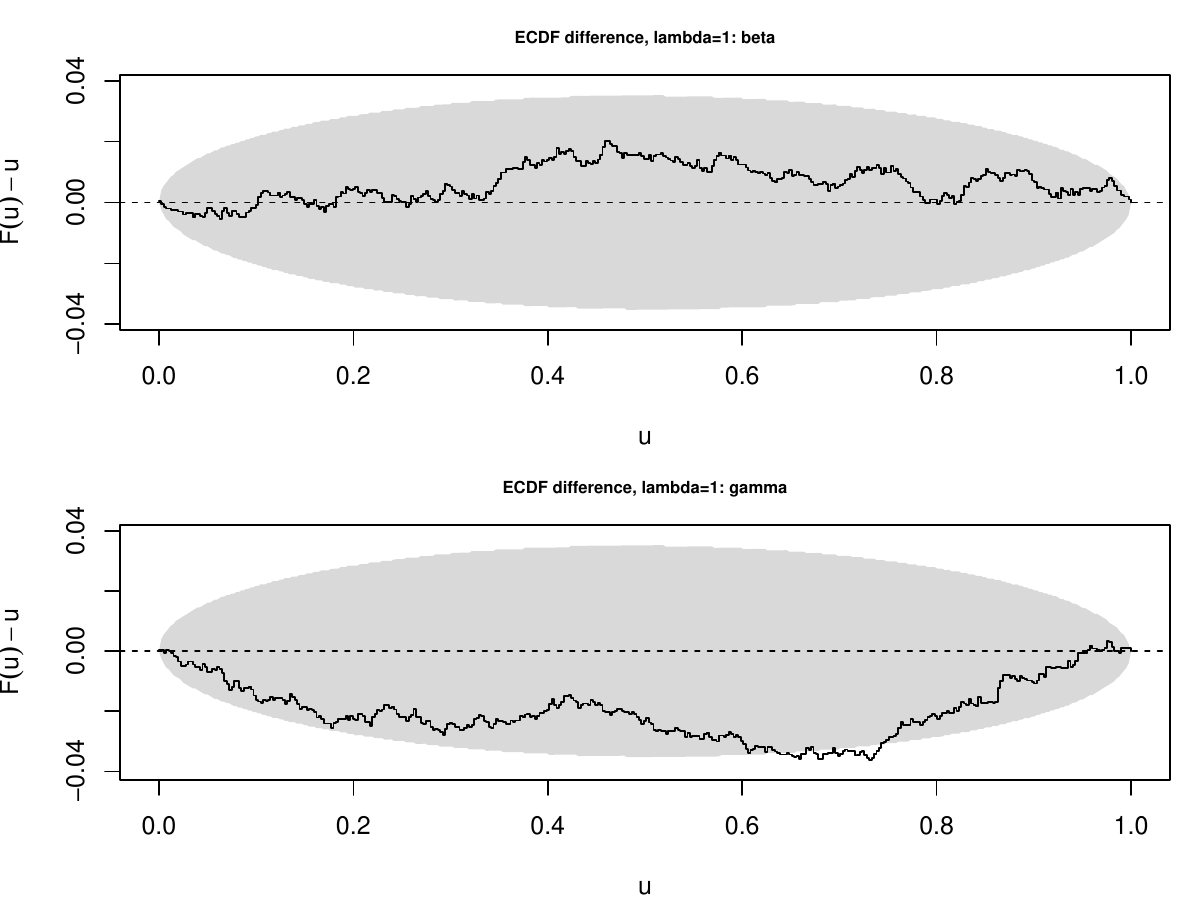}
        \caption{Difference ECDF of the PIT for $\beta$ (top) and $\gamma$ (bottom).}
    \end{subfigure}

    \caption{SIR: localized SBC when $\lambda=1$ for observation 3.}
    \label{fig:sir_obs3_sbc_lambda1}
\end{figure}

\clearpage
\newpage

\subsection{Lotka-Volterra model: further experimental results}

\begin{figure}[htbp]
    \centering

    \begin{subfigure}[t]{0.5\textwidth}
        \centering
        \includegraphics[width=\linewidth]{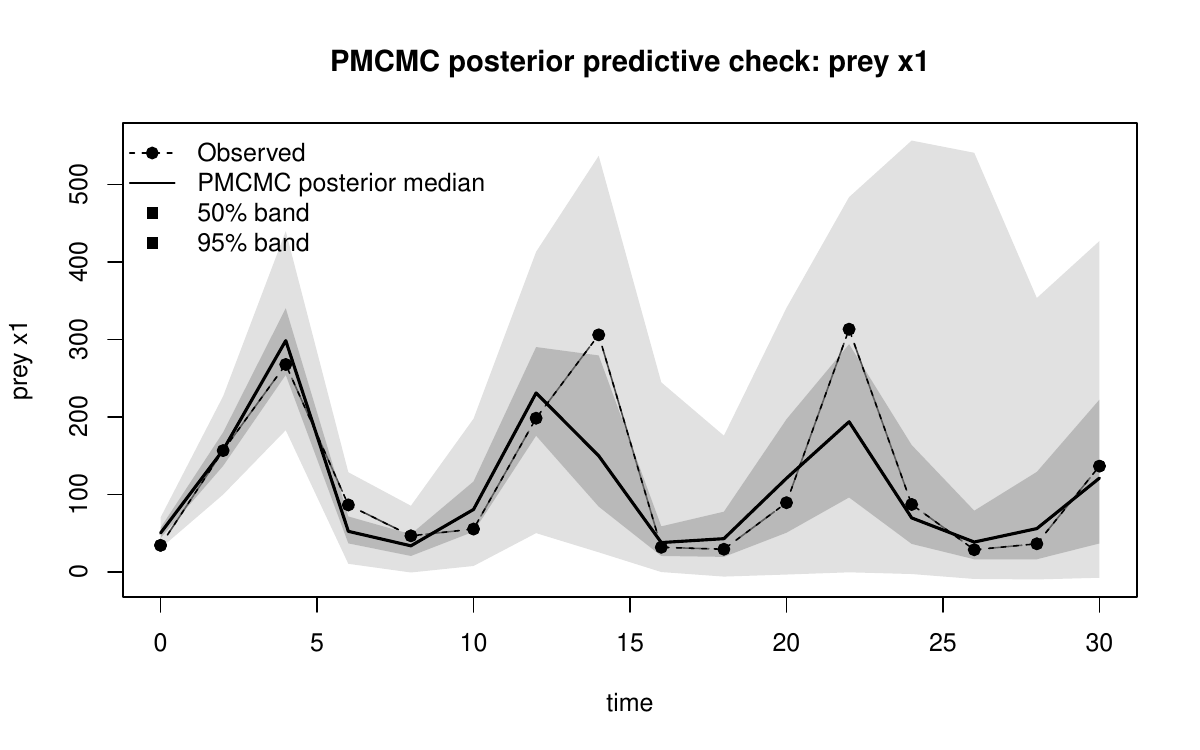}
        \caption{$X_1$.}
    \end{subfigure}
    \hfill
    \begin{subfigure}[t]{0.5\textwidth}
        \centering
        \includegraphics[width=\linewidth]{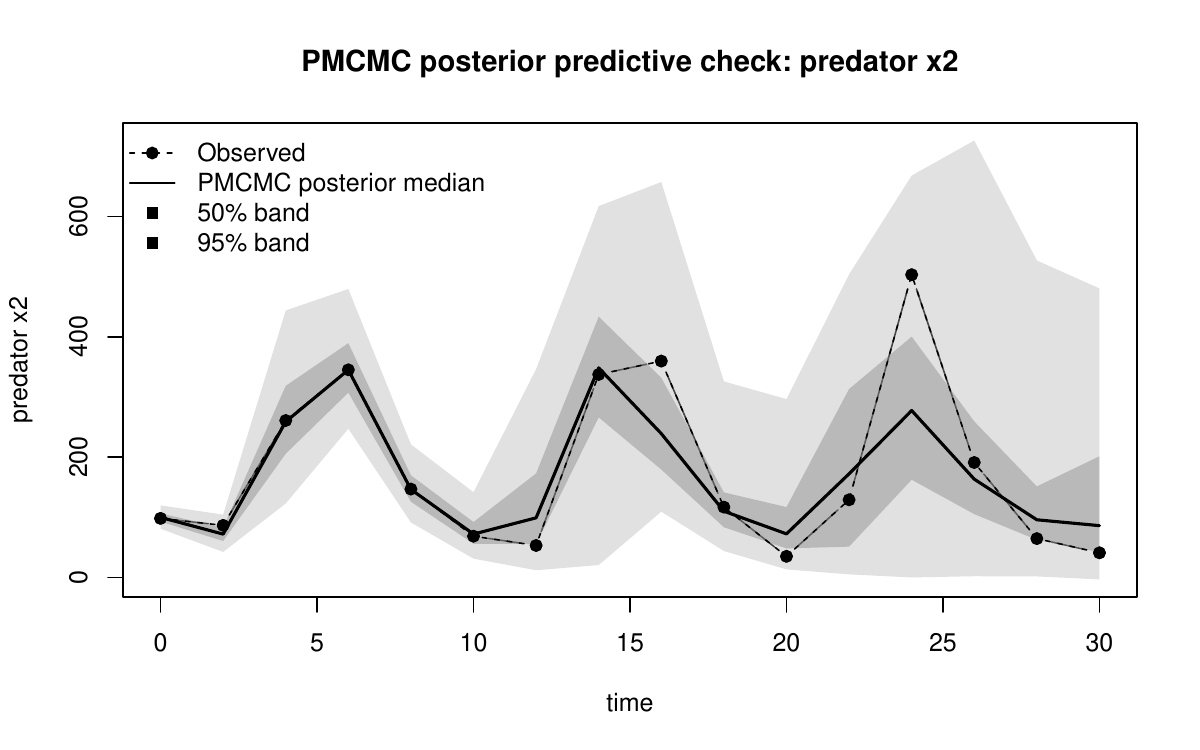}
        \caption{$X_2$.}
    \end{subfigure}

    \caption{Lotka-Volterra: posterior predictive trajectories produced with pseudomarginal (PMCMC) posterior samples. Circles are observed data, the bold solid line is the median of the simulated trajectories, the dark-gray region encloses 50\% posterior mass, and the light-gray region encloses 95\% mass.}
    \label{fig:lv_ppc_pmcmc}
\end{figure}

For Lotka-Volterra, the marginal posteriors are in the main paper. Figure \ref{fig:lv_ppc_pmcmc} shows posterior-predictive checks conditional on pseudomarginal MCMC inference (employing a bootstrap filter with 200 particles for each proposed parameter). The corresponding posterior-predictive checks based on the ratio-corrected posterior samples are in the main paper and they appear to be very similar to those in Figure \ref{fig:lv_ppc_pmcmc}.

\begin{figure}[htbp]
    \centering

    \begin{subfigure}[t]{0.4\textwidth}
        \centering
        \includegraphics[width=\linewidth]{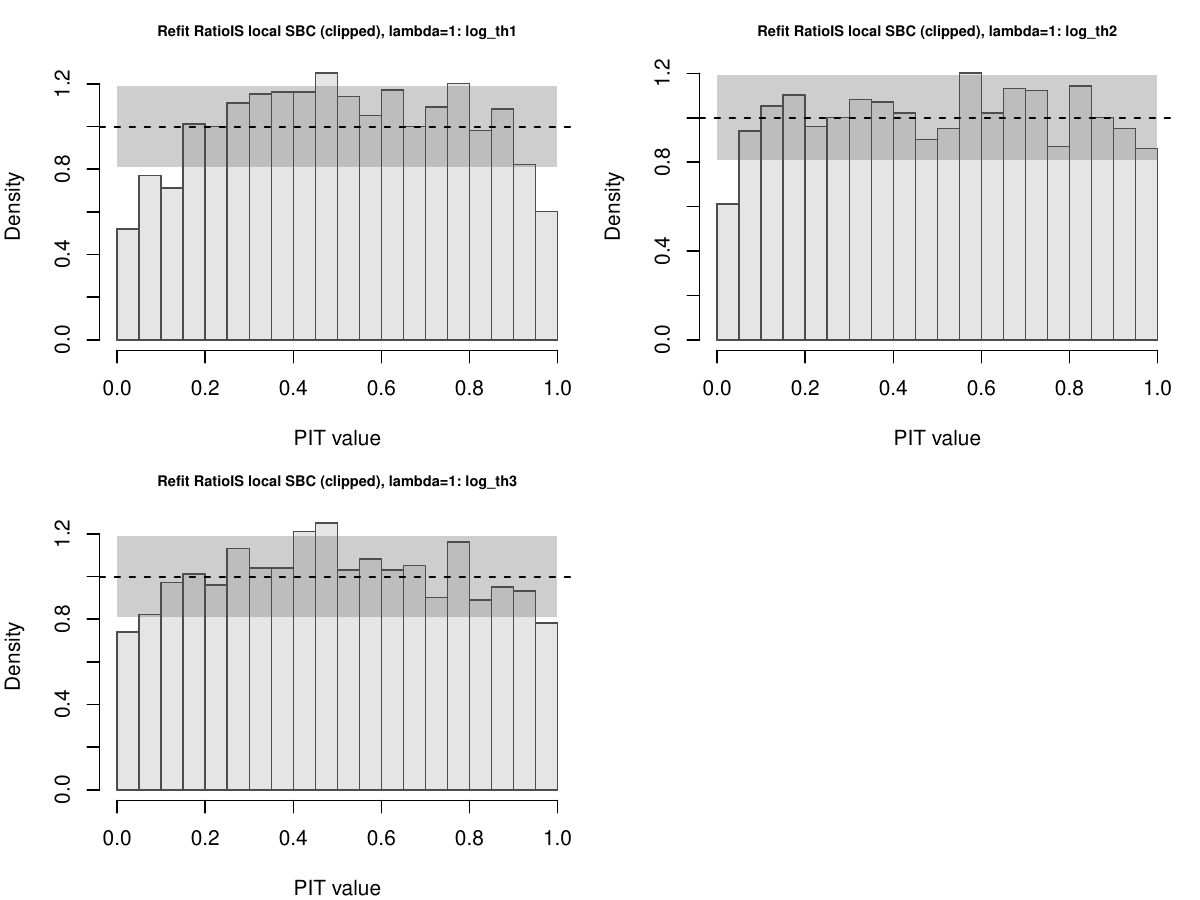}
        \caption{Histogram of the PIT, $\lambda=1$.}
    \end{subfigure}
    \begin{subfigure}[t]{0.4\textwidth}
        \centering
        \includegraphics[width=\linewidth]{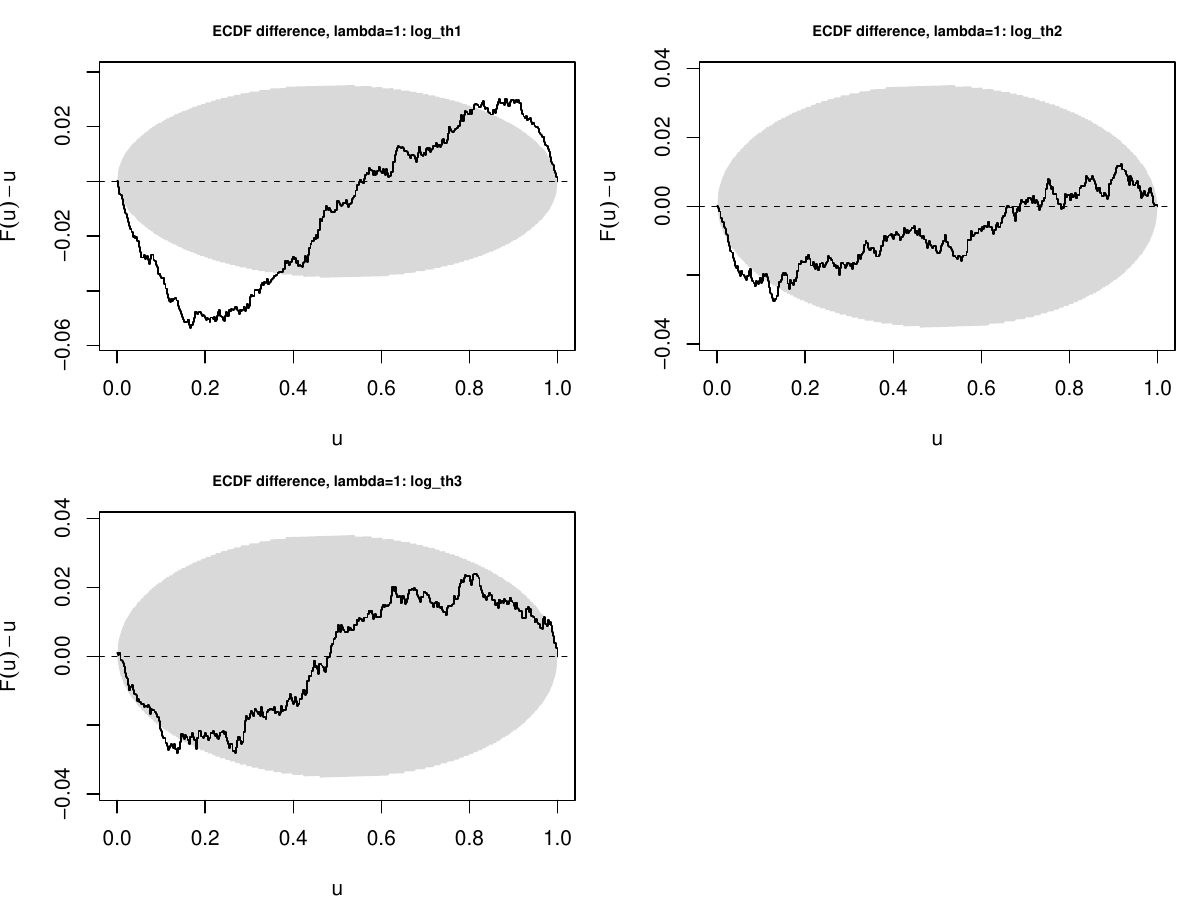}
        \caption{Difference ECDF of the PIT, $\lambda=1$.}
    \end{subfigure}

    \caption{Lotka-Volterra, localized SBC when $\lambda=1$.}
    \label{fig:lv_sbc_lambda1}
\end{figure}

When we run localized SBC, with $\lambda\in \{1,1.05,1.1,1.25,1.5,2\}$, despite the reassuring results above we observe miscalibration for $\theta_1$ even with $\lambda=1$, see Figure \ref{fig:lv_sbc_lambda1}. For $\lambda=1.05$ miscalibration extended to $\theta_2$. Results with larger $\lambda$ are not reported.

\begin{figure}[htbp]
    \centering

    \begin{subfigure}[t]{0.4\textwidth}
        \centering
        \includegraphics[width=\linewidth]{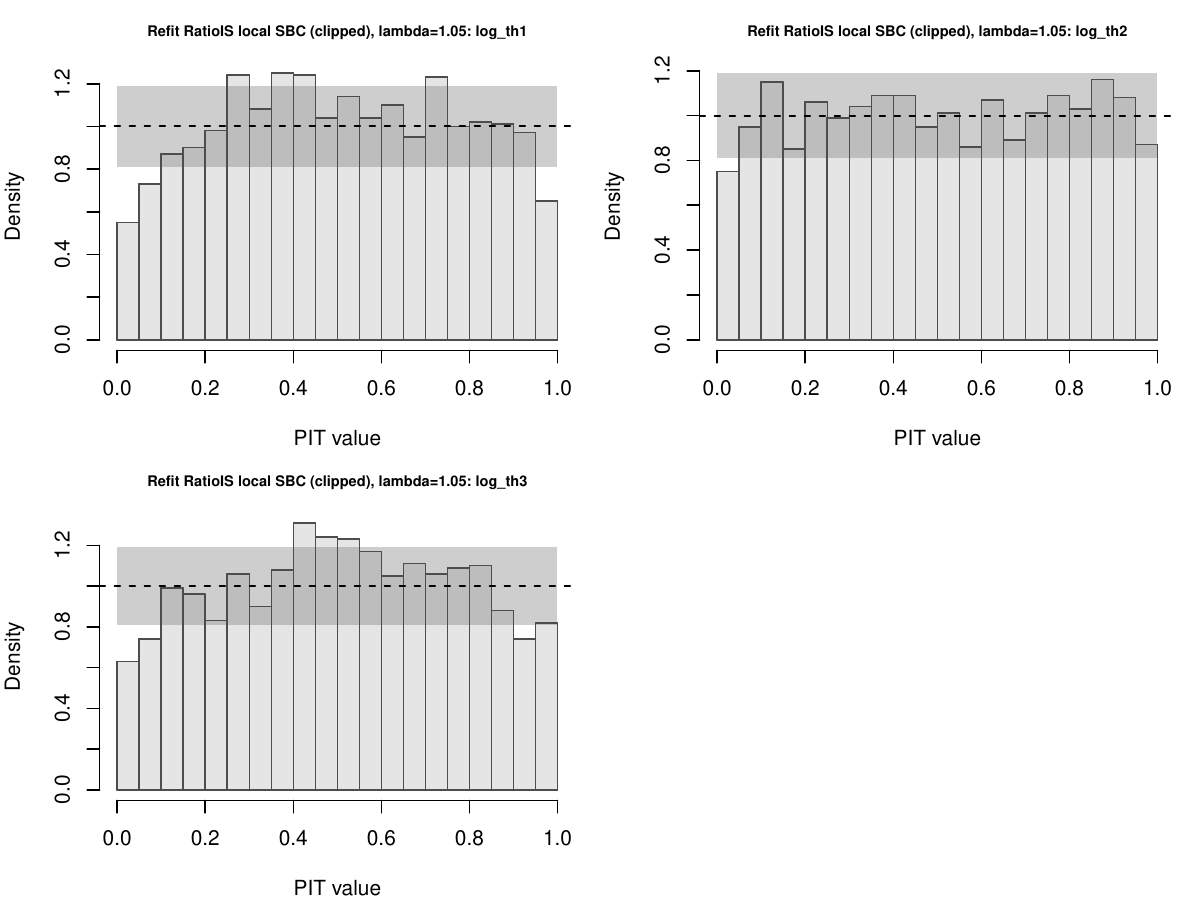}
        \caption{Histogram of the PIT, $\lambda=1.05$.}
    \end{subfigure}
    \begin{subfigure}[t]{0.4\textwidth}
        \centering
        \includegraphics[width=\linewidth]{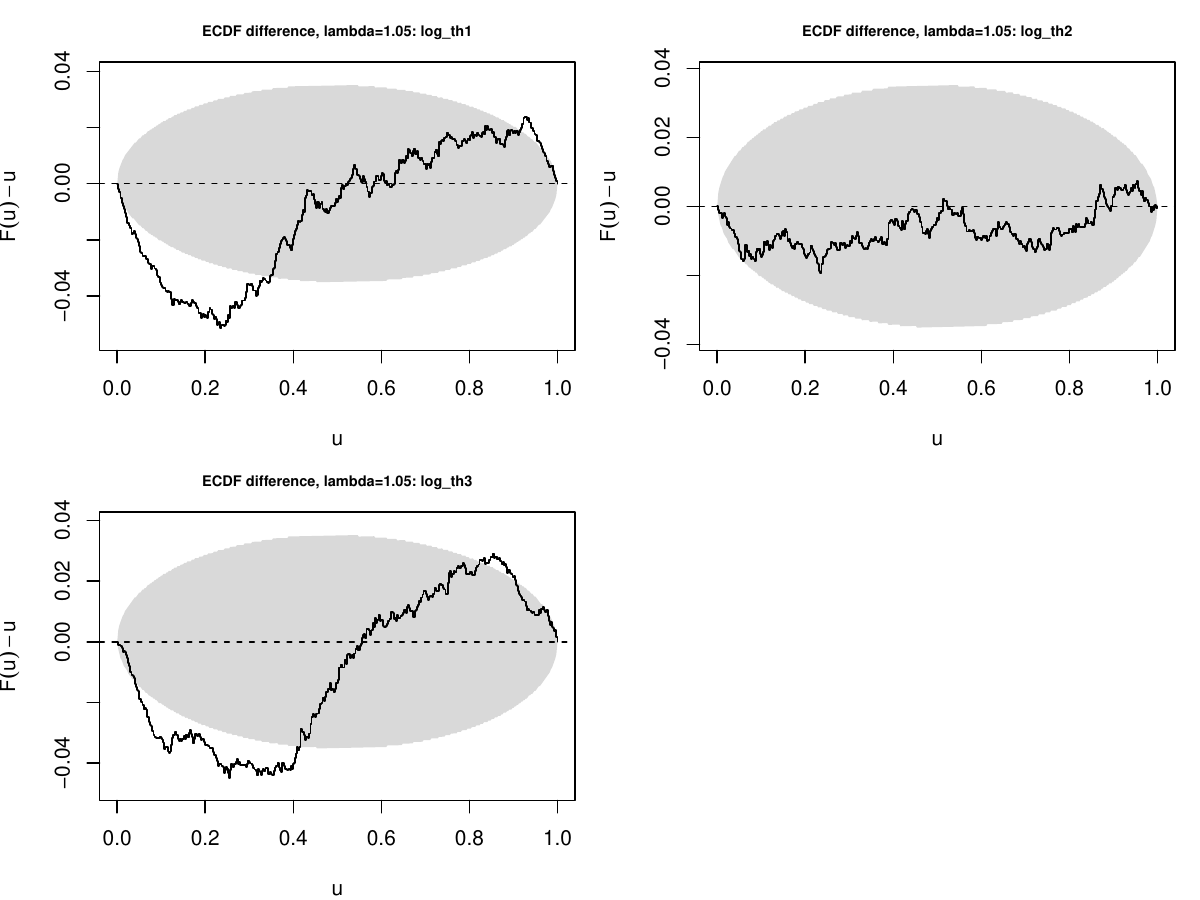}
        \caption{Difference ECDF of the PIT, $\lambda=1.05$.}
    \end{subfigure}

    \caption{Lotka-Volterra, localized SBC when $\lambda=1.05$.}
    \label{fig:lv_sbc_lambda1p05}
\end{figure}

\clearpage
\newpage

\subsection{Epidemiological model}

\subsubsection{Further model-specific details}

The latent colonization dynamics are described by the same SIR model as in \eqref{eq:sir-ode}, with parameters $(\beta,\gamma)$, again for a population size $P=10^6$, and set
$I(0)=1$, $R(0)=0$ and $S(0)=P-I(0)-R(0)$.
Following \cite{ojala2025basic}, rather than working directly with $(\beta,\gamma)$, inference is
performed using the epidemiologically more interpretable parametrization
$
R_0=\beta/\gamma$,
$\tau=\beta-\gamma,
$
where $R_0$ is the basic reproduction number and $\tau$ is the net
transmission rate. Therefore,
$
\gamma=\tau/(R_0-1)$,
$
\beta=\tau R_0/(R_0-1)$.
\\

\noindent
\textbf{Observational model:} the latent colonization trajectory is simulated at weekly resolution.
Let $I(t)$ denote the number of individuals colonized by ST131-A (which could also be ST131-C1 or ST131-C2) at week
$t$. The colonization process is linked to BSI incidence through the
clade-specific relative invasiveness parameter $\rho$, which is treated
as fixed and is obtained from the same external epidemiological
information used by \cite{ojala2025basic} (for clade A: $\rho=0.6931$, for clade C1 is 3.5985 and for clade C2 is 4.3577). Conditional on $I(t)$, a weekly
BSI count is generated according to
\[
Y(t)^{\mathrm{BSI}}
\sim
\operatorname{Poisson}(\mu(t)),
\]
with
\[
\mu(t)
=
0.1+
P\theta_{\mathrm{BSI}}
\frac{\rho I(t)}
     {(P-I(t))+\rho I(t)},
\]
where $\theta_{\mathrm{BSI}}$ denotes the population-level rate of
\textit{E.~coli} bloodstream infections. 
 The resulting weekly counts are subsequently
aggregated into yearly BSI counts: for example, for clade ST131-A which has 14 years of observations,
\[
Y_j
=
\sum_{t\in\mathcal T_j}
Y_t^{\mathrm{BSI}},
\qquad
j=1,\ldots,14,
\]
where $\mathcal T_j$ contains the 52 weeks belonging to year $j$.
A further unknown parameter, denoted by $D_t$, controls the temporal alignment
between the simulated BSI trajectory and the observation window. More
precisely, $D_t$ represents the delay, measured in weeks, between the
beginning of the simulated colonization trajectory and the beginning
of the observation window. The weekly BSI trajectory is shifted by
$D_t$ before the annual observations are extracted. Since the simulator
operates on a weekly time grid, the implemented shift is determined by
$\lfloor D_t\rfloor$. 
In summary, the data-generating model $\mathcal{M}(\thetaVec)$ we consider operates as follows
\[
(\tau,R_0)
\;\longrightarrow\;
(\beta,\gamma)
\;\longrightarrow\;
I(t)
\;\longrightarrow\;
Y^{\mathrm{BSI}}(t)
\;\xrightarrow{\;\text{shift by }D_t\;}
Y^{\mathrm{BSI}}_{\mathrm{shifted}}(t)
\;\longrightarrow\;
\text{annual observations}.
\]
Following 
\cite{ojala2025basic}, we employ a rescaled version of the delay
parameter, where they define
$
\widetilde D_t = 0.001 D_t,
$
and since the prior on the delay on its original scientific scale is
$
D_t\sim U(0,26),
$
the corresponding prior under the transformed parametrization is
$
\widetilde D_t\sim U(0,0.026).
$
The parameter vector to infer is therefore
$
\thetaVec=(\tau,R_0,\widetilde D_t).
$
Whenever the epidemiological simulator is evaluated, the delay is first
transformed back to its original scale in weeks according to
$
D_t = 1000\,\widetilde D_t,
$
and consequently 
$
p(\yVec\mid\thetaVec)
=
p\!\left(
\yVec\mid
\tau,R_0,D_t=1000\,\widetilde D_t
\right).
$

\subsubsection{Posterior predictive checks}

\begin{figure}[htbp]
        \centering
        \includegraphics[width=0.6\linewidth]{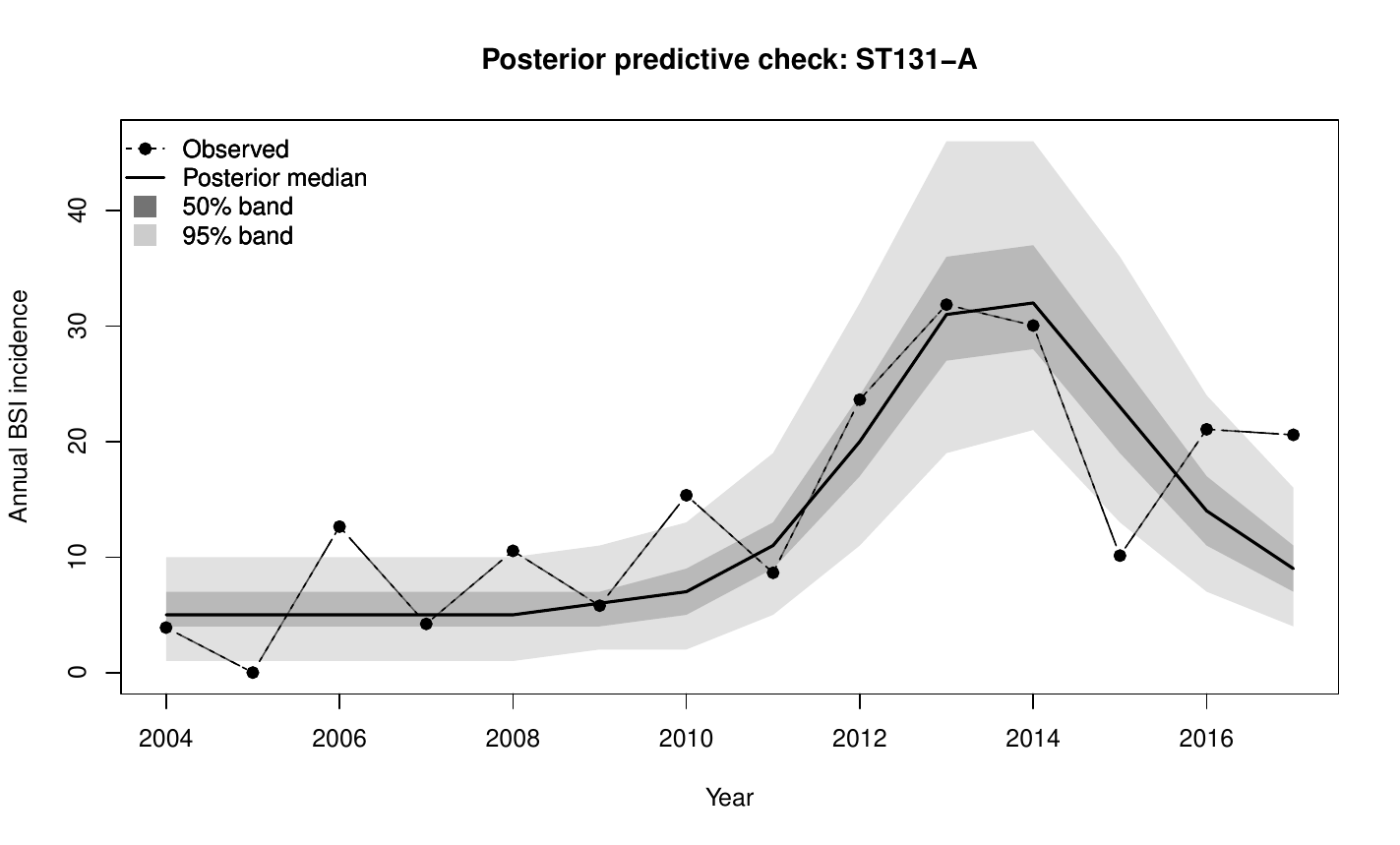}
    \caption{Epidemiological model, clade A: posterior predictive trajectories using the clipped ratio-corrected importance sampling. Circles are observed data, the bold solid line is the median of the simulated trajectories, the dark-gray region encloses 50\% posterior mass, and the light-gray region encloses 95\% mass.}
    \label{fig:epidem_ppc}
\end{figure}

We report posterior predictive simulations for clade A data, corresponding to clipped ratio-corrected posterior draws, in Figure \ref{fig:epidem_ppc}. These can be compared with Figure 4 (leftmost panel) in \cite{ojala2025basic}. While the fits are similar, we believe the one in Figure \ref{fig:epidem_ppc} to be slightly more compatible to the observed data.

\subsubsection{Marginal posteriors}

\begin{figure}[htbp]
    \centering
        \includegraphics[width=0.7\linewidth]{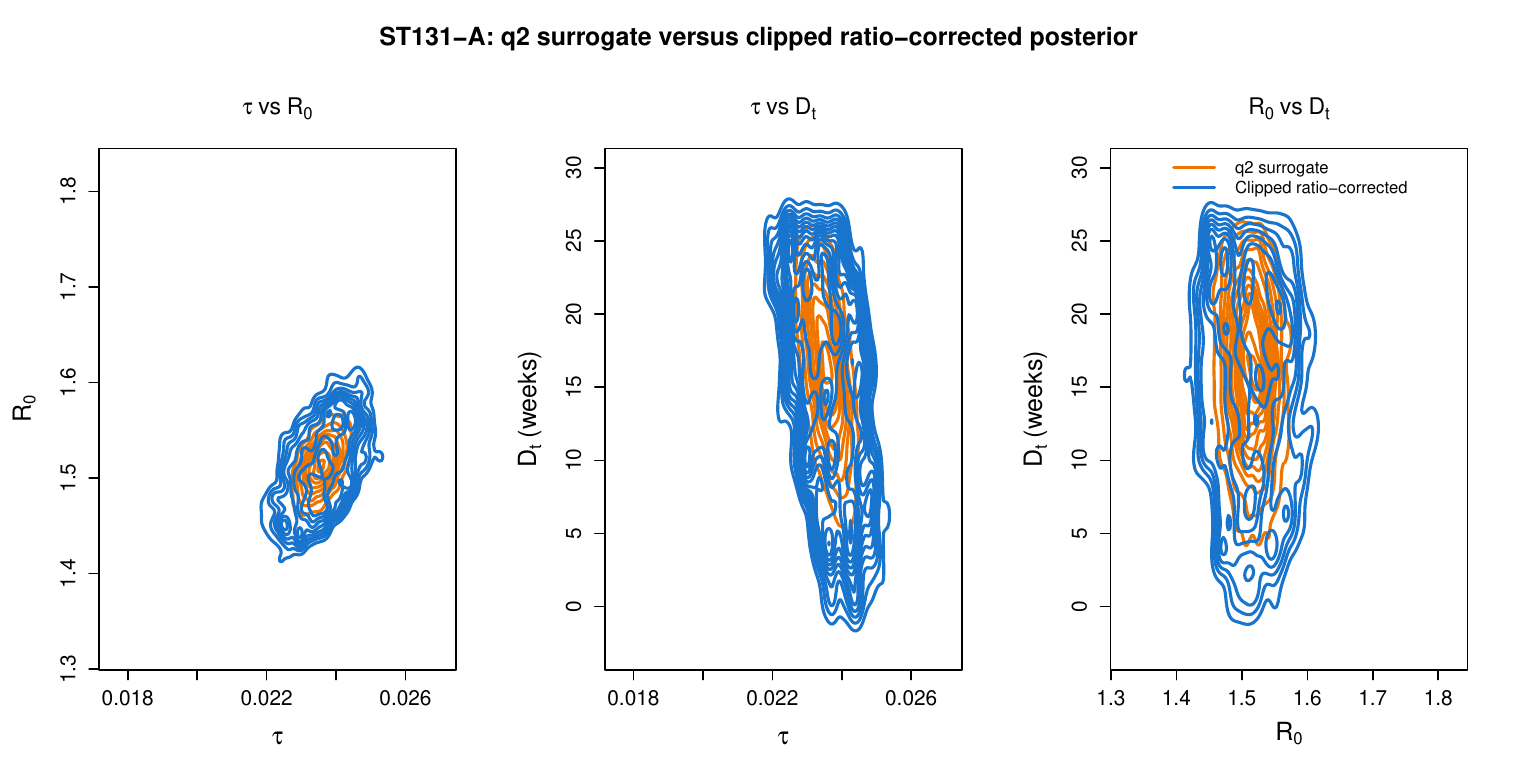}
    \caption{Epidemiological model, clade A: marginal posteriors, (left) $R_0$ vs $\tau$, (center) $D_t$ vs $\tau$, (right) $D_t$ vs $R_0$. Orange denotes  the uncorrected $\hat{q}_2$ surrogate. Blue denotes the clipped ratio-corrected posterior.}
    \label{fig:epidem_marginals}
\end{figure}

For the clade A data,  the clipped ratio-corrected marginal posteriors are in Figure \ref{fig:epidem_marginals} (notice that the posterior for $D_t$ does not actually reach negative values; that is merely an artifact of the Gaussian kernel smoothing approximation applied on posterior samples). Figure \ref{fig:epidem_marginals} is consistent with \cite{ojala2025basic} (compare with the leftmost panel in their Figure 3), that is inference for $D_t$ is uninformative, as its posterior covers the prior's support for the parameter. Therefore we recover their same broad scientific conclusion about the delay parameter: the data contain little information about \(D_t\).   Notice that, although the inference was conducted using $\tilde D_t$, for scientific interpretation we report graphical results by transforming back to the original $D_t$. Recall that the complete annual BSI trajectory is supplied to our GLLiM-based inference, whereas \cite{ojala2025basic} first compress the trajectory into a vector of three summary statistics for ABC inference. Such compression may discard information relevant to parameter inference and may consequently contribute to broader posterior distributions. 

\subsubsection{Localized SBC}

\begin{figure}[htbp]
    \centering

    \begin{subfigure}[t]{0.4\textwidth}
        \centering
        \includegraphics[width=\linewidth]{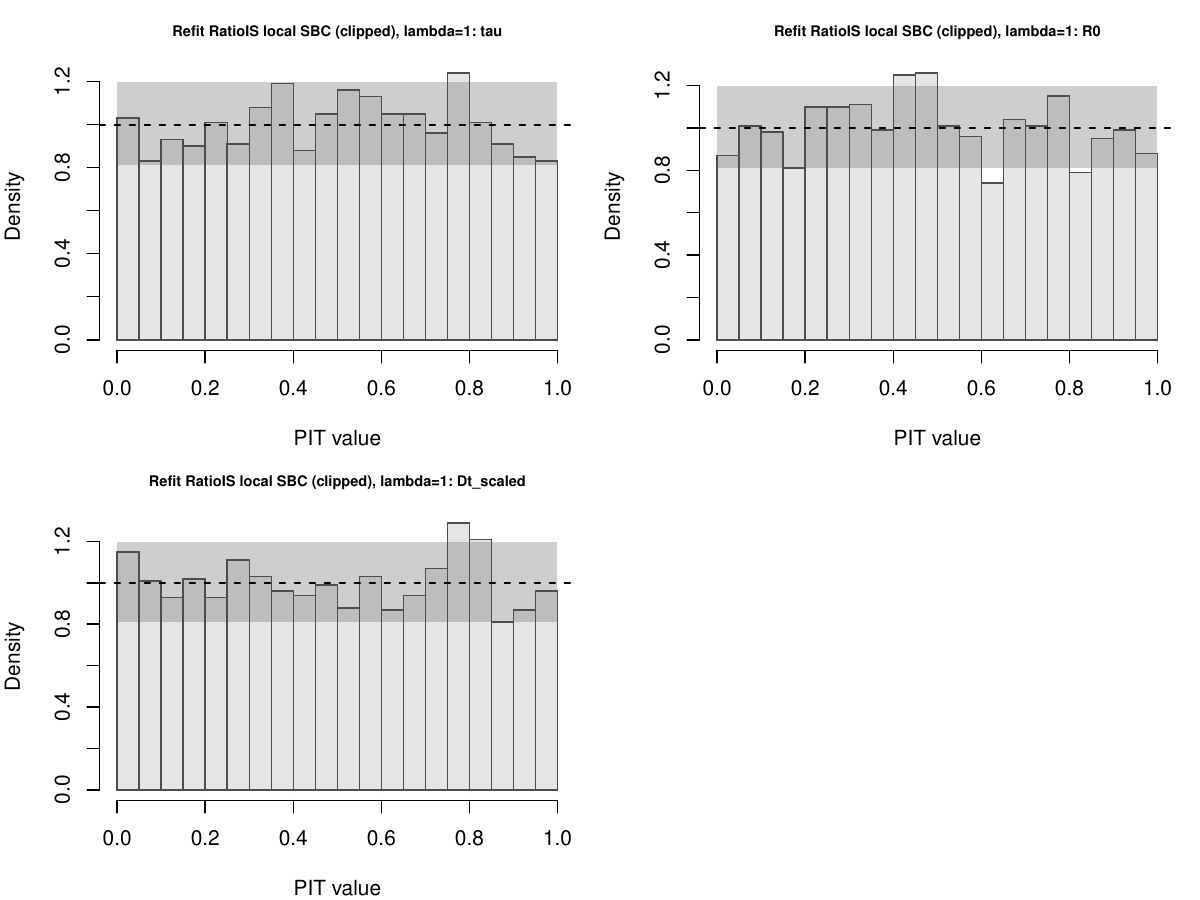}
        \caption{Histogram of the PIT, $\lambda=1$.}
    \end{subfigure}
    \begin{subfigure}[t]{0.4\textwidth}
        \centering
        \includegraphics[width=\linewidth]{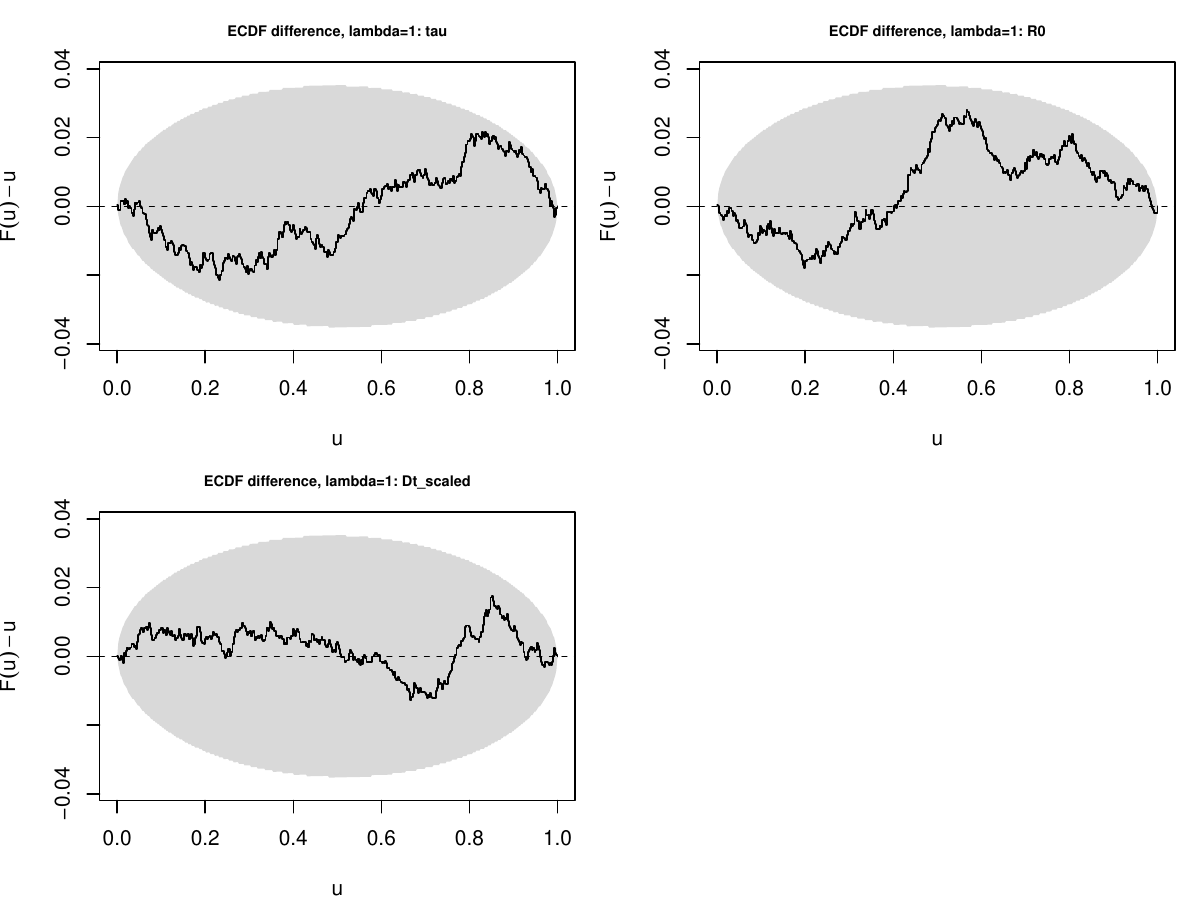}
        \caption{Difference ECDF of the PIT, $\lambda=1$.}
    \end{subfigure}

    \caption{Epidemiological model, localized SBC when $\lambda=1$ for clade A data. For both panels (a) and (b), the top left plot refers to $\tau$, the top right to $R_0$ and the bottom one to $\tilde{D}_t$.}
    \label{fig:epidem_sbc_lambda1}
\end{figure}

\begin{figure}[htbp]
    \centering

    \begin{subfigure}[t]{0.45\textwidth}
        \centering
        \includegraphics[width=\linewidth]{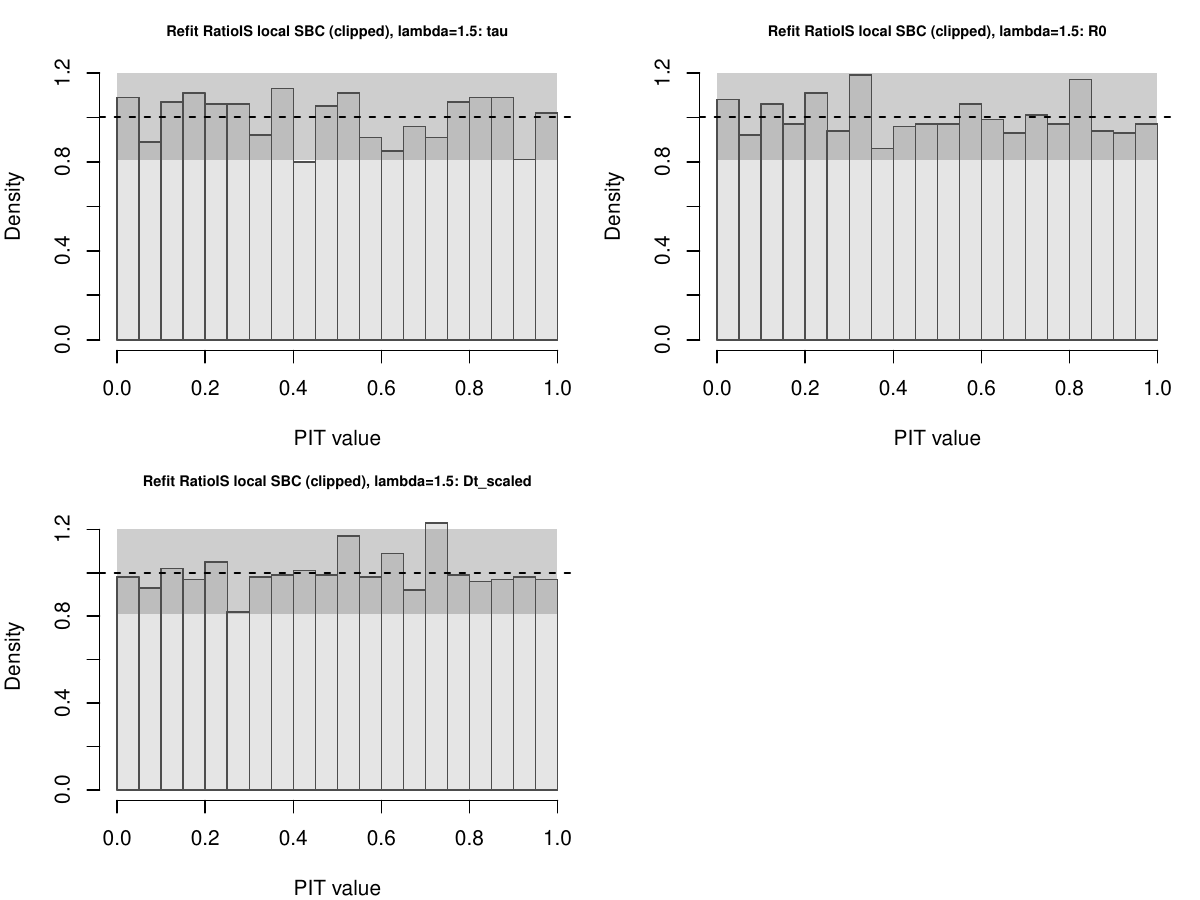}
        \caption{Histogram of the PIT, $\lambda=1.5$.}
    \end{subfigure}
    \begin{subfigure}[t]{0.45\textwidth}
        \centering
        \includegraphics[width=\linewidth]{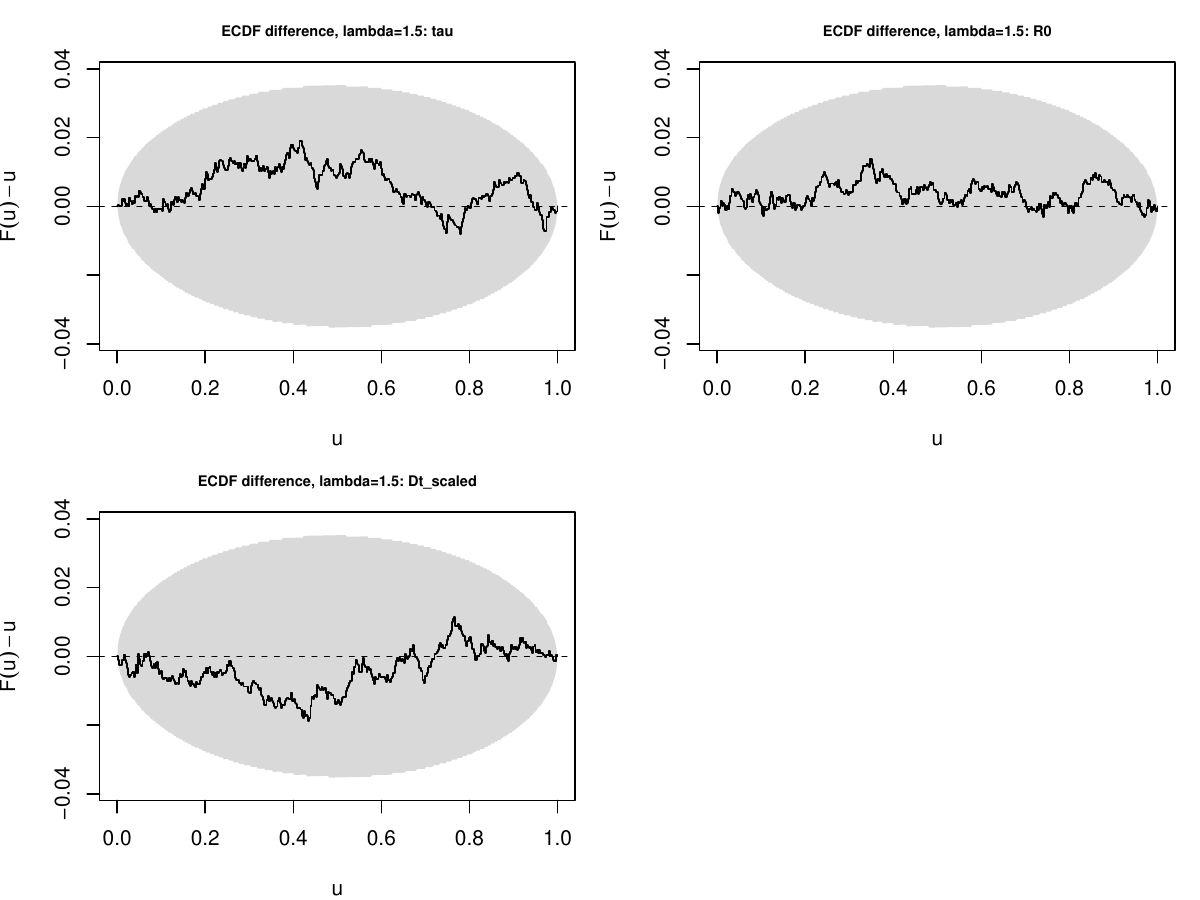}
        \caption{Difference ECDF of the PIT, $\lambda=1.5$.}
    \end{subfigure}

    \caption{Epidemiological model, localized SBC when $\lambda=1.5$ for clade A data. For both panels (a) and (b), the top left plot refers to $\tau$, the top right to $R_0$ and the bottom one to $\tilde{D}_t$.}
    \label{fig:epidem_sbc_lambda1dot5}
\end{figure}

\begin{figure}[htbp]
    \centering

    \begin{subfigure}[t]{0.45\textwidth}
        \centering
        \includegraphics[width=\linewidth]{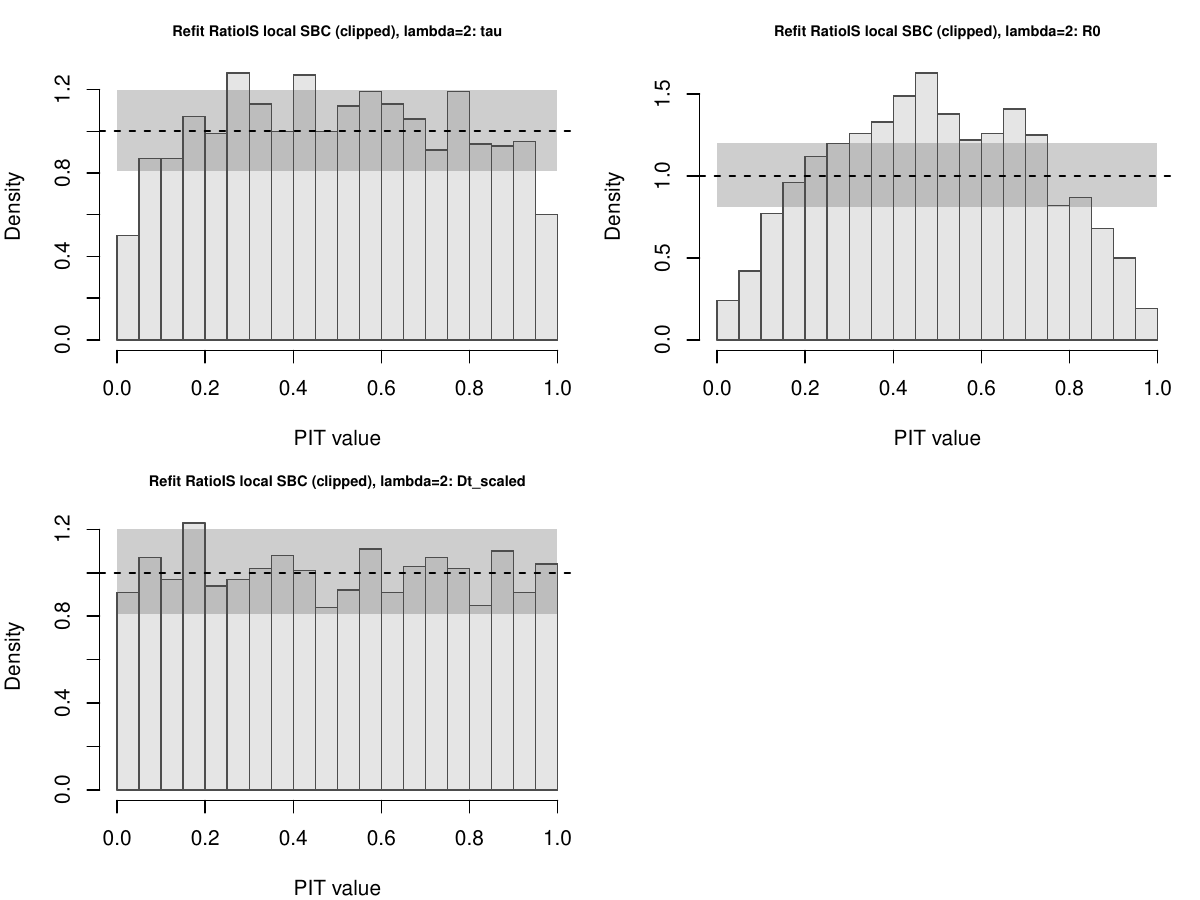}
        \caption{Histogram of the PIT, $\lambda=2$.}
    \end{subfigure}
    \begin{subfigure}[t]{0.45\textwidth}
        \centering
        \includegraphics[width=\linewidth]{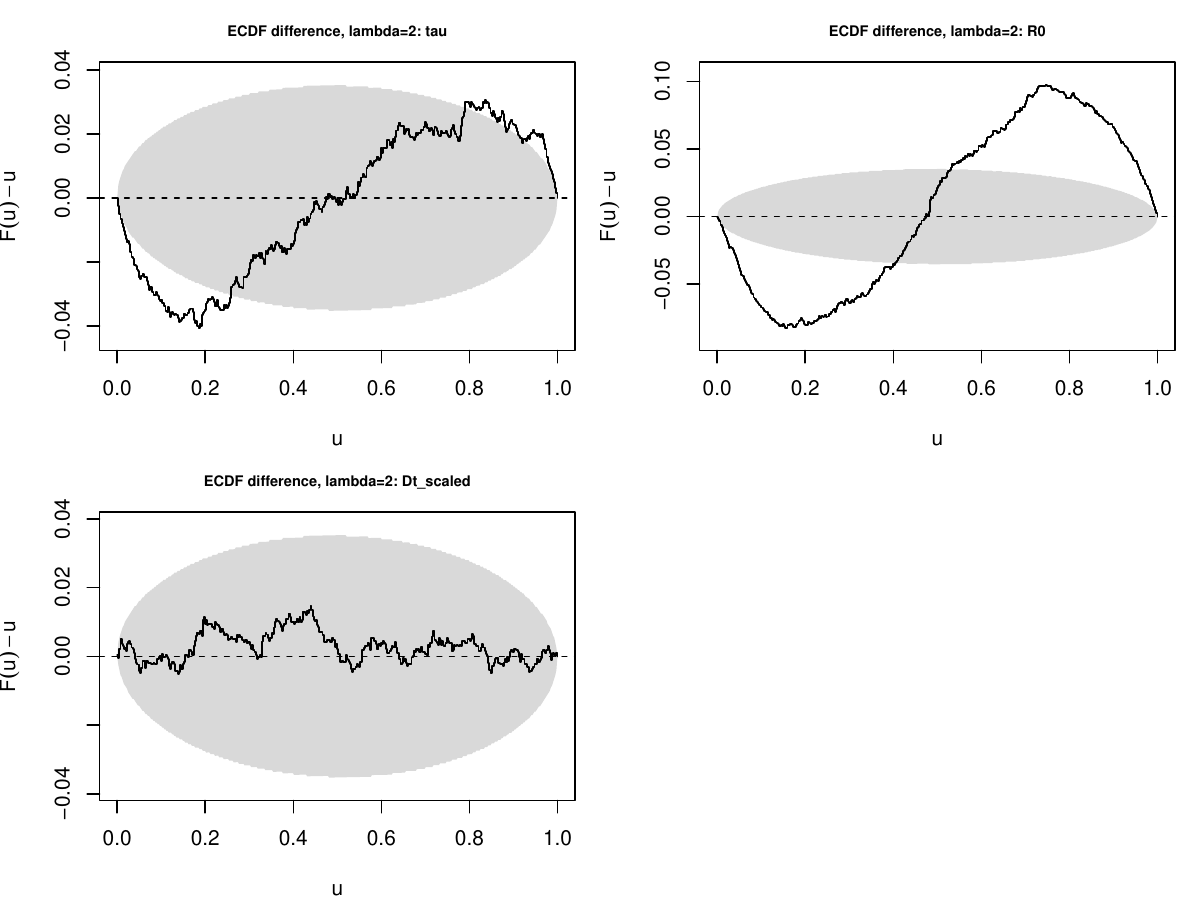}
        \caption{Difference ECDF of the PIT, $\lambda=2$.}
    \end{subfigure}

    \caption{Epidemiological model: localized SBC when $\lambda=2$ for clade A data. For both panels (a) and (b), the top left plot refers to $\tau$, the top right to $R_0$ and the bottom one to $\tilde{D}_t$.}
    \label{fig:epidem_sbc_lambda2}
\end{figure}

For the clade A data, we performed localized SBC for $\lambda\in\{1,1.05,1.1, 1.25, 1.5,2\}$ and we observed calibration for  all tested $\lambda$ except for $\lambda=2$. We illustrate the results for $\lambda=1,1.5$ and 2 in Figure \ref{fig:epidem_sbc_lambda1}, Figure \ref{fig:epidem_sbc_lambda1dot5} and Figure \ref{fig:epidem_sbc_lambda2} respectively.

\end{document}